%% file: M31_paper.tex
\RequirePackage[left]{lineno}
\documentclass[aps,prd,superscriptaddress,floatfix,nofootinbib,notitlepage]{revtex4}

\usepackage{graphicx}
\usepackage{subfigure}
\usepackage{amsmath}
\usepackage{longtable}
\usepackage{aas_macros}
\usepackage{xcolor}
\usepackage{multirow}

\graphicspath{{./figures/}}

\usepackage[sort&compress]{natbib}
\usepackage{hyperref}
\newcommand{\newtext}{\textcolor{black}}  

\input{fermiStyle.tex}

\begin{document}

\title{Search for $\gamma$-ray emission from dark matter particle interactions from Andromeda and Triangulum Galaxies with the \fermi\ Large Area Telescope}

\date{\today}

\author{Mattia Di Mauro}
\email{mdimauro@slac.stanford.edu}
\affiliation{NASA Goddard Space Flight Center, Greenbelt, MD 20771, USA}
\affiliation{Catholic University of America, Department of Physics, Washington DC 20064, USA}
\author{Xian Hou}
\email{xianhou.astro@gmail.com}
\affiliation{Yunnan Observatories, Chinese Academy of Sciences, 396 Yangfangwang, Guandu District, Kunming 650216, P. R. China}
\affiliation{Key Laboratory for the Structure and Evolution of Celestial Objects, Chinese Academy of Sciences, 396 Yangfangwang, Guandu District, Kunming 650216, P. R. China|}
\affiliation{Center for Astronomical Mega-Science, Chinese Academy of Sciences, 20A Datun Road, Chaoyang District, Beijing 100012, P. R. China}
\author{Christopher Eckner}
\affiliation{Center for Astrophysics and Cosmology, University of Nova Gorica, Vipavska 13, SI-5000 Nova Gorica, Slovenia}
\author{Gabrijela Zaharijas}
\affiliation{Center for Astrophysics and Cosmology, University of Nova Gorica, Vipavska 13, SI-5000 Nova Gorica, Slovenia}
\author{Eric Charles}
\affiliation{SLAC National Accelerator Laboratory, Stanford University, Stanford, CA 94305, USA}

\begin{abstract}
The Andromeda (M31) and Triangulum (M33) galaxies are the closest Local Group galaxies to the Milky Way, being only 785 and 870 kpc away.
These two galaxies provide an independent view of high-energy processes that are often obscured in our own Galaxy, including possible signals of dark matter (DM) particle interactions.
The {\it Fermi} Large Area Telescope ({\it Fermi}-LAT) preliminary eight year list of sources includes both M31, which is detected as extended with a size of about 0.4$^\circ$, and M33, which is detected as a point-like source.
The spatial morphology of M31 $\gamma$-ray emission could trace a population of unresolved sources and energetic particles originating in sources not related to massive star formation.  Alternatively, the $\gamma$-ray emission could also be an indication of annihilation or decay of DM particles.  
We investigate these two possibilities using almost 10 years of data from the {\it Fermi} LAT.
An interpretation that involves only a DM $\gamma$-ray emission is in tension with the current limits from other searches, such as those targeting Milky Way dwarf spheroidal galaxies.
When we include a template of astrophysical emission, tuned on $\gamma$-ray data or from observations of these galaxies in other wavelengths, we do not find any significant evidence for a DM contribution and we set limits for the annihilation cross section that probe the thermal cross section for DM masses up to a few tens of GeV in the $b\bar{b}$ and $\tau^+\tau^-$ channels. 
For models where the DM substructures have masses above $10^{-6}$ solar masses our limits probe the DM interpretation of the {\it Fermi} LAT Galactic center excess.
We provide also the lower limit for the DM decay time assuming the same spatial models of the DM distribution in M31 and M33.

\end{abstract}
\maketitle

\section{Introduction}\label{sec:intro}

It has been clear for many decades that the Universe includes a significant component of matter which is not made of any known particles of the Standard Model of particle physics. 
This form of matter is called dark matter (DM) because, though solid observational evidence exists for its gravitational influence from the earliest moments of the Universe's history \cite{Zwicky:1933gu,Rubin:1980zd,Olive:2003iq,Ade:2013zuv}, no direct measurements have been made so far of its particle nature. 

Though by no means the only possibility, a theoretically well motivated class of DM models include interactions between DM and Standard Model particles that are approximately as strong as the weak nuclear force, and a mass of similar scale ($\sim$10--1000  GeV). Such weakly interacting massive particles (WIMPs) would generically attain the observed DM density after thermal freeze-out in the early Universe. The canonical ``WIMP'' is a $\sim$100 GeV particle interacting through the $SU(2)_L$ weak force, although many other candidates have been proposed with a wide range of masses and interaction strengths \cite{Feng:2008ya,Steigman:2012nb}.

The WIMP models provide a useful benchmark for DM searches designed to look for the present-day pair annihilation (or decay) of DM particles in regions of high density of DM. 
A thermally-averaged annihilation cross section of $\langle \sigma v\rangle \sim 3 \times 10^{-26}$~cm$^3$/s would provide approximately the correct WIMP relic density at present, and so experiments capable of seeing the present-day annihilation of DM with cross sections near this value have the sensitivity to either \newtext{confirm} or exclude a large number of theoretically interesting models.

In terms of couplings to Standard Model particles, there are many possibilities for dominant annihilation (or decaying) channels. Of particular interest is annihilation or decay resulting in monochromatic $\gamma$ rays, as this signature is more easily distinguished from other astrophysical sources. However, as this is a loop interaction (DM does not couple to photons directly), this channel is suppressed; thus, searches for this signature are challenging. 
In addition to the direct annihilation to pairs of $\gamma$-ray photons, if DM annihilates into pairs of other Standard Model particles, 
the resulting hadronization and/or decay will result in a continuum of $\gamma$ rays observable from Earth with an energy distribution that extends up to the rest mass of the DM particle. 
$\gamma$ rays produced in the local Universe are relatively unaffected by their propagation in the interstellar medium; thus they arrive at the Earth unscattered and unattenuated.
This allows us to trace the $\gamma$-ray emission spatial distribution and spectral information back to its original source.   Thus, $\gamma$-ray observations together with separate information or assumptions about the distribution of DM in the regions under study and models for the hadronization allow for measurement of, or determination of upper limits for, the annihilation cross section. 

With this motivation in mind, $\gamma$-ray data measured by the Large Area Telescope (LAT) carried by the {\it Fermi Gamma-ray Space Telescope} ({\it Fermi}-LAT) are of great interest. 
The {\it Fermi}-LAT is a pair-conversion telescope. 
Incoming $\gamma$ rays pass through the anti-coincidence detector and convert in a Silicon strip tracker to $e^{+}/e^{-}$ pairs. 
The charged particle direction is reconstructed using the information in the tracker, and the energy is estimated from depositions in a CsI calorimeter.  
Detailed descriptions of the LAT and its performance can be found in dedicated papers~\cite{Atwood:2009ez,2013arXiv1303.3514A}.
At the present time, the {\it Fermi}-LAT is one of the most sensitive instruments to DM particles with weak-scale mass and producing $\gamma$ rays. 
Analysis of the LAT $\gamma$-ray data can place strong limits on---or discover---DM  
annihilation with cross sections near the canonical thermal value into a wide variety of Standard Model particles \cite{Charles:2016pgz}. 

%
%

A large number of DM searches have been performed using {\it Fermi}-LAT data. 
Since annihilation rates are proportional to the square of the DM density, lower annihilation cross sections can be probed by targeting regions in the local Universe with the greatest densities of DM, such as the center of the Milky Way (MW)~\cite{Goodenough:2009gk,Hooper:2010mq,Boyarsky:2010dr,Abazajian:2012pn,Huang:2013pda,Gordon:2013vta,Daylan:2014rsa}, 
satellite dwarf spheroidal galaxies of the MW~\cite{Ackermann:2011wa,Ackermann:2013yva,GeringerSameth:2011iw,Geringer-Sameth:2014qqa,Ackermann:2015zua,Geringer-Sameth:2015lua,Drlica-Wagner:2015xua}, 
unresolved halo substructures~\cite{Ackermann:2012nb,Belikov:2011pu,Berlin:2013dva,Buckley:2010vg}, 
galaxy clusters~\cite{Ackermann:2010rg,Dugger:2010ys}, 
and the Large and Small Magellanic Clouds (LMC and SMC)~\cite{Buckley:2015doa, Caputo:2016ryl}. 

The direction with the predicted brightest $\gamma$-ray emission from DM is towards the Galactic center.
The LAT observations of the Galactic center indicate that the region is brighter than expected from standard models for Galactic diffuse emission at GeV energies, and the spatial distribution is broadly consistent with our expectations for a DM signal (see, e.g., \cite{TheFermi-LAT:2017vmf}). 
However, previously unconsidered astrophysical backgrounds could match the observed morphology and spectrum and the true source of the $\gamma$ rays remains a subject of much debate, with an unresolved population of millisecond pulsars (MSPs) or a past transient event at the Galactic center being some of the most popular interpretations~\cite{Abazajian:2010zy,Wharton:2011dv,Hooper:2013nhl,Carlson:2014cwa,Petrovic:2014xra,Petrovic:2014uda,2013MNRAS.436.2461M}.
Considering both the broad interest in indirect searches for DM, and the current questions raised by the Galactic center excess, it is important to identify new high-density targets for DM annihilation indirect searches. 

Nearby galaxies, such as Andromeda (M31) and Triangulum (M33) offer a great opportunity to test the origin of the Galactic center excess and to look for signals of DM particles.
They are close enough (approximately 785 kpc and 870 kpc away respectively\footnote{https://ned.ipac.caltech.edu}) so that their stellar disks and bulges can be resolved as two separate components between radio and X-ray energies which is not possible in our Galaxy, since our bulge is obscured by the bright disk emission. It is worth noting that Andromeda in particular was one of the astrophysical objects where compelling evidence for the existence of DM was first brought forward, and shaped our understanding of the Universe (see e.g. \cite{Bertone:2016nfn}).

M31 was first detected in $\gamma$ rays by the LAT at a significance of 5.3 standard deviations ($\sigma$) with a marginal spatial extension (significance of $\sim$1.8$\sigma$)~\cite{2010M31}. 
M31, with its disk, has an angular size of over 3$^{\circ}$, and is therefore one of the rare nearby galaxies that can be spatially resolved by {\it Fermi}-LAT.
A more recent analysis~\cite{M31Fermi} revisited M31 with more than 7 years of Pass 8 observations detected the galaxy at a significance of nearly 10$\sigma$ and confirmed the spatial extension at 4$\sigma$ significance with a size of about $0\fdg4$. M31 has been detected with a similar size of extension also in \cite{2018ApJS..237...32A}.
Its spectrum is consistent with a power law with an index of $2.4\pm0.1$. The spatial distribution of the emission is consistent with a uniform-brightness disk over the plane of sky with a radius of 0$^{\circ}$.4 and no offset from the center of M31. 
The flux from M31 appears confined to the inner regions of the galaxy and does not fill the disk of the galaxy or extend far from it. 
Since the spatial morphology of the $\gamma$-ray signal is not compatible with the M31 disk, which is a region rich of gas and star formation activity, the emission probably is not interstellar in origin, unless the energetic particles radiating in $\gamma$ rays do not originate in recent star formation activities. 
Alternative interpretations are that the emission results from a population of MSPs located in the bulge of M31 \cite{Eckner:2017oul} or from the decay or annihilation of DM particles, similar to what has been proposed to account for the Galactic center excess found in {\it Fermi}-LAT observations of the MW.

On the other hand, M33 was not detected significantly by the LAT team in~\cite{M31Fermi} in an analysis using only $\gamma$ rays with energies above 1 GeV, and has been detected as point-like in the {\it Fermi}-LAT 8 year source list\footnote{\url{https://fermi.gsfc.nasa.gov/ssc/data/access/lat/fl8y/gll_psc_8year_v5.fit}} (FL8Y) with a significance of about $4.1\sigma$.


M31 and M33 are thus natural targets for DM indirect detection searches. As a general strategy of this work, we will apply the techniques developed in the search for DM in the LMC and SMC~\cite{Buckley:2015doa,Caputo:2016ryl}, which are both extended $\gamma$-ray sources as is M31.

In this paper we analyze almost 10 years of {\it Fermi}-LAT data, about two more years than previous analysis, and we dedicate our search to any evidence of a possible DM contribution.

In Sec.~\ref{sec:dm}, we describe the DM distribution in M31 and M33 and how it relates to searches for indirect signals of DM annihilation or decay.
In Sec.~\ref{sec:analysis}, we discuss the analysis setup and technique that we apply, we show how the DM signal from these galaxies would be detected by the LAT and we describe a search for such a signal in the M31 and M33 regions.
Finally in Sec.~\ref{sec:results} we report the results for the annihilation and decay of DM particles and we conclude in Sec.~\ref{sec:conclusion}.

\section{M31 and M33 dark matter flux and spatial distribution}
\label{sec:dm}

The flux spectrum $\textrm{d}\phi/\textrm{d}E_\gamma$ of $\gamma$ rays originating in DM decay/annihilation processes can be factored into  astrophysics- and particle physics-dependent terms~\cite{Ullio:2002pj}:
\begin{eqnarray}
\textrm{decay}:\quad & \frac{\textrm{d}\phi}{\textrm{d}E_\gamma} = \left(\frac{x}{4\pi\tau_{\chi}} \frac{\textrm{d}N_\gamma}{\textrm{d}E_\gamma} \frac{1}{m_{\chi}} \right)\left(\int_{\Delta\Omega} \textrm{d}\Omega	\int_{\rm l.o.s.}\textrm{d}\ell\;\rho_{\chi}(\vec{\ell}) \right)\textrm{,}\\
\textrm{annihilation}:\quad & \frac{\textrm{d}\phi}{\textrm{d}E_\gamma} = \left(\frac{x \langle \sigma v\rangle}{8\pi} \frac{\textrm{d}N_\gamma}{\textrm{d}E_\gamma} \frac{1}{m_{\chi}^2} \right)\left(\int_{\Delta\Omega} \textrm{d}\Omega	\int_{\rm l.o.s.}\textrm{d}\ell\;\rho^2_{\chi}(\vec{\ell}) \right)\textrm{.} \label{eq:dmflux}
\end{eqnarray}
The quantities in the first parentheses are the DM decay rate $\tau_{\chi}$ or the thermally-averaged annihilation cross section with respect to the velocity distribution of DM particles $\langle \sigma v\rangle$, respectively. Moreover, there is the differential yield of $\gamma$ rays from a single DM annihilation $\textrm{d}N_\gamma/\textrm{d}E_\gamma$, the mass of the DM particle $m_\chi$, and a normalization factor $x$ which is unity if the DM is its own antiparticle and $1/2$ otherwise. All of these depend on the unknown particle physics model that includes DM particles. The typical approach for DM indirect detection searches, as we will follow here, is to set an upper (lower) bound, if no excess is observed, on $\langle \sigma v\rangle$ ($\tau_{\chi}$) as a function of the DM mass $m_\chi$ while assuming a particular annihilation (decay) channel and its associated spectrum $\textrm{d}N_\gamma/\textrm{d}E_\gamma$.

In this paper, we assume $x=1$ and consider the final states $b\bar{b}$ and $\tau^+ \tau^-$, which have been of particular interest given the Galactic center excess.
Other sets of Standard Model final states are possible, but have sufficient similarity to the channels selected that bounds can be extrapolated reasonably. In this work, we calculate the spectrum $\textrm{d}N_\gamma/\textrm{d}E_\gamma$ for each chosen final state and DM mass using a code available as part of the {\it Fermi}-LAT {\em ScienceTools}.\footnote{The {\tt DMFitFunction} spectral model described at 
\url{http://fermi.gsfc.nasa.gov/ssc/data/analysis/documentation/Cicerone/Cicerone_Likelihood/Model_Selection.html}, see also Ref.~\cite{Jeltema:2008hf}.}  
Note that our implementation does not include electroweak corrections~\cite{2007PhRvD..76f3516K,2002PhRvL..89q1802B,2009PhRvD..80l3533K,2010PhRvD..82d3512C,Ciafaloni:2010ti}. 
Such corrections can be important for heavy DM ($m_\chi \gtrsim 1$~TeV); in any case, they would increase the resulting flux and, thus, strengthen the resulting bounds~\cite{Ciafaloni:2010ti,Cirelli:2010xx, Ackermann:2015tah}.  

In order to describe experimental results in terms of the particle physics parameters, the astrophysical quantities in the second set of parentheses in Eq.~\ref{eq:dmflux} must be known. The integral of the DM density along the line of sight and over a solid angle $\Delta\Omega$ 
corresponding to the region under study, is known as the $J-$factor (or $D-$factor in the case of decaying DM), and encapsulates the dependence of an indirect detection search on the distribution of DM in the search target. Of particular interest is the case of annihilating DM where the $J-$factor depends on the DM density squared and also implicitly on the inverse distance squared. Hence, targeting nearby overdensities of DM yields larger values of the $J-$factor. Such targets are, thus, very well suited to probe smaller annihilation cross sections $\langle \sigma v\rangle$. 

To apply the indirect search pipeline that has been developed to study the DM content of the LMC and SMC, we must first determine the DM density distribution of M31 and M33; that is to calculate their expected $J-$factor ($D-$factor). This task is of a complex nature as $N-$body simulations of the formation and evolution of MW-sized galaxies predict a hierarchical formation scenario. The DM halo of spiral galaxies, like the one of M31 or M33, is expected to form by mergers of small overdensities which are referred to as subhalos. Depending on the particle resolution of the respective $N-$body simulation (\cite{2008MNRAS.391.1685S, 2008JPhCS.125a2008K, 2016ApJ...818...10G}), around 10 - 20 $\%$ of the mass of a MW-sized galaxy's DM halo has been found to be present in the form of substructure. An extrapolation of these results to less-massive, and yet unresolved, subhalos seems to predict that in the most extreme scenario about 50$\%$ of a DM halo's mass stems from substructure. This strongly affects $\gamma$-ray indirect DM searches because a population of DM subhalos can boost the $J-$factor of the parent halo substantially \cite{2008ApJ...686..262K,Sanchez-Conde:2013yxa}. The $D-$factor of a DM halo, on the other side, is mostly unaffected by the presence of substructure since it grows linear in the DM density. Nonetheless, the region exhibiting the largest $J-$/$D-$factor in a galaxy is its center where the DM density is dominated by the profile of the smooth parent halo. The observed origin of the extended gamma-ray emission from M31 is coinciding with its central region so that we are required to carefully select smooth DM halos for M31 (and M33) that, on one side, cover the full variety of existing DM profiles types and, on the other side, are in accordance with the available stellar data.   

As a matter of fact, M31 seems to be the only well-studied galaxy which was argued to require the effect of adiabatic contraction around its central region \cite{Seigar:2006ia, Gnedin:2011uj}. Adiabatic contraction is caused by baryonic physics and gravitational interactions between baryons and DM in galaxies. During the formation of a galaxy, typical processes like gas dissipation, supernova feedback and star formation lead to substantial energy losses of a sizable fraction of galactic baryons which hence fall into the central region of their host galaxy. As first reported in \cite{zel1980astrophysical}, these particles deepen the gravitational potential of the galactic center so that the surrounding DM follows the baryonic pull creating a compressed DM halo in the central region. Subsequent hydrodynamical simulations of galaxy formation (\cite{Gustafsson:2006gr, Colin:2005rr, 2011arXiv1108.5736G, 2010MNRAS.406..922T, 2012ApJ...748...54Z, 2010MNRAS.408.1998S}) confirm the prediction of adiabatic contraction of DM halo profiles obtained from DM-only simulations. In fact, an adiabatically contracted Navarro-Frenk-White (NFW) profile \cite{1996ApJ...462..563N, Navarro:1996gj} enhances the $J-$factor in the center of M31 which is remarkable since, following a DM interpretation, a large $J-$factor could be the source of the observed extended gamma-ray emission from M31 in this region. The NFW profile is a particular instance (with $\left(\alpha, \beta, \gamma\right) = \left(1,3,1\right)$) of the general Hernquist-Zhao profile \cite{1990ApJ...356..359H, 1996MNRAS.278..488Z}:

\begin{equation}
\rho_{{\rm ZHAO}}\!\left(r\right)=\frac{\rho_{s}}{\left(\frac{r}{r_s}\vphantom{\left(\frac{r}{r_s}\right)^{\alpha}}\right)^{\gamma}\left[1+\left(\frac{r}{r_{s}}\right)^{\alpha}\right]^{\frac{\beta - \gamma}{\alpha}}}\textrm{,}
\end{equation}
where $\rho_s$ is a density normalization, $r_s$ refers to the profiles' scale radius and $\alpha, \beta$ and $\gamma$ determine the inner and outer slope of $\rho_{{\rm ZHAO}}$ as well as the transition between both regimes. These parameter labels are also used in the definitions of other DM density profiles considered in this analysis.\\
We adopt the smooth adiabatically contracted NFW profile from \cite{Seigar:2006ia} where it is called ``M1 B86''. 
In detail, we read off the ``Halo'' mass-to-radius curve in their Fig.~6. Afterwards, we convert it into a radial density profile via $\rho(r)=\left(4\pi r^{2}\right)^{-1}\textrm{d}M/\textrm{d}r$ and interpolate the obtained data points. In fact, the resulting density profile cannot be described by a single set of parameters of the Hernquist-Zhao profile. However, for $r<25.65$ kpc (the profile's scale radius) it provides an acceptable -- albeit far from good -- approximation with the parameters $\left(\alpha, \beta, \gamma\right) = \left(0.38, 3.84, 1.54\right)$. 

Alongside this non-standard DM density distribution, we consider two distinct but frequently-used profiles, namely an Einasto profile that provides a better fit to the profile of DM halos derived from $N-$body simulations \cite{einasto1968constructing, Navarro:2003ew} (representing the family of cuspy profiles)
\begin{equation}
\rho_{{\rm Ein}}\!\left(r\right)=\rho_{s}\exp\!\left(-\frac{2}{\alpha}\left[\left(\frac{r}{r_{s}}\right)^{\alpha}-1\right]\right),
\end{equation}
and a Burkert profile \cite{Burkert:1995yz} (representing the family of cored profiles)
\begin{equation}
\rho_{{\rm Burkert}}\!\left(r\right)=\frac{\rho_{s}}{\left(\vphantom{\left.\frac{1}{2}\right)^2}1+\frac{r}{r_{s}}\right)\left[1+\left(\frac{r}{r_{s}}\right)^{2}\right]}{\rm .}
\end{equation}
These three DM profiles bracket the range of cosmologically and astrophysically viable DM halo morphologies according to the current understanding of large structure formation and baryonic feedback. As a matter of fact, baryonic physics has been identified not only to be the driving force of an adiabatic contraction of the innermost region of a DM halo but it can also have the opposite effect leading to the formation of a DM core \cite{2012ApJ...744L...9M}. The infall of baryons into the center of a DM halo is described to trigger a large number of enhanced star formation periods which each time create a massive outflow of baryons from the central region. The DM follows the baryonic flow as this flow causes a shallower gravitational potential so that the inner cusp is successively washed out by the cycles of baryonic infall and a burst in star formation \cite{Mashchenko:2006dm, Mashchenko:2007jp, 2012MNRAS.421.3464P}. There have been attempts to implement this kind of baryonic feedback in cosmological simulations of structure formation. They confirm a flattening of the inner DM cusp observed in DM-only simulations \cite{Governato:2009bg, 2012ApJ...744L...9M}. However, as for example pointed out in \cite{DiCintio:2013qxa}, the impact of baryons on the evolution of a DM halo depends on its properties like the stellar-to-halo mass ratio. Hence, the ultimate effect of baryonic feedback can vary from galaxy to galaxy. 

Regarding the Einasto and Burkert profile of M31, we adopt the set of parameters that has been derived in \cite{Tamm:2012hw} to match its available kinematical stellar data. As concerns M33, we make the same distinction between cored and cuspy profiles by selecting a Burkert profile\footnote{The admissible Burkert core radii of M33 seem to span a rather wide range \cite{Corbelli:2014lga, Fune:2016uvn}. The difference is most likely attributed to the chosen analysis approach, rotation curve data selection and fitting scheme.} and an NFW profile where the respective parameters are taken from \cite{Fune:2016uvn}. A summary of the adopted profile parameters is found in Table~\ref{tab:substructure_para}. In case of M33's NFW profile and M31's adiabatically contracted NFW profile, we had to calculate $r_s$ and $\rho_s$ from the given quantities in the respective reference. 

\begin{figure}
\centering
\subfigure[$J-$factor M31]{\includegraphics[width=0.49\linewidth]{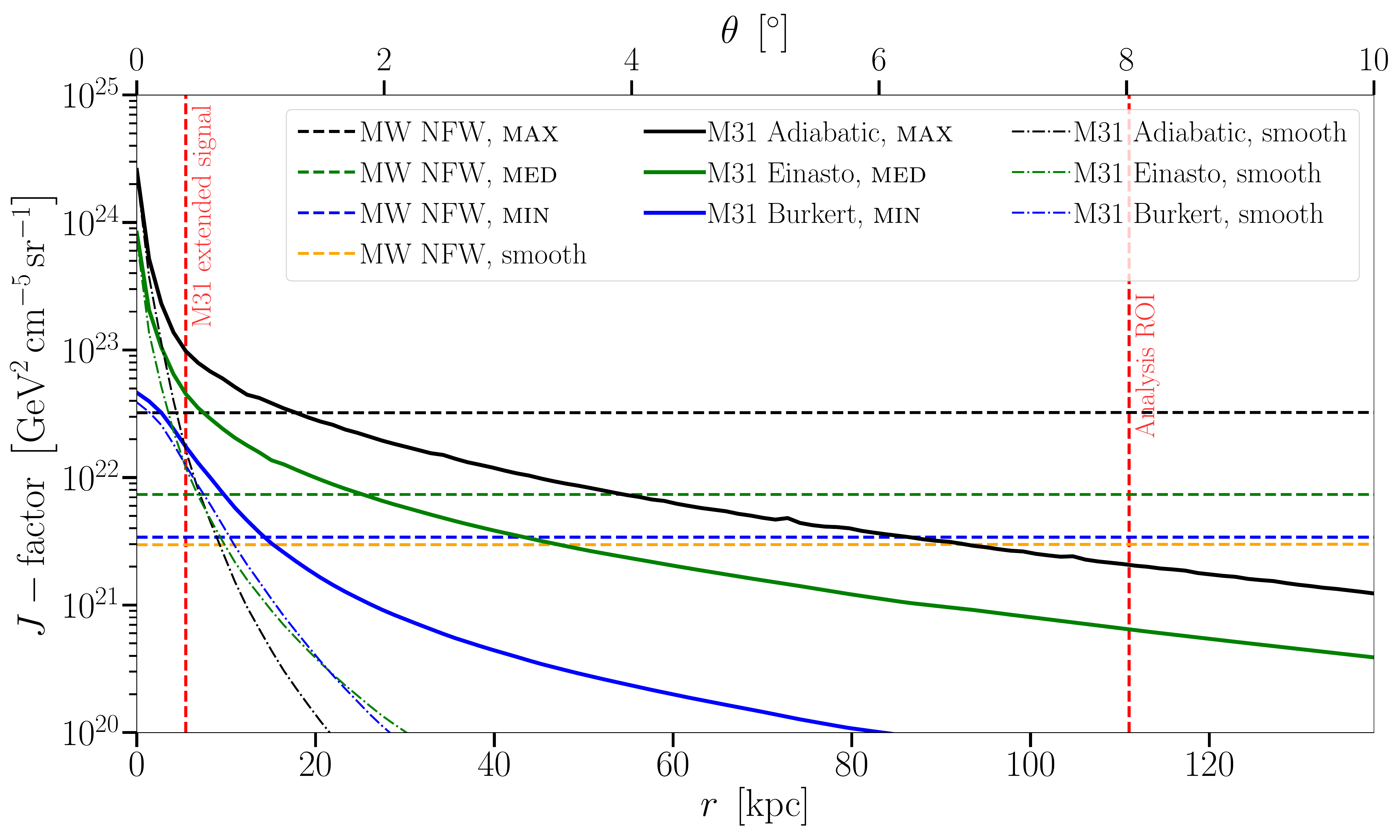} \label{fig:d-and-j-factors_a}}
\subfigure[$D-$factor M31]{\includegraphics[width=0.49\linewidth]{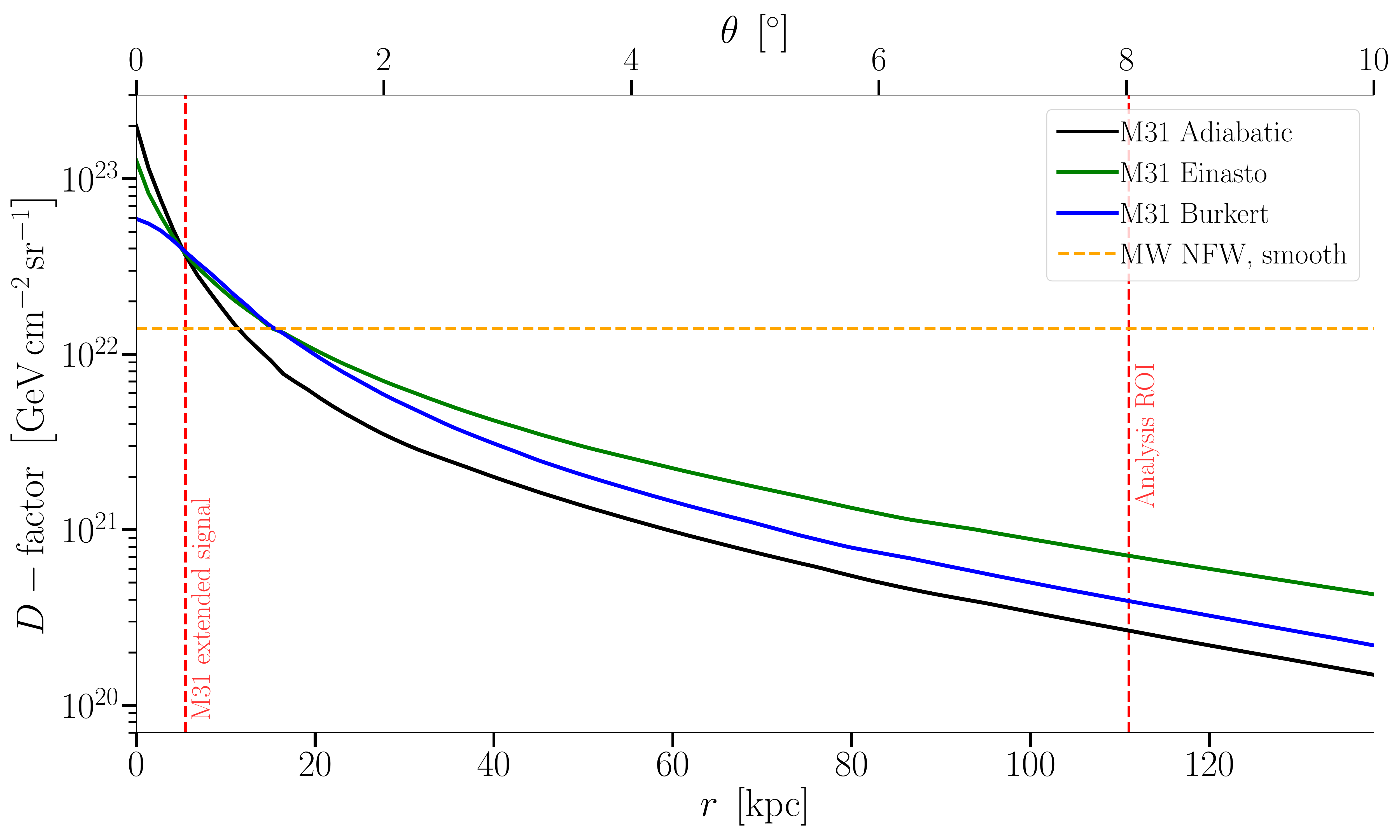}}
\subfigure[$J-$factor M33]{\includegraphics[width=0.49\linewidth]{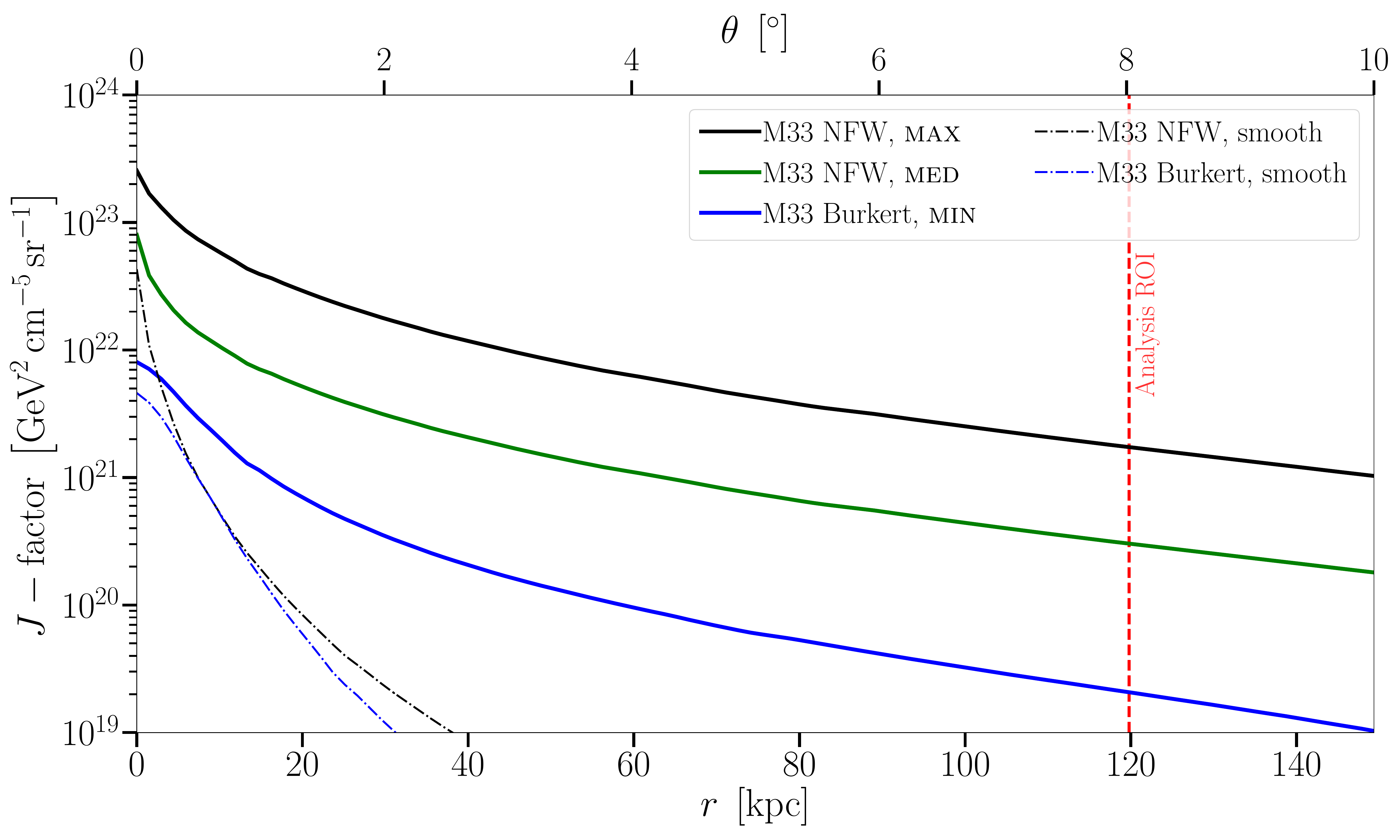}}
\subfigure[$D-$factor M33]{\includegraphics[width=0.49\linewidth]{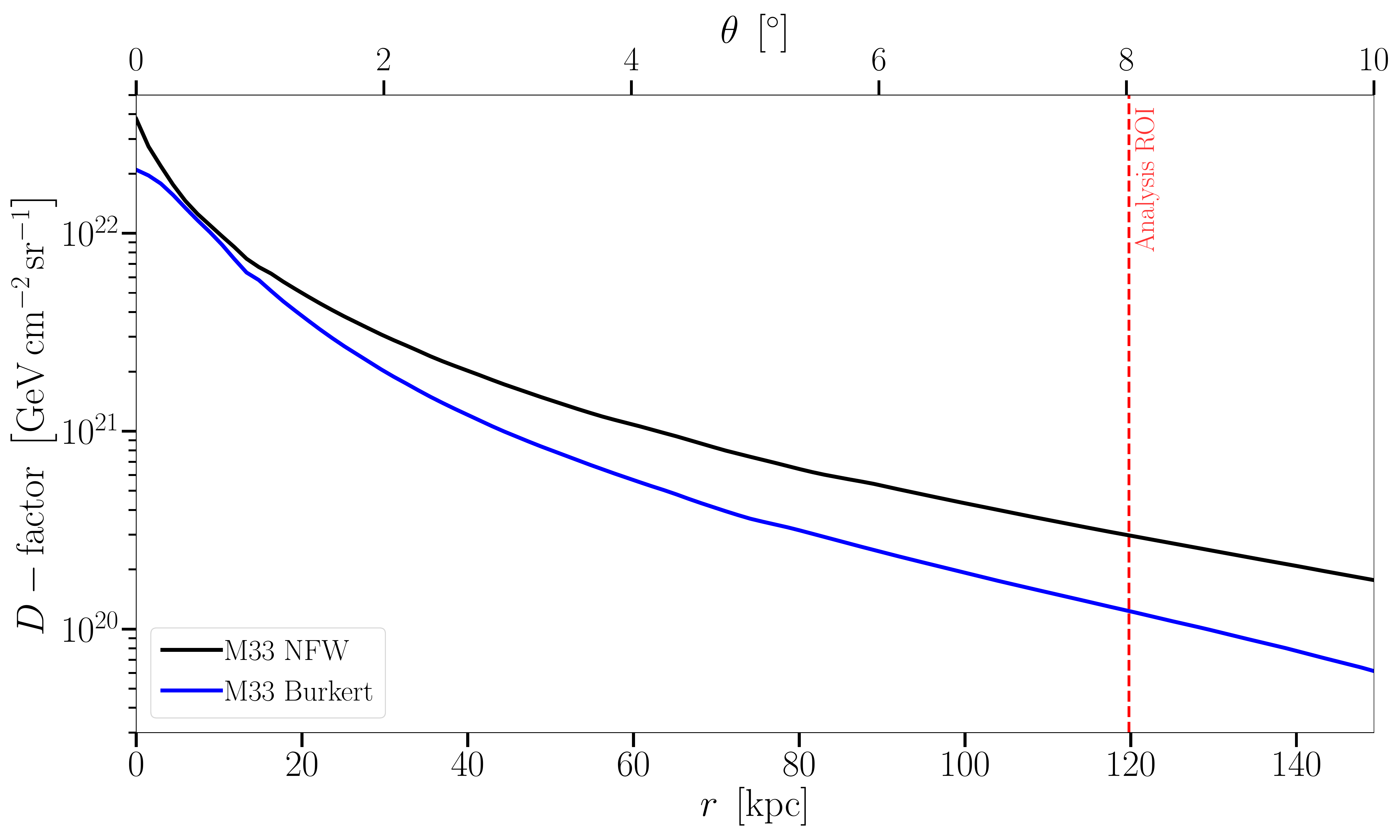}}
\caption{Radial profiles of the $J-$/$D-$factors of M31 and M33 derived for annuli of width $0.1^{\circ}$ centered on the respective galaxy. In case of the $J-$factor panels, the solid lines represent the full DM distribution taking into account both the smooth and substructure DM components (black: {\tt MAX}, green: {\tt MED}, blue: {\tt MIN}) whereas the dash-dotted lines show the respective smooth DM component without subhalos. With dashed lines, figure (a), in particular, displays the expected $J-$factor contribution from the MW DM halo in the direction of M31 which is likewise distinguished in four distinct cases featuring assuming one of the three substructure realizations or a smooth DM component only. The radial $D-$factor profiles where derived using the smooth DM halo component standalone since the boost to substructure is not expected to yield a sizeable enhancement of the $\gamma$-ray flux. A detailed list of the chosen DM halo profiles and substructure parameters is given in Tab.~\ref{tab:substructure_para}. The vertical dashed lines in red indicate the spatial extension of the M31 $\gamma$-ray signal (if applicable) or our analysis ROI, respectively. \label{fig:d-and-j-factors}}
\end{figure}

Since Eq.~\ref{eq:dmflux} involves a line of sight integral, we also have to consider the DM distribution in the MW when pointing towards M31 or M33. Moreover, it is not possible to exclude a priori that the DM halo of the MW has left no traces in the $\gamma$-ray data of {\it Fermi}-LAT. Classically, the MW DM halo is fit by either an NFW or a Burkert profile to cover profiles featuring either a central cusp or core. However, the morphological difference between both profiles in the Galactic center is a marginal aspect
with respect to the sky position of M31 ($(\textrm{RA},\textrm{DEC})=(10\fdg685, 41\fdg269)$) or M33 ($(\textrm{RA},\textrm{DEC})=(23\fdg462, 30\fdg660)$) so that we choose the NFW profile from \cite{Nesti:2013uwa} as the smooth MW DM halo. 
Comparing the MW $J-$factor within a circular region of interest (ROI) of radius $1^{\circ}$ centered on M31 or M33, the relative difference between a Burkert and an NFW profile is about $15\%$ in both cases. 
For instance, while the average $J-$factor of M31 from an adiabatically contracted NFW profile inside a circular ROI of $0.4^{\circ}$ (corresponding to the spatial extension of M31's $\gamma$-ray emission) is about a factor of 100 larger than the respective $J-$factor from the MW NFW halo, the situation reverses not far outside this particular ROI due to the almost perfect isotropy of MW's $J-$factor (cf.~Fig.~\ref{fig:d-and-j-factors_a}). In another extreme case, namely choosing the same circular ROI of  $0.4^{\circ}$ and centering it on M33, the $J-$factor from M33's DM halo following a Burkert profile is only about $90\%$ of the corresponding $J-$factor due to the MW's DM halo.

We aim to go beyond the zeroth order $J-$factor estimates by accounting for the effect of substructure inside the smooth halos of the MW, M31 and M33. However, quantitative statements about the net effect of the DM substructure boost are hard to formulate as a precise prediction of the present-day subhalo population properties -- e.g., their radial distribution in the host halo or the subhalo survival probability until present time -- of MW-sized galaxies remains an objective of ongoing research. To account for those uncertainties, we define two limiting substructure scenarios ({\tt MAX} and {\tt MIN}) that model the expected substructure of the galaxies under study according to the extreme cases still allowed by $N-$body simulations. Moreover, we also create a benchmark scenario ({\tt MED}) that features a DM substructure distribution based on the best-fit parameters of recent observations and numerical simulations of structure formation. To this end, we use the public code \textsc{\textcolor{black}{CLUMPY}} \cite{Hutten:2018aix, Bonnivard:2015pia, 2012CoPhC.183..656C} to generate 2D $J-$/$D-$factor sky maps of M31 and M33 as source model for our analysis pipeline as well as MW DM templates as an additional background source.

The parameters governing the substructure distribution in a galaxy that have the largest impact on the expected $J-$factors are:
\begin{itemize}
\item [-] The index $\alpha_{\textrm{sub}}$ of the subhalo mass function $\textrm{d}n/\textrm{d}M$ which was found to follow a power-law \cite{2008JPhCS.125a2008K,2008MNRAS.391.1685S}, 
\item [-] the fraction of the DM halo mass which is stored in substructure $f_{\textrm{sub}}$,
\item [-] the minimal mass of DM subhalos $M_{\textrm{min}}$ and
\item [-] the subhalo concentration parameter $c_{200,\,\textrm{sub}}$ \cite{2001MNRAS.321..559B}. 
\end{itemize}
We rely on the most recent model of the concentration parameter of subhalos \cite{Moline:2016pbm}. 
We make use of a developer's version of \textsc{\textcolor{black}{CLUMPYv3}} which features this concentration model for extended extragalactic objects.
This model reports a flattening of the concentration of subhalos towards the low mass tail of the relation and, furthermore, it includes a dependence on the position of the subhalo within its host halo\footnote{Neither the release version nor the developer's version of \textsc{\textcolor{black}{CLUMPYv3}} implement the spatial dependence of the subhalo concentration parameter so that this model can only be considered as a slight improvement with respect to the previous parametrization given in \cite{Sanchez-Conde:2013yxa}}. Using this description of the subhalo concentration relation implicitely assumes that the DM profiles of subhalos follow an NFW profile which is a fair assumption given the large uncertainty on this quantity as obtained from $N-$body simulations \cite{2008MNRAS.391.1685S, 2008JPhCS.125a2008K, 2016ApJ...818...10G}.

The remaining mutually dependent parameters are chosen as to be consistent with the findings of DM-only $N-$body simulations of MW-sized DM halos. In fact, it has been theoretically established that the minimal subhalo mass $M_{\textrm{min}}$ depends on the particle physics nature of DM so that it might cover orders of magnitude down to values like $10^{-12}\,M_{\odot}$ \cite{Binder:2017rgn,Bringmann:2009vf}. In fact, $M_{\textrm{min}}$ cannot be constrained very well even with astronomical data from current-generation instruments. A natural upper limit on $M_{\textrm{min}}$ are the masses of dSphs that have already been detected and are resolved in the MW halo. The faintest of them possess masses as low as $\approx 10^7\,M_{\odot}$ \cite{Strigari:2007ma,Walker:2007ju}. From this range, we fix $M_{\textrm{min}} = 10^6\,M_{\odot}$ in the context of the {\tt MIN} scenario (being very conservative) and $M_{\textrm{min}} = 10^{-12}\,M_{\odot}$ with respect to the {\tt MAX} scenario (being overly optimistic)\footnote{\textsc{\textcolor{black}{CLUMPY}} restricts the user to at most $M_{\textrm{min}} = 10^6\,M_{\odot}$ for extragalactic objects.}. Our benchmark case {\tt MED} assumes a typical value of $M_{\textrm{min}} = 10^{-6}\,M_{\odot}$ which is often used in the context of a $\Lambda$CDM cosmology with cold, thermal WIMP DM \cite{Green:2003un}. From these values of the minimal subhalo mass we can infer the expected fraction of subhalos in a galaxy for a given subhalo mass function index $\alpha_{\textrm{sub}}$ using the reported behavior found, e.g., in the Aquarius project or the Via Lactea simulation \cite{2008JPhCS.125a2008K,2008MNRAS.391.1685S}. We summarize the definitions of our three substructure scenarios and their assignments to the smooth DM halos in Table \ref{tab:substructure_para}. Those assignments were made in order to bracket the theoretically and observationally allowed $J-$factors from M31 and M33 according to the combination of smooth halo and substructure parameter set.     

\begin{table}
\begin{centering}
\begin{tabular}{|c|c|c|c|c|c|c|}
\hline 
 & smooth profile M31 & smooth profile M33 & $\alpha_{\textrm{sub}}$ & $f_{\textrm{sub}}$ & $M_{\textrm{min}}\;\left[M_{\odot}\right]$ & $c_{200,\,\textrm{sub}}$ \tabularnewline
\hline 
\hline
\multirow{4}{*}{$\hphantom{0.1}$\textsc{min}$\hphantom{0.1}$}  & {Burkert \cite{2012A&A...546A...4T}} & Burkert \cite{Fune:2016uvn} & \multirow{4}{*}{$1.9$} & \multirow{4}{*}{$0.12$} & \multirow{4}{*}{$10^{6}$}  & \multirow{12}{*}{$\hphantom{1}$\cite{Moline:2016pbm}$\hphantom{1}$}\tabularnewline
 & $\hphantom{0.1}r_s = 9.06\;\textrm{kpc,}$ & $\hphantom{0.1}r_s = 9.6\;\textrm{kpc,}$ &  &  & &\tabularnewline 
 & $\hphantom{0.1}\rho_s = 3.68\times10^7\;M_{\odot}\textrm{/kpc}^3\hphantom{0.1}$ & $\hphantom{0.1}\rho_s = 1.23\times10^7\;M_{\odot}\textrm{/kpc}^3\hphantom{0.1}$ & & & &\tabularnewline
& $\hphantom{0.1}M_{200} = 7.9\times10^{11}\;M_{\odot}\hphantom{0.1}$ & $\hphantom{0.1}M_{97.2} = 3.0\times10^{11}\;M_{\odot}\hphantom{0.1}$ & & & &\tabularnewline
\cline{1-6}  
\multirow{4}{*}{\textsc{med}}  & {Einasto \cite{2012A&A...546A...4T}} &  & \multirow{4}{*}{$\hphantom{0.1}1.9\hphantom{0.1}$} & \multirow{4}{*}{$\hphantom{0.1}0.19\hphantom{0.1}$} & \multirow{4}{*}{$10^{-6}$} & \tabularnewline
 & $r_s = 178\;\textrm{kpc,}$ &  &  &  & &\tabularnewline
 & $\rho_s = 8.12\times10^3\;M_{\odot}\textrm{/kpc}^3$ & NFW \cite{Fune:2016uvn} & & & &\tabularnewline
 & $M_{200} = 1.13\times10^{12}\;M_{\odot}$ & $r_s = 22.41\;\textrm{kpc,}$ & & & &\tabularnewline
\cline{1-2}
\cline{4-6}
\multirow{4}{*}{\textsc{max}}  & adiabatically contracted NFW \cite{Seigar:2006ia} & $\rho_s = 2.64\times10^6\;M_{\odot}\textrm{/kpc}^3$ & \multirow{4}{*}{$2.0$} & \multirow{4}{*}{$0.45$} & \multirow{4}{*}{$10^{-12}$} & \tabularnewline
 & $r_s = 25.65\;\textrm{kpc,}$ & $M_{97.2} = 5.4\times10^{11}\;M_{\odot}$ & &  & &\tabularnewline
 & $\rho_s = 4.44\times10^7\;M_{\odot}\textrm{/kpc}^3$ & & & & &\tabularnewline
  & $M_{200} = 5.7\times10^{11}\;M_{\odot}$ & & & & &\tabularnewline

\hline
\end{tabular}
\par\end{centering}
\caption{Summary of the most important parameters of \textsc{\textcolor{black}{CLUMPY}} to model the substructure contribution to the total $J-$/$D-$factor in M31 and M33. Note that the virial mass of M33 assuming either an NFW or Burkert profile has been derived with respect to $\Delta = 97.2$ times the critical density of the universe instead of $\Delta = 200$ as for all other profiles. Moreover, $r_s$ and $\rho_s$ of M33's NFW profile have been derived with respect to $\Delta = 97.2$, too. The choice of $\Delta = 97.2$ has been made by the authors of \cite{Corbelli:2014lga, Fune:2016uvn}.  \label{tab:substructure_para}}
\end{table}

To complete our model definition, we assume that the radial distribution of subhalos follows the smooth DM density profile of their host halo. As a consequence of this choice, subhalos are expected to populate even the central region of the host DM halo despite the impact of tidal forces and other effects, like baryonic physics, in this part of the galaxy. Nonetheless, DM-only $N-$body simulations, e.g.~the aforementioned Aquarius and Via Lactea II simulations \cite{2008JPhCS.125a2008K,2008MNRAS.391.1685S}, poorly constrain the subhalo population in this particular region \cite{Pieri:2009je}. In fact, while certain simulations including baryonic feedback during the galaxy formation process reveal a strong depletion of subhalos in a large volume around the center of a galaxy \cite{Bullock:2017xww}, there are opposing works like \cite{vandenBosch:2018tyt, 2010MNRAS.406.1290P} arguing that the observed tidal disruption of subhalos is a numerical artifact due to the particle resolution of the simulations. In the context of this ongoing debate, our choice seems as justified as any other. 

Having established a full description of our models and substructure parameter sets, we show in Fig.~\ref{fig:d-and-j-factors} the generated radial profiles of $J-$/$D-$factors. To stress it again, in the case of decaying DM, we only use the here discussed smooth DM profiles since the $D-$factor is proportional to the DM density and does not feel the boost due to subhalos.

\section{Data analysis}
\label{sec:analysis}

The analysis technique we apply in this paper closely follows those used in the DM searches from dwarf spheroidal galaxies~\cite{Ackermann:2015zua, Drlica-Wagner:2015xua}, Smith High-Velocity Cloud~\cite{Drlica-Wagner:2014yca}, and LMC and SMC~\cite{Buckley:2015doa,Caputo:2016ryl}. In the next sections we describe the data selection and the different steps of the pipeline.

\subsection{Data selection}

We use 115 months of {\it Fermi}-LAT data (from 2008 August 4 to 2018 March 4) selecting \irf{Pass 8} SOURCE class events and using the corresponding instrument response functions (IRFs) {\tt P8R3\_SOURCE\_V2}.
We choose an energy range from 300 MeV to 1 TeV and select events with reconstructed directions within a 16$\dg\times$16$\dg$ region of interest (ROI) centered at the infrared position of M31 and M33.
We test also different lower bound values (e.g., 100 MeV) for the energy range and different sizes of the ROI to see how the results are affected by these parameters (see Sec.~\ref{sec:sys}).
We bin the data using 8 energy bins per decade in energy and 0$\fdg$08 pixel size.

The Pass8 event reconstruction and selection called Pass 8 introduces a generalization of the conversion type classification in the form of event types.
PSF event types are event-level quantities indicating the quality of the reconstructed direction. The data are divided into quartiles, from the lowest quality quartile (PSF0) to the best quality quartile (PSF3) \cite{Atwood:2013rka}.

We apply zenith angle cuts to the data in order to reduce the contamination from the low-energy Earth limb emission.
We select for $E=[0.1,0.3]$ GeV PSF2 and PSF3 event types with zenith angles $\theta_z > 90^{\circ}$, for $E=[0.3,1.0]$  PSF1, PSF2 and PSF3 event types with zenith angles larger than $\theta_z < 100^{\circ}$ and finally above 1 GeV we keep all PSF types with zenith angles $\theta_z > 105^{\circ}$.
We apply the same cuts used in the construction of FL8Y (and 4FGL, see later in the text) source list; these reduce the contribution of the Earth limb at that zenith angles to less than $10\%$ of the total background. 
See Table \ref{tab:data} for a summary of the analysis setup.

We construct a background model of each region that includes the FL8Y sources in the region, as well as an interstellar emission model and an isotropic emission template.
Very recently, the {\it Fermi}-LAT 8-year Point Source Catalog (4FGL) has been created using a new IEM and isotropic templates \cite{Fermi-LAT:2019yla}. 
The FL8Y and 4FGL catalogs have been created with the same years of data and no significant difference in the characteristics of the sources present in M31 and M33 ROIs are present thus we do not find any relevant difference in our result by using the 4FGL instead of FL8Y (see Sec.~\ref{sec:sys}).
Specifically, we use the interstellar emission model (IEM) released with Pass 8 data \citep{Acero:2016qlg} (i.e., {\tt gll\_iem\_v06.fits}) since this is the model routinely used in Pass~8 analyses. 
We will label this IEM as Official (Off).
This model is derived by performing a template fitting to {\it Fermi}-LAT $\gamma$-ray data. 
It is thus based on the spatial correlations between $\gamma$-ray data and a linear combination of gas and inverse Compton scattering maps.
This model contains patches to account for extended excess emissions of unknown origin. 
However, the M31 and M33 regions do not contain any of these patch components.
We use for the isotropic emission the template associated to this IEM ({\tt iso\_P8R3\_SOURCE\_V2.txt})\footnote{For descriptions of these templates, see \url{http://fermi.gsfc.nasa.gov/ssc/data/access/lat/BackgroundModels.html}.}. 

To approximately study the systematic uncertainties from the mismodelling of the diffuse emission, we also run our analysis using the 8 alternative IEM models and corresponding isotropic templates used in the first {\it Fermi}-LAT supernova remnant (SNR) catalog \cite{Acero:2015prw}.
These models were generated by varying the cosmic-ray (CR) source distribution, height of the CR propagation halo, and HI spin temperature in order to test the effect of the choice of the IEM in the flux and spatial distribution of SNRs. 
These 8 models, which are all based on the GALPROP\footnote{\url{http://galprop.stanford.edu/}} CR propagation and interaction code, have been used in the SNR catalog to explore the systematic effects on SNRs fitted properties, including the size and morphology of the extension, caused by IEM modeling. 
We will label these models as alternatives (Alt).

It is important to stress that the Off and Alt IEMs have been designed to model the diffuse background for analysis of point and small extended sources.
Because they are fit to the data, they are not suited for studies of very extended sources and/or large-scale diffuse emissions. 
Since both M31 and M33 signal are extended at $\leq$ degree scales, and are not correlated with any other diffuse template, these diffuse models are applicable for our analysis.

In fact, the extended sources studied in the {\it Fermi}-LAT SNR catalog \cite{Acero:2015prw} have similar spatial extension as M31 and M33.
Finally, we note that the Off IEM and isotropic templates have been routinely used in previous DM analysis from {\it Fermi}-LAT.  
We also stress that if the excess signal would have been found in our ROI with any of these models, then a dedicated diffuse analysis would be required to determine its properties, but it will be shown below that this is not the case here\footnote{This approach is complementary to the recent work \cite{Karwin:2019jpy}, in which an almost isotropic emission on the 10 degree scales from M31, degenerate with the isotropic component  of the MW was studied. In that case a careful modeling of the diffuse emission was necessary and was indeed undertaken in that work.}.

We employ {\tt FermiPy} to perform our analysis of {\it Fermi}-LAT data.
{\tt FermiPy} is a Python package that automates analyses with the {\it Fermi} Science Tools \citep{2017ICRC...35..824W}\footnote{See \url{http://fermipy.readthedocs.io/en/latest/}.}\footnote{We use version 17.3 of {\tt FermiPy} and 11-07-00 of the \textit{ScienceTools}.}.
We will explain in detail the analysis pipeline in the next sections.

\begin{table}[ht]
\begin{tabular}{cc}
\hline \hline
Selection & Criteria \\ \hline
Observation Period & 2008 August 4 to 2018 March 4\\
Mission Elapsed Time (s)\footnote{$Fermi$ Mission Elapsed Time is defined as seconds since 2001 January 1, 00:00:00 UTC} & 239557417 to 541779795 \\ 
Energy Range (GeV) & 0.3--1000\\
Fit Region (M31) & 16$\dg\times$16$\dg$ centered on $(\alpha,\delta)=(10\fdg685, 41\fdg269)$ \\ 
Fit Region (M33) & 16$\dg\times$16$\dg$ centered on $(\alpha,\delta)=(23\fdg462, 30\fdg660)$ \\ 
Zenith Range & $\theta_z<$90$\dg$ and PSF2 and PSF3 for $E\in [0.1,0.3]$ GeV \\
&			$\theta_z<$100$\dg$ and PSF1, PSF2 and PSF3 for $E\in [0.3,1.0]$ GeV \\
&		$\theta_z<$105$\dg$ all PSF types for $E >1$ GeV\\
Data Quality Cut\footnote{Standard data quality selection: \texttt{DATA\_QUAL$>$1 $\&\&$ LAT\_CONFIG==1 $\&\&$ roicut==yes} with the $gtmktime$ Science Tool} &  yes \\ \hline
\end{tabular}
\caption{Summary table of {\it Fermi}-LAT data selection criteria used for this paper's DM analysis. }
\label{tab:data} 
\end{table}




\subsection{Baseline fit}
\label{sec:fitting}

We first perform a broadband fit to our ROIs using the sources from the catalog, the Off IEM and isotropic template. The results of this baseline fit (without the DM template) will be used as an input to the DM dedicated analysis described below. 
The size of the ROI has been taken to be much larger than the DM contribution. 
Indeed, we see from Fig.~\ref{fig:d-and-j-factors} that for an angular distance $>5^{\circ}$ from the center of the ROIs the contribution to the $J$ and $D$ factors are negligible.
During the broadband fit the spectral energy distribution (SED) parameters of all the point sources in the ROI, the normalization and spectral index of the IEM and the normalization of the isotropic templates are free to vary.   
At this stage in the analysis M31 and M33 are modeled as point-like sources.

Then, we relocalize all the sources in the ROIs, including M31 and M33.
Since we are using more years of data than FL8Y, we identify new sources detected with a test statistic\footnote{The $TS$ is defined as twice the difference in maximum log-likelihood between the null hypothesis (i.e., no source present) and the test hypothesis: $TS = 2 ( \log\mathcal{L}_{\rm test} -\log\mathcal{L}_{\rm null} )$~\cite{1996ApJ...461..396M}.} $TS>25$.
In the last step of the procedure to define the baseline model for the ROI we refine the astrophysical model for M31 and M33. 
M31 has been detected in \cite{M31Fermi}, using about 7 years of data above 100 MeV, as extended with $TS_{\rm{EXT}}=16$ and size $0\fdg38 \pm 0\fdg05$.
We re-run the extension analysis at $E>0.3$ GeV finding that $TS_{\rm{EXT}}=13.6$ and the size (i.e., the 68\% containment radius) is $0\fdg33 \pm 0\fdg04$ for a disk template and $TS_{\rm{EXT}}=12.8$ with a size of $0\fdg42 \pm 0\fdg10$ for a Gaussian template.
We also run this analysis for $E>0.1/0.5/1$ GeV to see if there is an energy dependence of the spatial extension. 
We report the results for the detection and the spatial morphology of M31 in Table \ref{tab:ext} for the disk and Gaussian spatial templates. The sizes of extension for M31 in the different energy ranges are all compatible within $1\sigma$.
We significantly detect M31 as extended also at $E>0.1$ GeV with a similar size of extension and $TS$.
This justifies our choice of using an extended template that is energy independent.
Since the disk morphology is slightly preferred we use this geometry for M31 as the benchmark case in the rest of our analysis (see Sec.~\ref{sec:dmplusastro}, \ref{sec:dminterp} and \ref{sec:sim}).

\begin{table}[ht]
\begin{tabular}{c|ccc|ccc}
\hline \hline
$E$ & $TS$ & $TS_{EXT}$ & $\theta_{\rm{EXT}}$ [deg] & $TS$ & $TS_{EXT}$ & $\theta_{\rm{EXT}}$ [deg] \\ \hline
       & \multicolumn{3}{c|}{Disk template} & \multicolumn{3}{c}{Gaussian template} \\ \hline
$>0.1$ GeV & 110 & 15.3 &  $0\fdg33 \pm 0\fdg03$  & 109 & 13.9 &  $0\fdg41 \pm 0\fdg09$ \\
$>0.3$ GeV & 98 & 13.6 &  $0\fdg33 \pm 0\fdg04$  & 97 & 12.8 &  $0\fdg42 \pm 0\fdg10$  \\
$>0.5$ GeV & 82 & 9.6 &  $0\fdg32 \pm 0\fdg04$  & 81 & 8.6 &  $0\fdg37 \pm 0\fdg09$ \\
$>1.0$ GeV & 58 & 9.3 &  $0\fdg31 \pm 0\fdg05$  & 58 & 8.2 &  $0\fdg31 \pm 0\fdg09$ \\
\end{tabular}
\caption{Summary table for the $TS$ of detection ($TS$) and extension ($TS_{EXT}$) in our analysis of {\it Fermi}-LAT data in the M31 ROI for the disk (left side) and Gaussian templates (right side) and using in the fit the Off IEM. }
\label{tab:ext}
\end{table}

\begin{table}[ht]
\begin{tabular}{c|ccc|cc}
\hline \hline
IEM & $TS$ & $TS_{EXT}$ & $\theta_{\rm{EXT}} [deg]$ & $TS$ \\ \hline
    & \multicolumn{3}{c|}{M31} & M33 \\ \hline
Off & 110 & 15.3 &  $0\fdg33 \pm 0\fdg03$ & $39$ \\
Alt 1 & $90.1$ & $12.4$ & $0\fdg32 \pm 0\fdg04$ & $42$ \\
Alt 2 & $100$ & $10.2$ & $0\fdg37 \pm 0\fdg06$ & $36$ \\
Alt 3 & $89$ & $12.3$ & $0\fdg32 \pm 0\fdg04$ & $42$ \\
Alt 4 & $85$ & $9.8$ & $0\fdg42 \pm 0\fdg10$ & $37$ \\
Alt 5 & $86$ & $12.1$ & $0\fdg32 \pm 0\fdg04$ & $39$ \\
Alt 6 & $94$ & $9.6$ & $0\fdg36 \pm 0\fdg07$ & $34$ \\
Alt 7 & $106$ & $11.1$ & $0\fdg43 \pm 0\fdg07$ & $40$ \\
Alt 8 & $84$ & $10.0$ & $0\fdg32 \pm 0\fdg04$ & $35$ \\
\end{tabular}
\caption{Summary table for the significance of detection and extension in our analysis of {\it Fermi}-LAT data in the M31 ROI and M33 ROIs with the Off and Alt IEMs. We assume here a uniform disk spatial template.}
\label{tab:extAlt}
\end{table}

M33 is detected as a point-like source ($TS_{\rm{EXT}}\approx 0$) with a $TS=41.9$ ($TS=39.4$) with the analysis performed in the energy range $E\in[0.1,1000]$ GeV ($E\in[0.3,1000]$ GeV).

\begin{figure}[ht]
\includegraphics[scale=0.37]{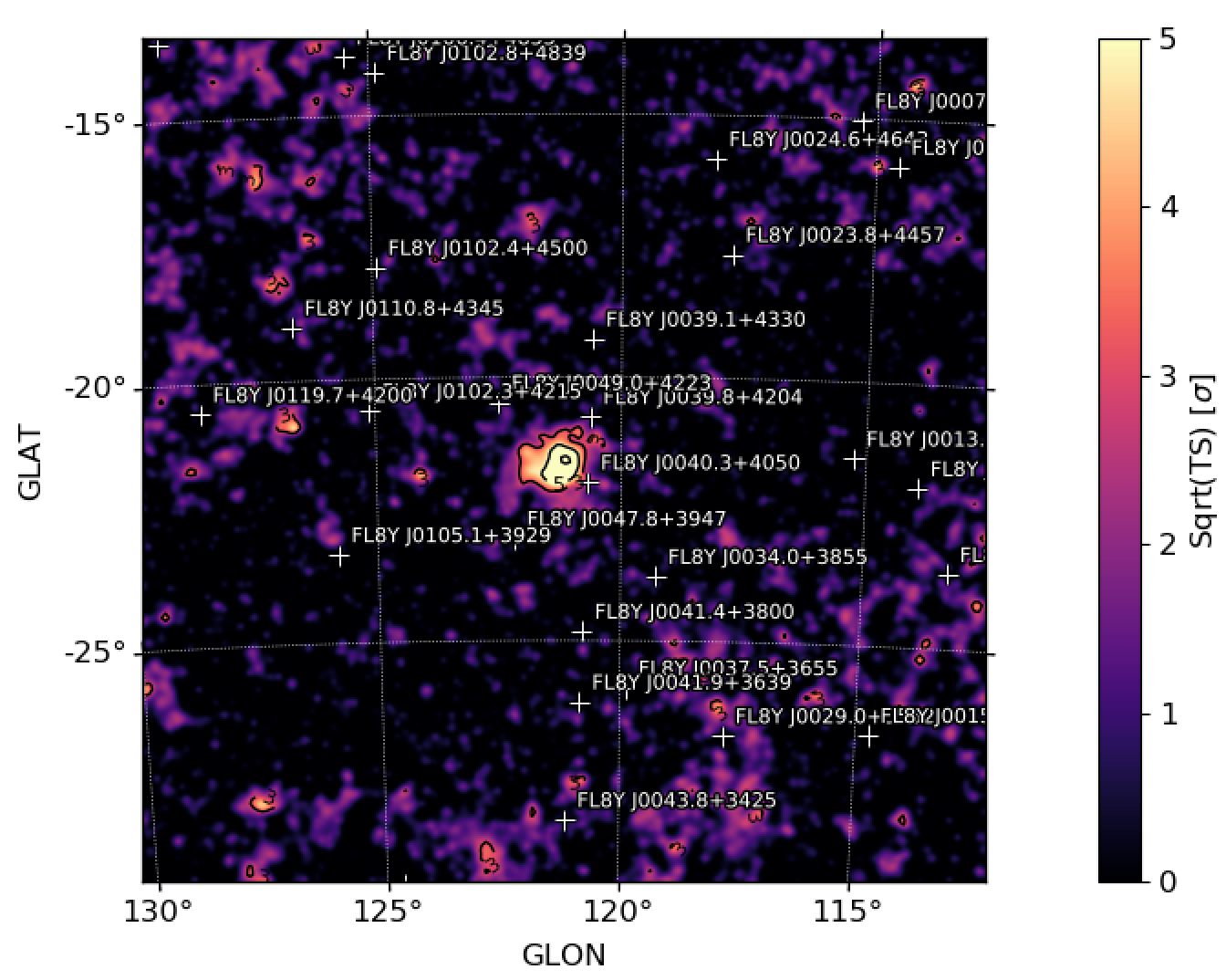}
\includegraphics[scale=0.37]{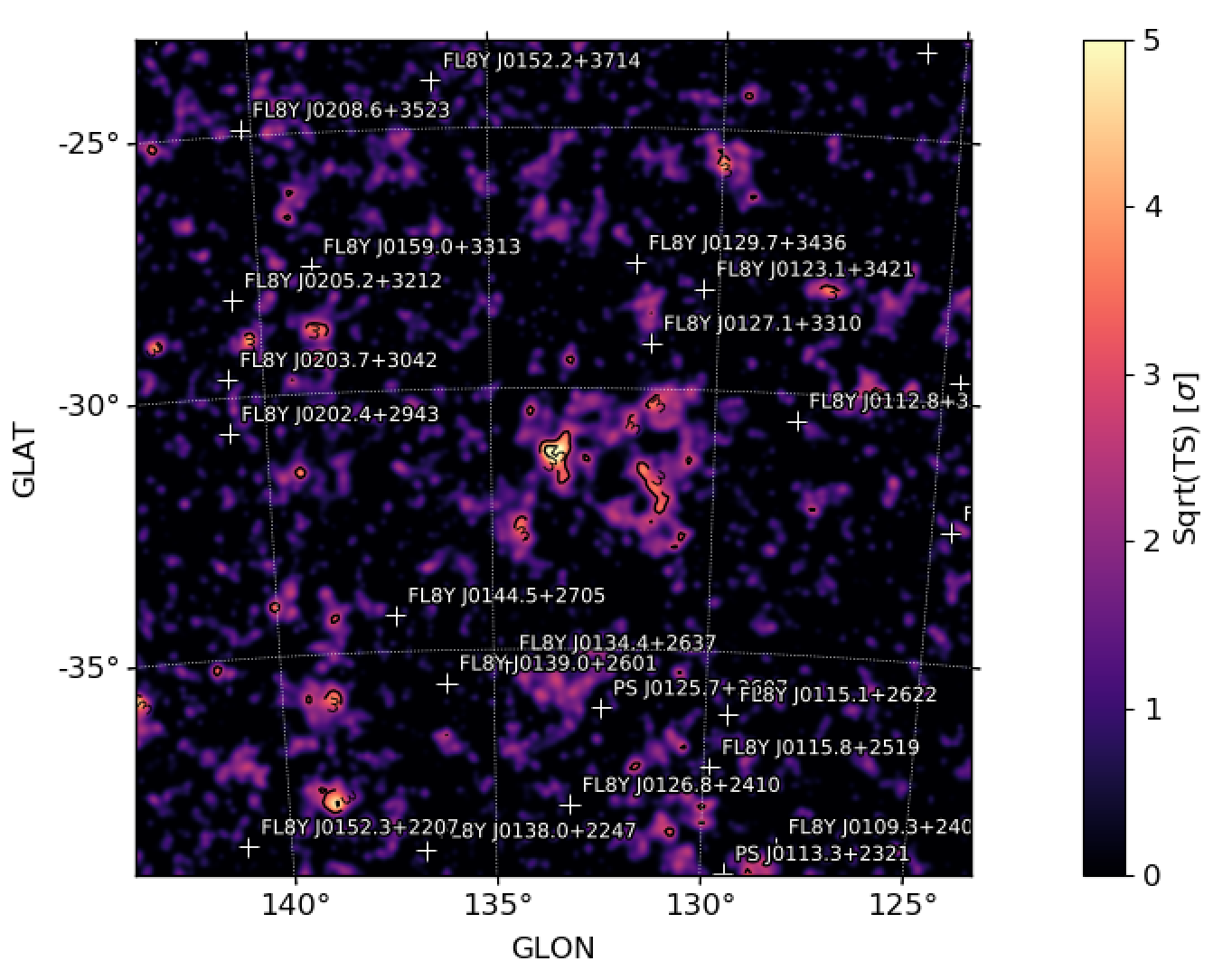} 
\includegraphics[scale=0.37]{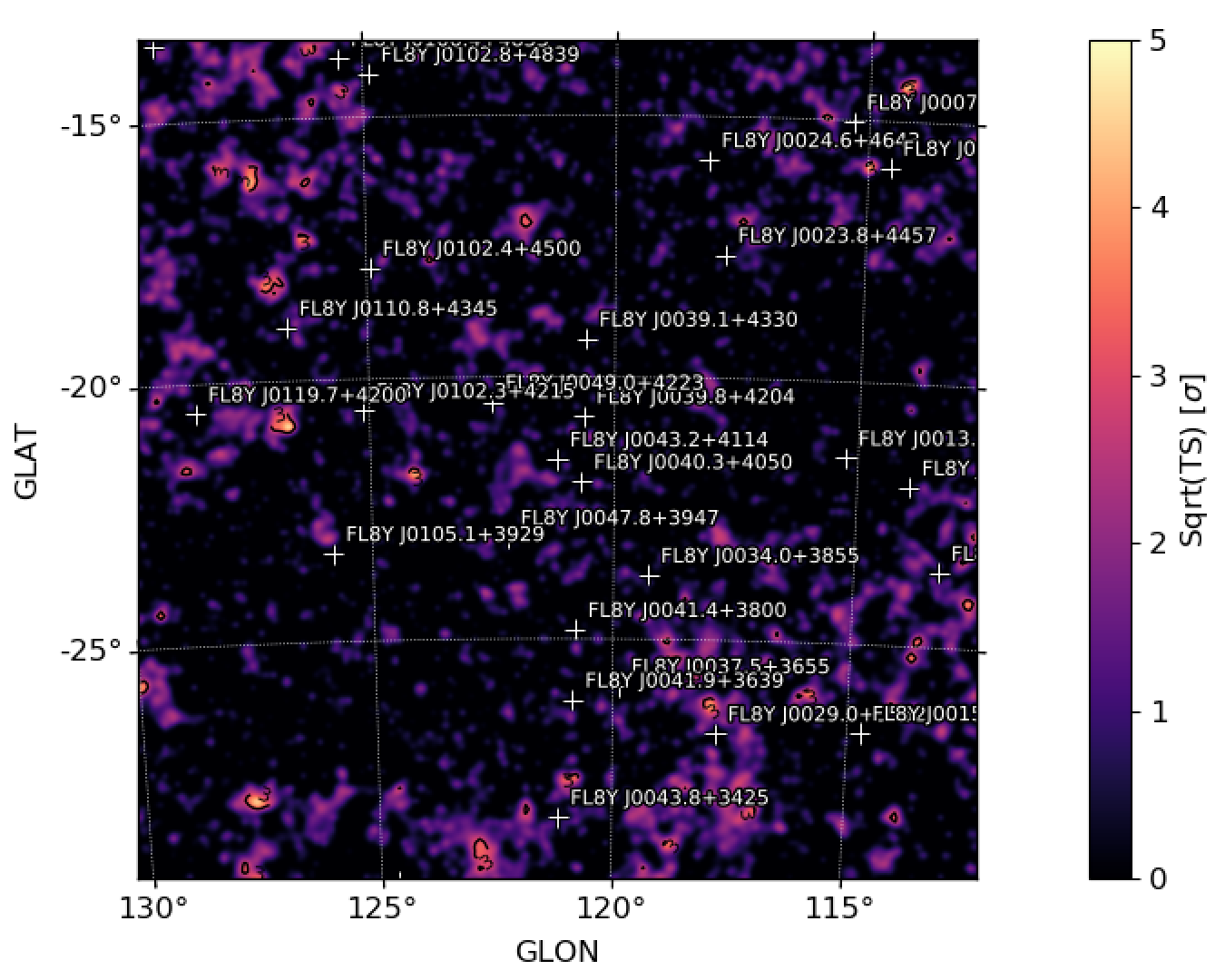}
\includegraphics[scale=0.37]{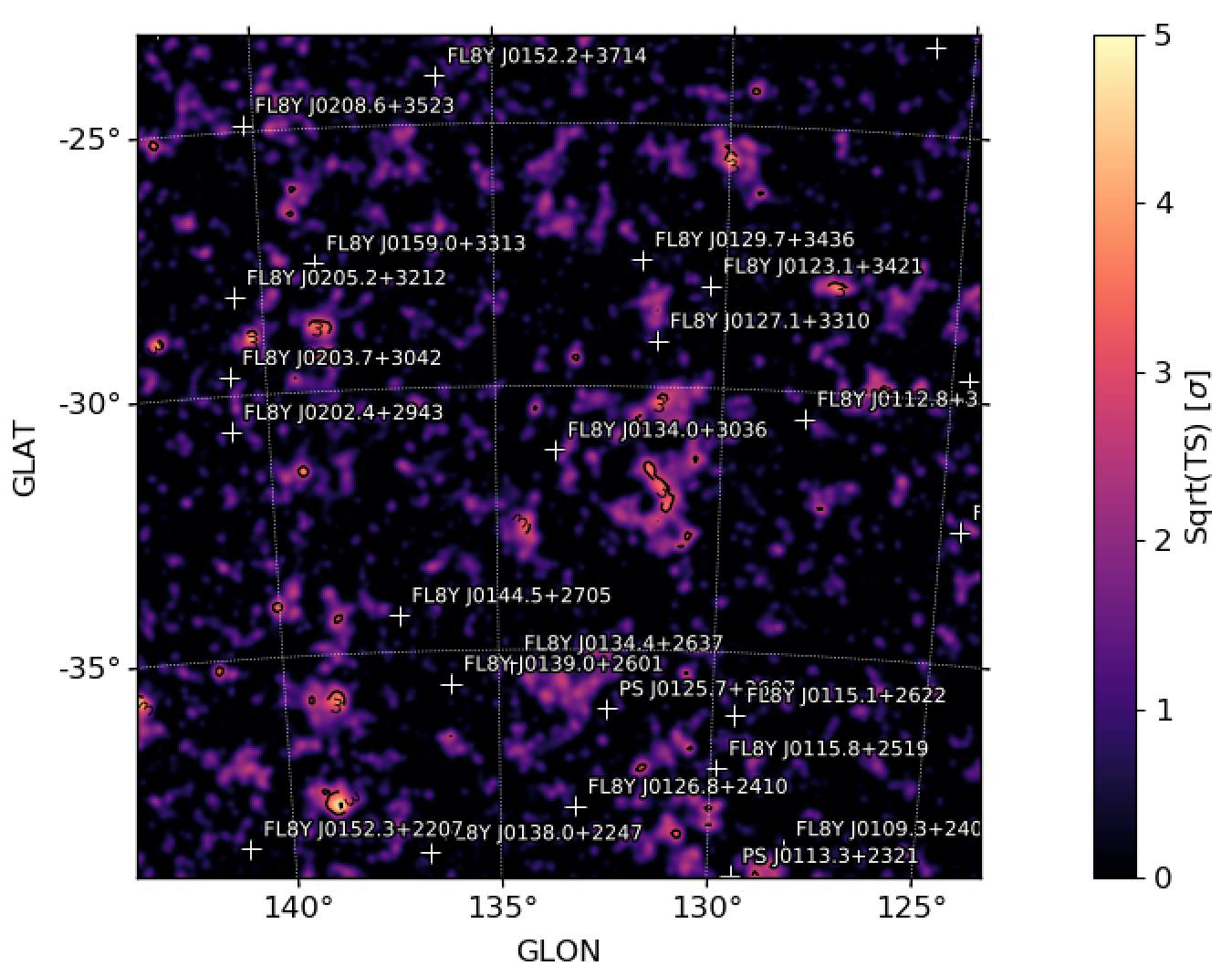} 
\caption{$TS$ map of M31 (left panel) and M33 (right panel) for $E\in[0.3,1000]$ GeV and for a pixel size of $0\fdg08$. The color scale represents values of $\sqrt{TS}$ in the range between 0 and 5 that for a point source corresponds approximatively to a $TS$ between 0 and 25. The bottom (top) panels are $TS$ maps where M31 and M33 are included (are not included) in the model. We also display the name and position of sources from FL8Y catalog included in the model.}
\label{fig:TSmap}
\end{figure}

Fig.~\ref{fig:TSmap} shows the $TS$ map of the ROI of M31 and M33 without these galaxies included in the model. It is clear from these plots that M31 is much brighter than M33 and that it has an extension of the order of the size detected in our analysis.
In the same figure we also show the $TS$ map for the baseline model, i.e.~with all sources included and with their positions refined. There are no significant residuals in these maps meaning that the baseline model is an appropriate fit to the two ROIs. The highest $TS$ peaks are of the order of $2-3\sigma$ significance. We use this model as a baseline for the search of a DM signal.

We also run the baseline analysis for M31 and M33 using the 8 Alt IEM models and corresponding isotropic templates used in the first {\it Fermi}-LAT SNR catalog \cite{Acero:2015prw}.
Depending on the Alt IEMs, M31 is detected with $TS\in[84-110]$, $TS_{\rm{EXT}}\in [9-15]$, disk size $\theta_{\rm{EXT}} = [0\fdg30,0\fdg45]$ and Gaussian template size $\theta_{\rm{EXT}} \in [0\fdg32,0\fdg43]$ (see Table \ref{tab:extAlt}).  In short, no significant changes are found by using different IEMs.
Similarly, M33 is detected as a point source with all the IEMs ($TS_{\rm{EXT}}\approx 0$) with a $TS$ in the range $34-42$ (see Table \ref{tab:extAlt}).

We test a power-law shape for the SED and also a log-parabola (LP) or power-law with an exponential cutoff (PLEC). 
There is no significant preference for the LP or PLEC over the simple power-law so we decide to assume this SED shape in the rest of the paper when we consider the models used in this section.

We test a power-law shape for the SED and also a log-parabola (LP) or power-law with an exponential cutoff (PLEC). 
There is no significant preference for the LP or PLEC over the simple power-law so we decide to assume this SED shape in the rest of the paper when we consider the models used in this section.

\subsection{Baseline fit using astrophysical models from the observations of M31 and M33 in other wavelengths}
\label{sec:altmodels}

\begin{figure}[ht]
\includegraphics[scale=0.35]{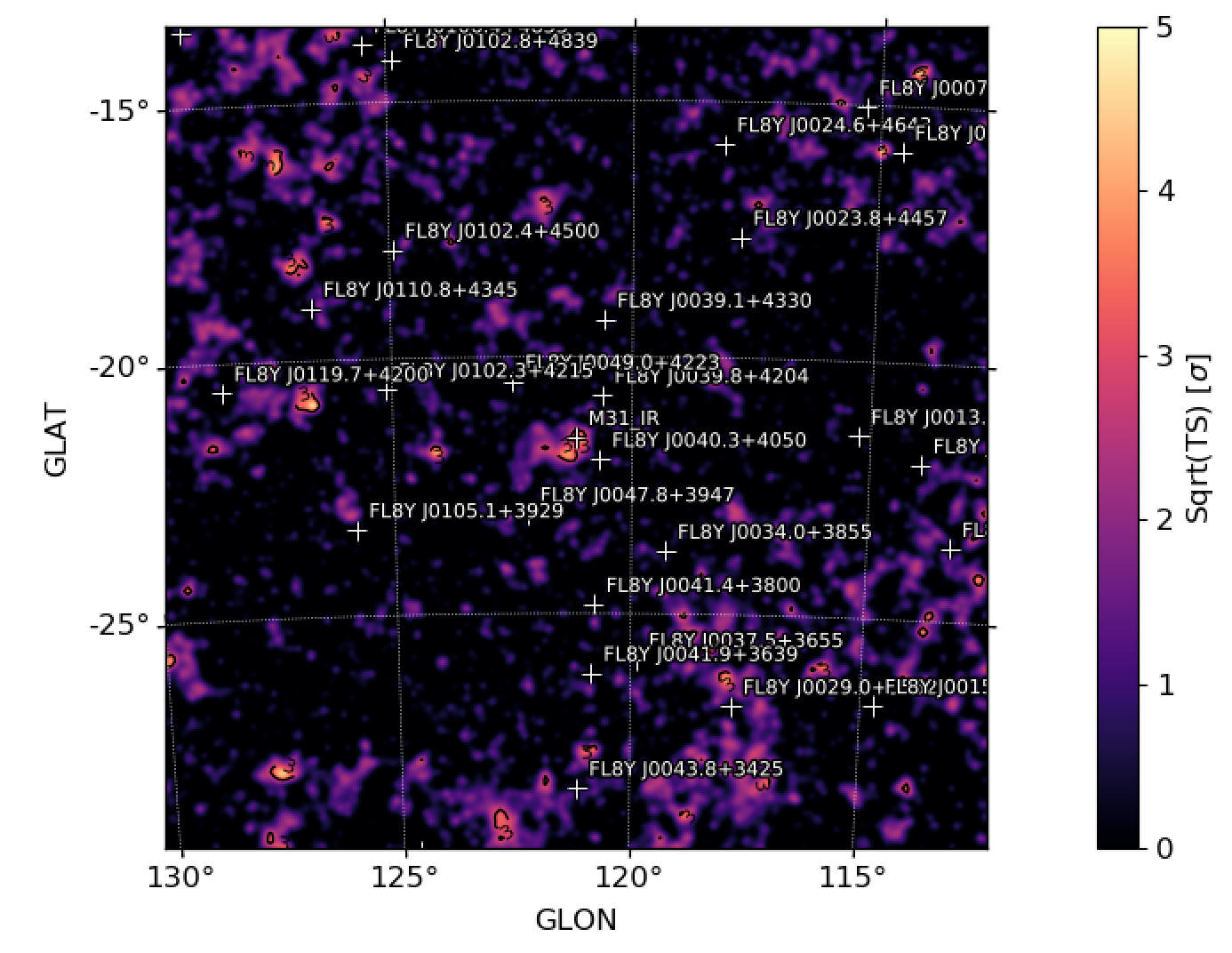}
\includegraphics[scale=0.35]{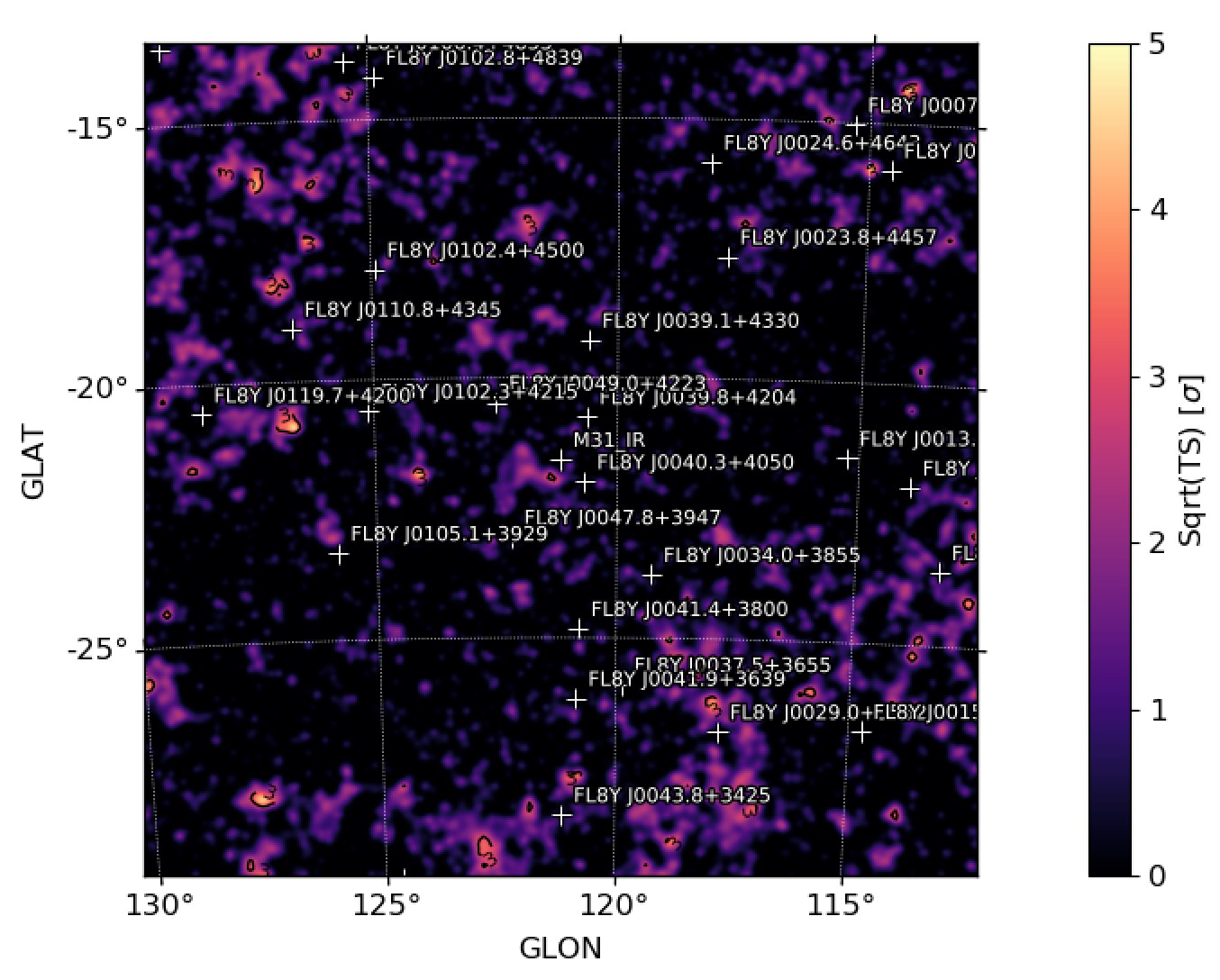} 
\caption{$TS$ map for the baseline model for M31 with atomic gas column density $N_{H}$ map (left panel) and {\it Spitzer}/IRAC map (right panel) for $E\in[0.3,1000]$ GeV and for a pixel size of $0\fdg08$.}
\label{fig:TSmapalt}
\end{figure}

\begin{table}[ht]
\begin{tabular}{c|c|c||c|c|c}
\hline \hline
\multicolumn{3}{c|}{M31} & \multicolumn{3}{c}{M33} \\
\hline
Model & $TS$ & $\Delta \log{L}$ & Model & $TS$ & $\Delta \log{L}$  \\ 
\hline
Disk & 98 & 0 & Point & 41 & 0 \\
Gauss & 97 & 1 & Gauss & 41 & 0 \\
Point & 80 & 5 & Optical ($3552$ A) & 41 & 0 \\
{\it Spitzer} (3.6$\mu$m) & 94 & 3 & {\it Spitzer} (24$\mu$m) & 45 & -2 \\
{\it Herschel} (160$\mu$m) & 75 & 9 & {\it Spitzer} (160$\mu$m) & 42 & 0 \\
$N_H$ & 65 & 18 & 2MASS (2.2$\mu$m) & 41 & 0
\end{tabular}
\caption{Value of the $TS$ and difference of likelihood with respect to the disk model ($\Delta \log{L}$) for M31 and M33 spatial models considered in our analysis. We show the results for the point source, disk and Gaussian geometrical models and for the templates taken from other wavelengths.}
\label{tab:altmodel}
\end{table}

The baseline fit reported in the previous section, with M31 modeled with a disk template and M33 with a point-like source, is tuned directly on $\gamma$-ray data. It is thus a phenomenological way to explain the $\gamma$-ray emission from M31 and M33 without any direct relation with what is observed from these galaxies in other wavelengths.
$\gamma$ rays are produced predominantly by the so-called interstellar emission that is traced by atomic gas density, radio and infrared emissions.
Therefore, we used maps derived from observations in other wavelengths as templates for the astrophysical components of the $\gamma$-ray emission from M31 and M33.
For M31 we use the {\it Herschel}/PACS map at 160 $\mu$m (which traces the star formation), {\it Spitzer}/IRAC map at 3.6$\mu$m (that traces old stellar population), and an atomic gas column density $N_{H}$ map from \cite{2009ApJ...695..937B} (which  traces  gas densities).
On the other hand, for M33 we consider the 2MASS infrared map at 2.2$\mu$m, the {\it Spitzer}/IRAC infrared maps at 24$\mu$m and at 160$\mu$m, and the Mayall optical map at $3552$ A.
The analysis used here is the same as in Sec.~\ref{sec:fitting} and we assume for the SED shape of these templates a simple power-law.
We report the values of the $TS$ for the different components in comparison with the result of the point source, disk and Gaussian templates in Tab.~\ref{tab:altmodel}.

Among the templates considered for M31, the $N_H$ map yields the fit with the lowest likelihood and $TS$. 
Indeed, we see in Fig.~\ref{fig:TSmapalt} that the $TS$ map for the case with $N_H$ template has more residuals than the one with {\it Spitzer}/IRAC template. 
The reason is that the $N_H$ map traces the disk of M31 that is extended about $3^{\circ}$ across the sky while the $\gamma$-ray emission detected by {\it Fermi}-LAT is concentrated within about $0.^{\circ}4$ with a spherical symmetry.

For the same number of degrees of freedom, the {\it Herschel}/PACS provides a slightly better fit that however it is not favored compared to a simple point source at the center of M31.
Finally the {\it Spitzer}/IRAC map provides the best fit to the data among the templates considered from other wavelengths. 
The {\it Spitzer}/IRAC infrared map is the most similar to those obtained with geometrical models because it is dominated by the bulge component and it is thus similar with what is observed in $\gamma$-ray data.
Therefore, the $\gamma$-ray emission that we detect with {\it Fermi}-LAT from the M31 direction is dominated by the bulge emission.
Indeed, the stellar bulge of M31 has a total mass of about $3.1 \cdot 10^{10}$ $M_{\odot}$ and a size of about 5 kpc \cite{2012A&A...546A...4T}.
The size is perfectly compatible with the extension that we measure in $\gamma$ rays.
Moreover, the M31 stellar bulge is a factor of about 5 more massive than the MW one.
These results are compatible with what has been presented in \cite{M31Fermi}.

The astrophysical templates we try for M33 give almost all the same significance which is the of the same order of the point source scenario.
Indeed, the infrared and optical emission observed from this galaxy have an extension of about $0\fdg2$ that is of the order of the point source emission.
Since the simple point source emission gives the same $TS$ as the templates derived from observations in other wavelength we use directly this model in the rest of the analysis.

\subsection{M31 and M33 DM morphology and correlations with other background sources}
\label{sec:DMsignal}
As we have seen in the previous sections we detect with {\it Fermi}-LAT a $\gamma$-ray signal that is well fit with a radial disk of size $0\fdg33$ for M31 and with a point source for M33.
We have demonstrated that the $\gamma$-ray flux in the direction of M31 is compatible with the emission from the stellar bulge while M33 with infrared and optical maps.

\begin{figure}[ht]
\includegraphics[scale=0.42]{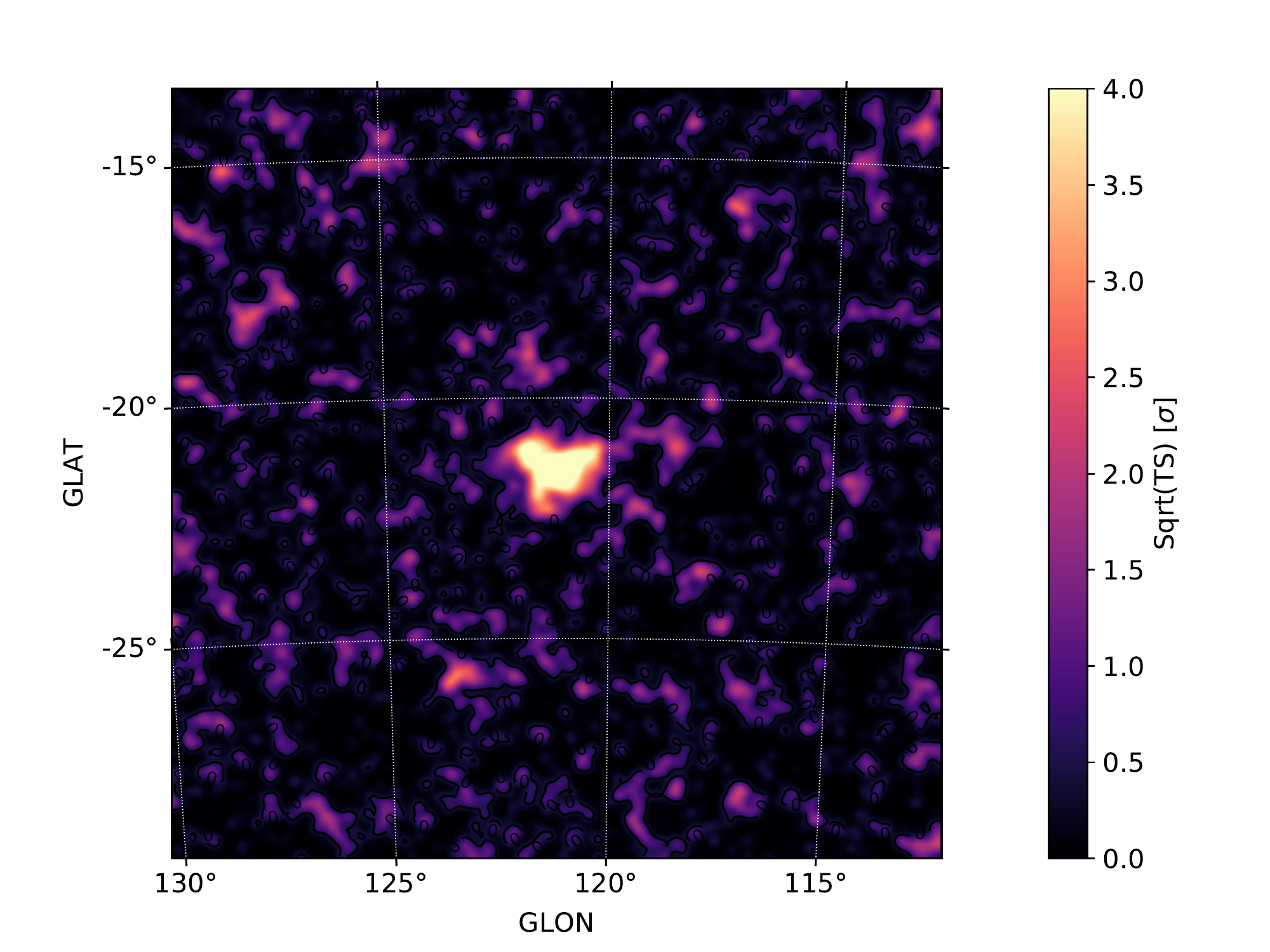}
\includegraphics[scale=0.42]{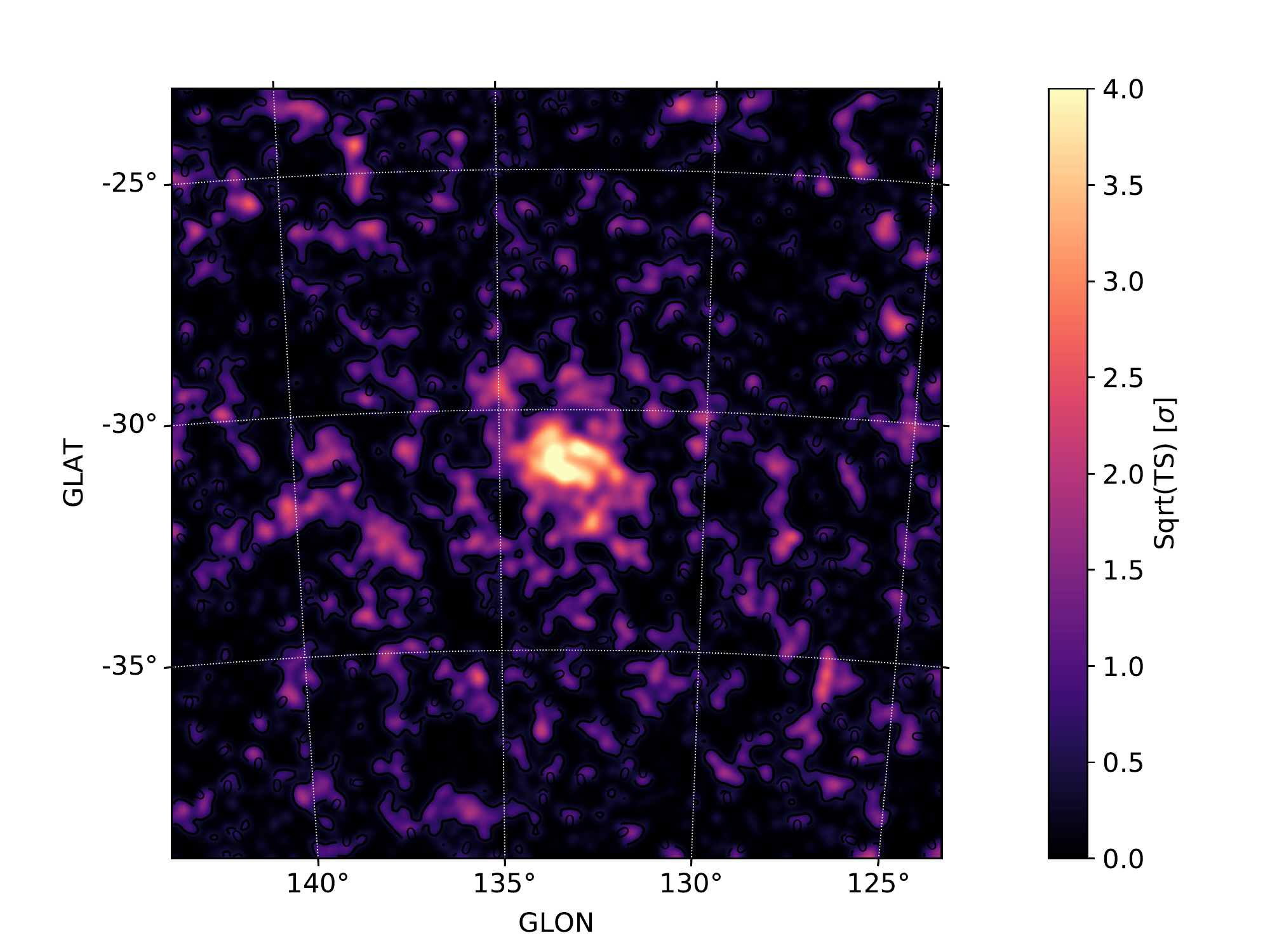} 
\caption{$TS$ maps for the simulations of a DM signal using the {\tt MED} DM model from M31 (left panel) and M33 (right panel). The peak of $TS$ present in the center of these maps is due to the DM flux.}
\label{fig:TSDMsim}
\end{figure}

A possible contribution might also come from DM so it is important to understand the {\it Fermi}-LAT sensitivity to such a signal.
In particular we are interested in determining the spatial morphology of a putative DM emission (i.e., determine whether it would be detected as point like or extended source and in the latter case estimate the size of extension) and calculating the correlations between this component and the other background sources. 
We use {\tt Fermipy} to simulate a DM signal from M31 and M33 DM templates and we derive its size and spatial morphology.
We take the baseline model from Sec.~\ref{sec:fitting}, we remove M31 and M33 sources from the model and we add a DM contribution using the templates reported in Sec.~\ref{sec:dm}.
We simulate a DM signal that would give a detection at $9\sigma$ significance with a power-law SED with index $2.0$.
We use such a simple power-law SED, as we are interested here only in finding the spatial morphology of the DM signal.
We show in Fig.~\ref{fig:TSDMsim} the $TS$ map for the $\gamma$-ray emission from this DM signal for M31 and M33 for the {\tt MED} DM model. A very similar $TS$ map is found when using the {\tt MIN} or the {\tt MAX} DM model.
Then, we fit this excess with an extended source finding that it is well fitted with a Gaussian template with size of about $0\fdg 5$ and $TS_{\rm{EXT}}=30$ for M31 and $0\fdg 9$ and $TS_{\rm{EXT}}=25$ for M33.
The Gaussian template has about the same likelihood value as the disk one.
The correlation coefficients between the M31 DM template normalization (spectral index) and the isotropic and IEM template normalizations are -0.07 (-0.14) and -0.23 (-0.12), respectively.
Instead, the correlation coefficients between the M33 DM template normalization (spectral index) and the isotropic and IEM template normalizations are -0.04 (-0.19) and -0.16 (-0.15), respectively. 
Therefore, no significant correlations are present between the DM template of M31 and M33 and IEM and isotropic components.
The correlation coefficients between the DM component normalization and the M31 disk normalization is -0.87; instead with the M33 point source SED normalization it is -0.61.
Therefore, SED parameters of the M31 and M33 DM templates are correlated with the SED of the disk template for M31 and M33, respectively.
We consider these correlations in our analysis as we will explain in Sec.~\ref{sec:DMsed}.
We find similar results using the {\tt MIN} or {\tt MAX} DM models.

In the line of sight of M31 and M33 a contribution of $\gamma$ rays could also come from DM present in the MW.
As shown in Fig.~\ref{fig:d-and-j-factors}, the $J$-factor for the MW is much smaller than the M31 and M33 component in the inner few degrees from M31. Moreover, the DM MW signal is almost isotropic with a variation of about a factor of $10\%$ across the M31 and M33 ROIs.
We test a possible effect of the presence of this additional DM component by taking the simulation done before, which includes the DM M31 component, and adding also the contribution of DM from the MW.
We then ran a fit and found that the MW contribution is almost completely absorbed by the isotropic template. Moreover, there is no correlation between the MW and the M31 DM templates.
Therefore, we decide in the rest of our analysis to not add this component to the model. 

\subsection{DM SED and significance}
\label{sec:DMsed}

\begin{figure}[ht]
\includegraphics[scale=0.37]{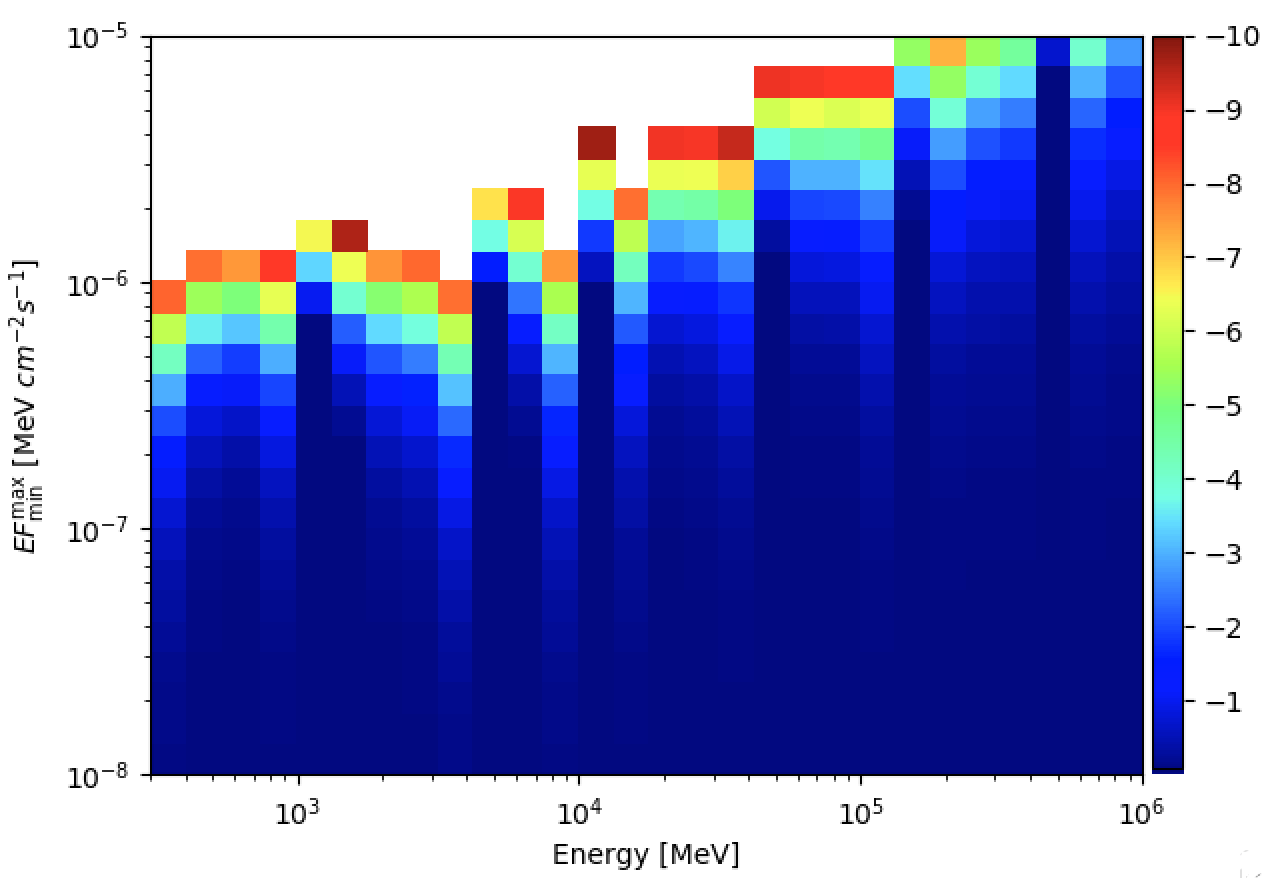}
\includegraphics[scale=0.37]{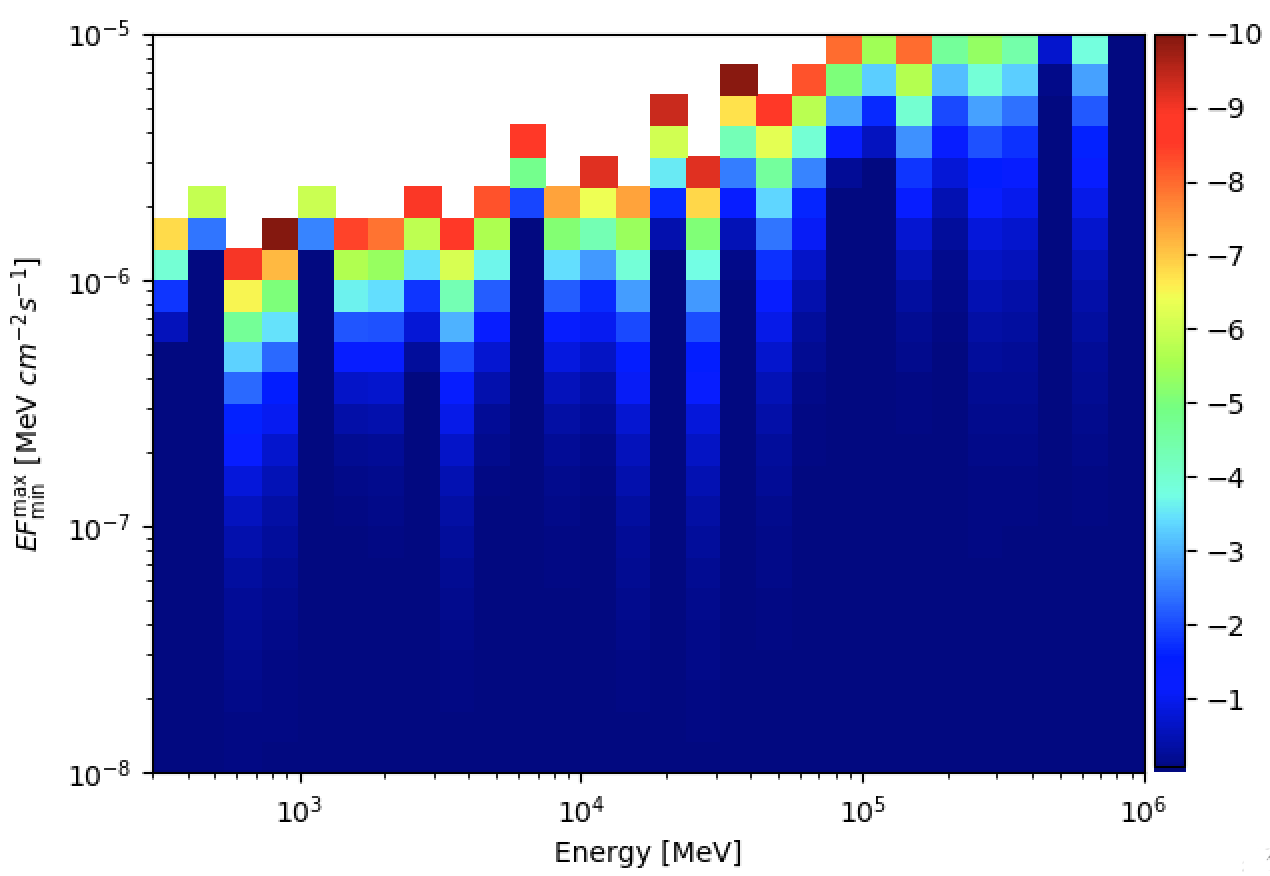} 
\caption{DM SED (reported as energy multiplied for the photon flux in each energy bin) for M31 (left panel) and M33 (right panel) for $E\in[0.3,1000]$ GeV. We consider here the {\tt MED} DM model for both sources. The different colors are related to values of the $\Delta \log{\mathcal{L}}$ (see Eq.~\ref{eq:ts}).}
\label{fig:sed}
\end{figure}

At this stage of the analysis we add to the baseline model, which includes the template for astrophysical emission from M31/M33 (i.e., a disk template for M31 and a point like source for M33), the DM template. We run a fit and then compute the likelihood profile as a function of energy and energy flux of DM.
We scan in each energy bin the likelihood as a function of the flux normalization for the assumed DM signal which is specified by the choice of decay or annihilation, by the channel and the DM mass. In the rest of the paper we will consider the decay and annihilation into $b\bar{b}$ and $\tau^+\tau^-$. The case of $b\bar{b}$ is representative of hadronic channels such as quarks and gauge bosons while $\tau^+\tau^-$ represents the leptonic channels $\mu^+\mu^-$ and $e^+e^-$.
For this bin-by-bin scan, we fix the SED parameters of the sources that have an absolute value of the correlation parameter smaller than 0.10. 
By analyzing each energy bin separately, we avoid selecting a single spectral shape to span the entire energy range 
at the expense of introducing additional degrees of freedom into the fit. 
For the fit in any given bin, the only free parameter describing the DM component is the normalization. 
In Fig.~\ref{fig:sed} we report the SED of the DM template for M31 and M33 with the {\tt MED} model.

\begin{figure}[ht]
\includegraphics[scale=0.37]{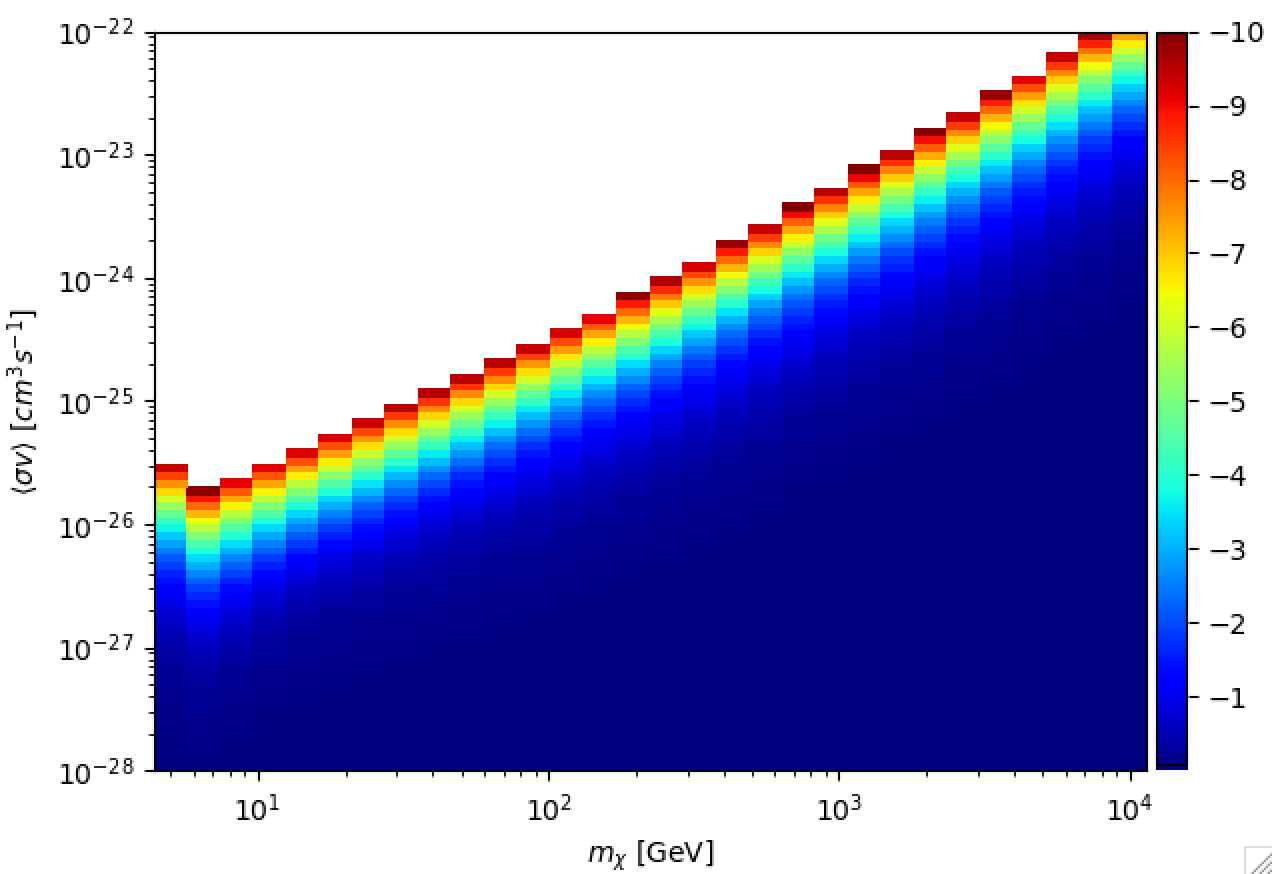}
\includegraphics[scale=0.37]{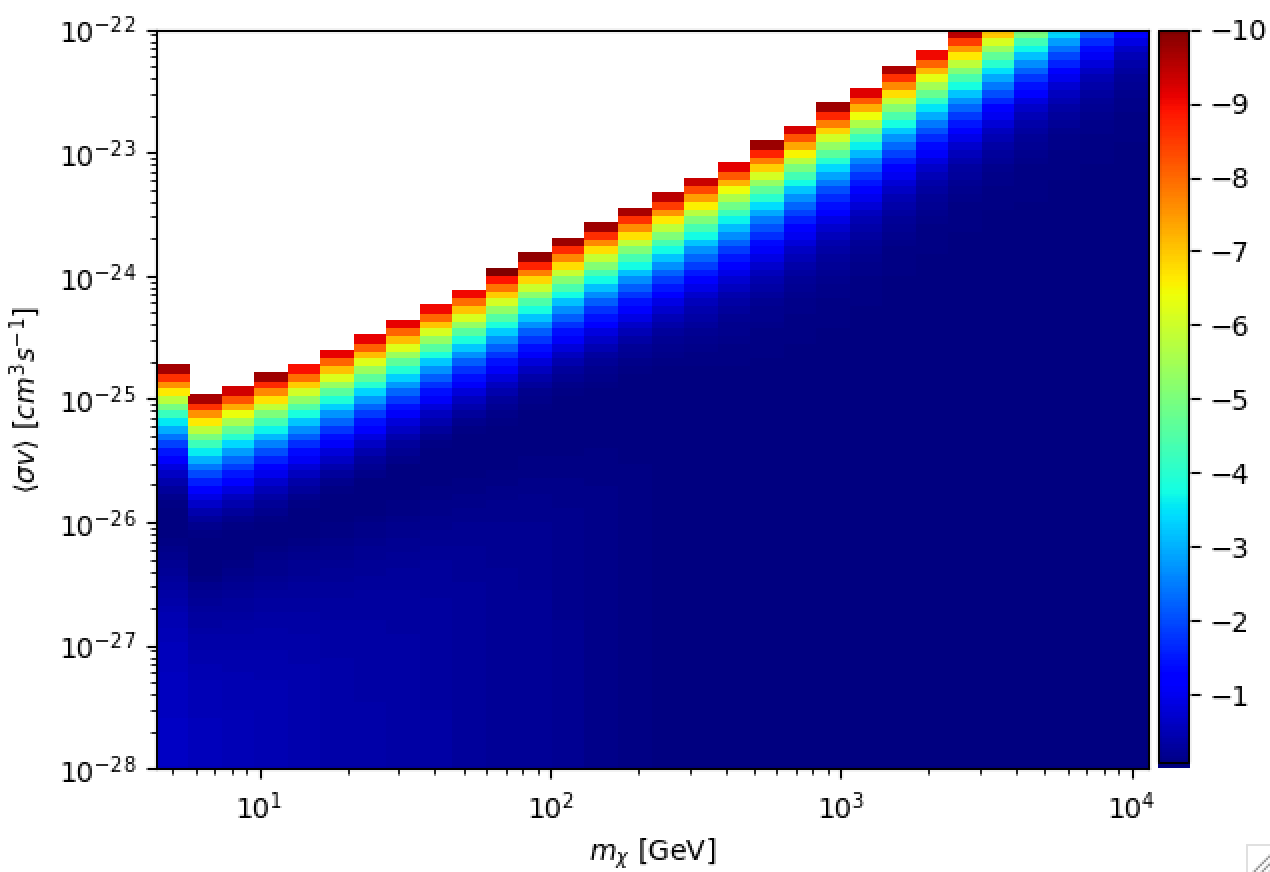} 
\caption{Likelihood profile for $\langle \sigma v \rangle$ as a function of the DM mass for M31 (left panel) and M33 (right panel) for $E\in[0.3,1000]$ GeV. These plots are for DM annihilating into $b\bar{b}$ quarks. The different colors are related to values of the $\Delta \log{\mathcal{L}}$.}
\label{fig:castro}
\end{figure}

In the final step of the fitting procedure we convert the bin-by-bin likelihood curve in flux into a likelihood curve in $\langle \sigma v \rangle$ for each spatial profile and annihilation or decay channel, which determines the spectrum. We scan DM masses ($m_{\chi}$) for 5--10000 GeV (when kinematically allowed in the annihilation or decay channel under consideration), and the pair-annihilation or decay final states $\tau^+\tau^-$ and $b\bar{b}$. 
For each DM spectrum, we extract the expected flux, $F_{j}$, in each energy bin and calculate
the likelihood of observing that flux value. The log-likelihood in each energy bin is summed to get the log-likelihood curve, defined as: 

\begin{equation}
\mathrm{ln}~\mathcal{L}(\mu, \theta | \mathcal{D}) = \sum_{j} \mathrm{ln}~\mathcal{L}_{j}(\mu, \theta_{j} | \mathcal{D}_{j})  \label{eq:llcurve},
\end{equation}
where $\mathcal{L}$ is the likelihood, $j$ runs over the energy bins of {\it Fermi}-LAT data ($\mathcal{D}$), $\mu$ are the DM parameters ($\langle \sigma v \rangle$ or $\tau$ and $m_{\chi}$) and $\theta$ are all the other parameters in the background model, i.e. the nuisance parameters.

Therefore, the DM SED has all the information needed to find if the presence of DM is significant or not for any possible DM annihilation channel and mass.
Indeed, we can choose a DM annihilation or decay channel and convert the SED into the likelihood profile as a function of the DM $\langle \sigma v \rangle$ and $m_{\chi}$.

We can also evaluate the significance of the DM hypothesis using the $TS$ defined as: 
\begin{equation}
TS= 2~\Delta \log{\mathcal{L}} = 2~\mathrm{ln}\frac{\mathcal{L}(\mu, \theta | \mathcal{D})}{\mathcal{L}_{\mathrm{null}}(\theta | \mathcal{D})}, \label{eq:ts}
\end{equation}
where $\mathcal{L}_{\mathrm{null}}$ is the likelihood for the null signal of DM, $\mathcal{L}$ is the likelihood for the presence of DM.
For the energy bins up to about 10 GeV the statistics are large enough that Chernoff’s theorem applies, and we expect the $TS$-distribution to follow a $\chi^2$ distribution \cite{REF:Chernoff}. At higher energies, the counts per bin are in the Poisson regime and the $\chi^2$ distribution moderately over-predicts the number of high $TS$ trials observed in simulated data.

From $\mathrm{ln}~\mathcal{L}(\mu, \theta | \mathcal{D})$ we can evaluate the one-sided 95\% confidence level (CL) exclusion limit on the flux as the point at which the p-value for a $\chi^{2}$ distribution with 1 degree of freedom is 0.05 when we take the maximum likelihood estimate as the null hypothesis. That is, the 95\% CL upper limit on the flux assigned to DM is the value at which the log-likelihood decreases by 1.35 with respect to its maximum value.
An example of this is shown in Fig.~\ref{fig:castro} for $b\bar{b}$ annihilation channel for M31 and M33.


\section{Results}
\label{sec:results}

In Sec.~\ref{sec:modind}, we first show a model-independent analysis where we search for an excess in M31 and M33 ROIs without using any DM model.
In Sec.~\ref{sec:sys} we show how the limits on the DM annihilation cross section change by considering a different energy range for the analysis, size of the ROI, assuming different templates for the astrophysical emission of M31 and M33.
Then, in Sec.~\ref{sec:sim} we use 100 simulations to calculate the expected limits on DM in the null signal hypothesis.
In Sec.~\ref{sec:dminterp} we assume that all signal comes from DM and we derive best-fit contours in cross section and mass parameters space. As there is no compelling evidence that the emission from these galaxies is solely due to DM and in fact the best fit region is in tension with the DM search from other targets (e.g., dSphs or the Galactic Center), in Sec.~\ref{sec:dmplusastro} we set DM limits.
We do this for the $b\bar{b}$ and $\tau^+\tau^-$ channels and for the {\tt MAX}, {\tt MED} and the {\tt MIN} DM distributions considered in Section~\ref{sec:dm}. These channels were previously considered in the dwarf spheroidal 
analyses~\cite{Ackermann:2015zua, Drlica-Wagner:2015xua}. 

\subsection{DM model independent search for an excess in M31 and M33 ROIs}
\label{sec:modind}

As discussed in Sec.~\ref{sec:dm} the DM spatial distribution can vary significantly by assuming a {\tt MAX}, {\tt MED} and the {\tt MIN} model or considering annihilation or decay of DM particles.
In addition to this, the DM SED is also uncertain and can vary for the different annihilation or decay channels.

In this section we describe a search for a radial dependent excess by adding to the M31 and M33 ROIs three uniform annuli with radial shapes: $r\in[0.4^{\circ},3.5^{\circ}]$, $r\in[3.5^{\circ},6.0^{\circ}]$ and $r\in[6.0^{\circ},8.0^{\circ}]$.
For this analysis we use a ROI width of $20^{\circ}$ to avoid edge effects in the farthest annulus and we select an energy range between $0.3-1000$ GeV.
The analysis pipeline is the same presented for the baseline fit in Sec.~\ref{sec:fitting} where we leave free to vary in the fit the SED parameters of point sources, the normalization of the isotropic template and the normalization and slope of the IEM. We model the emission from M31/M33 with templates described in Sec.~\ref{sec:fitting}.

\begin{figure}[ht]
\includegraphics[scale=0.43]{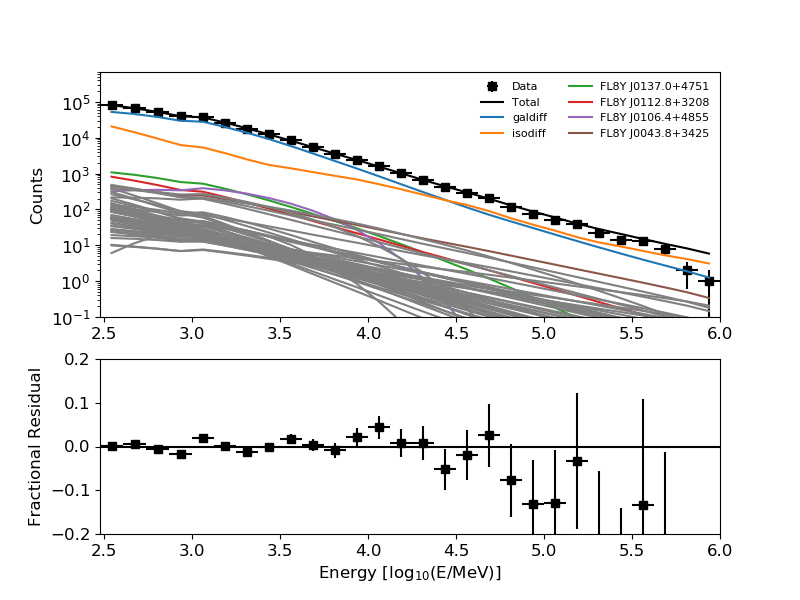}
\includegraphics[scale=0.43]{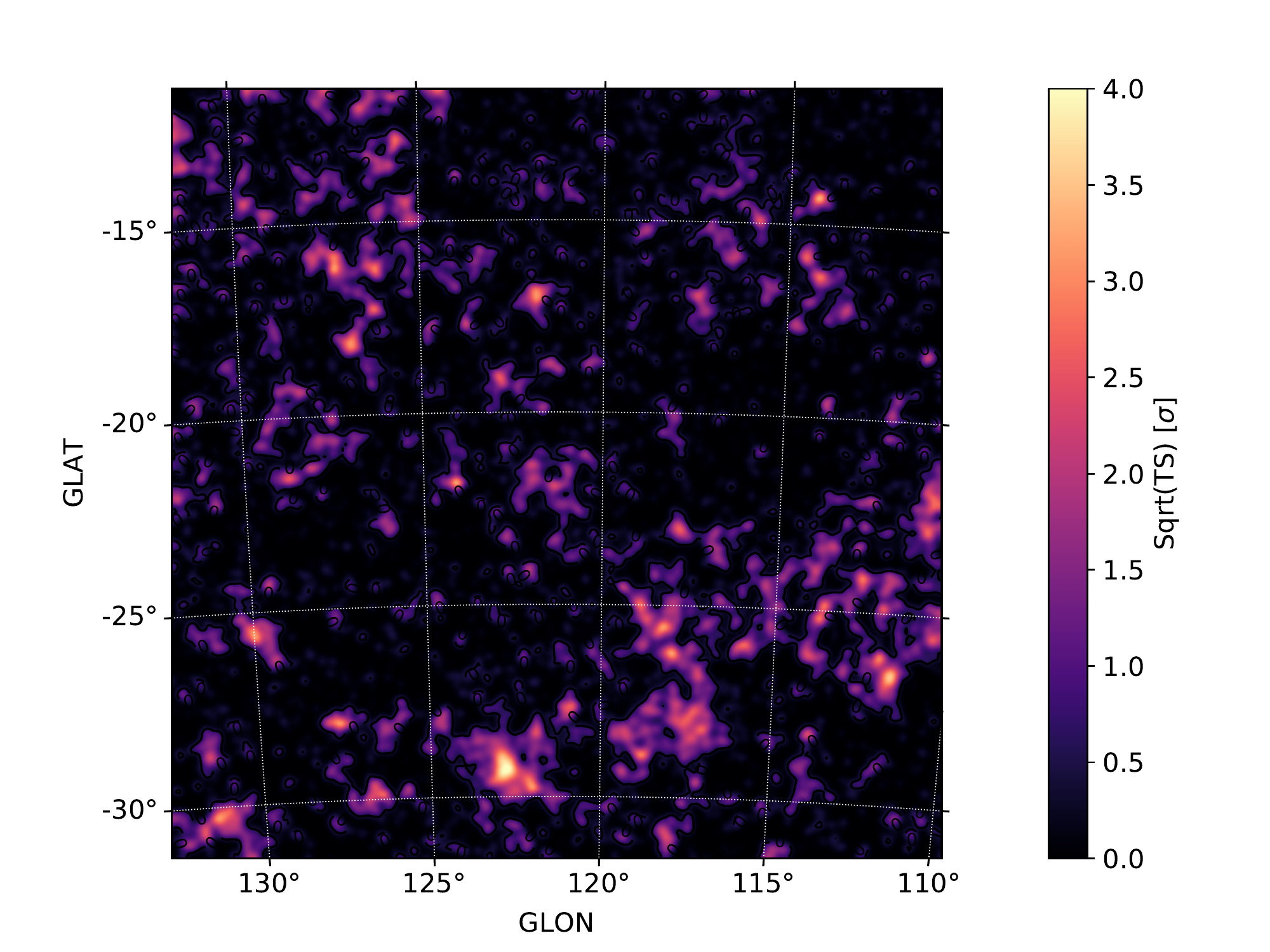} 
\includegraphics[scale=0.43]{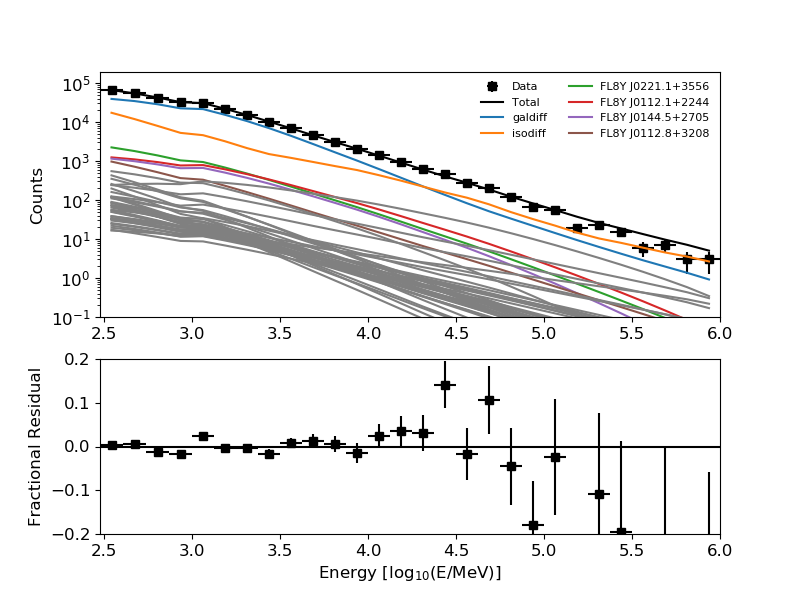}
\includegraphics[scale=0.43]{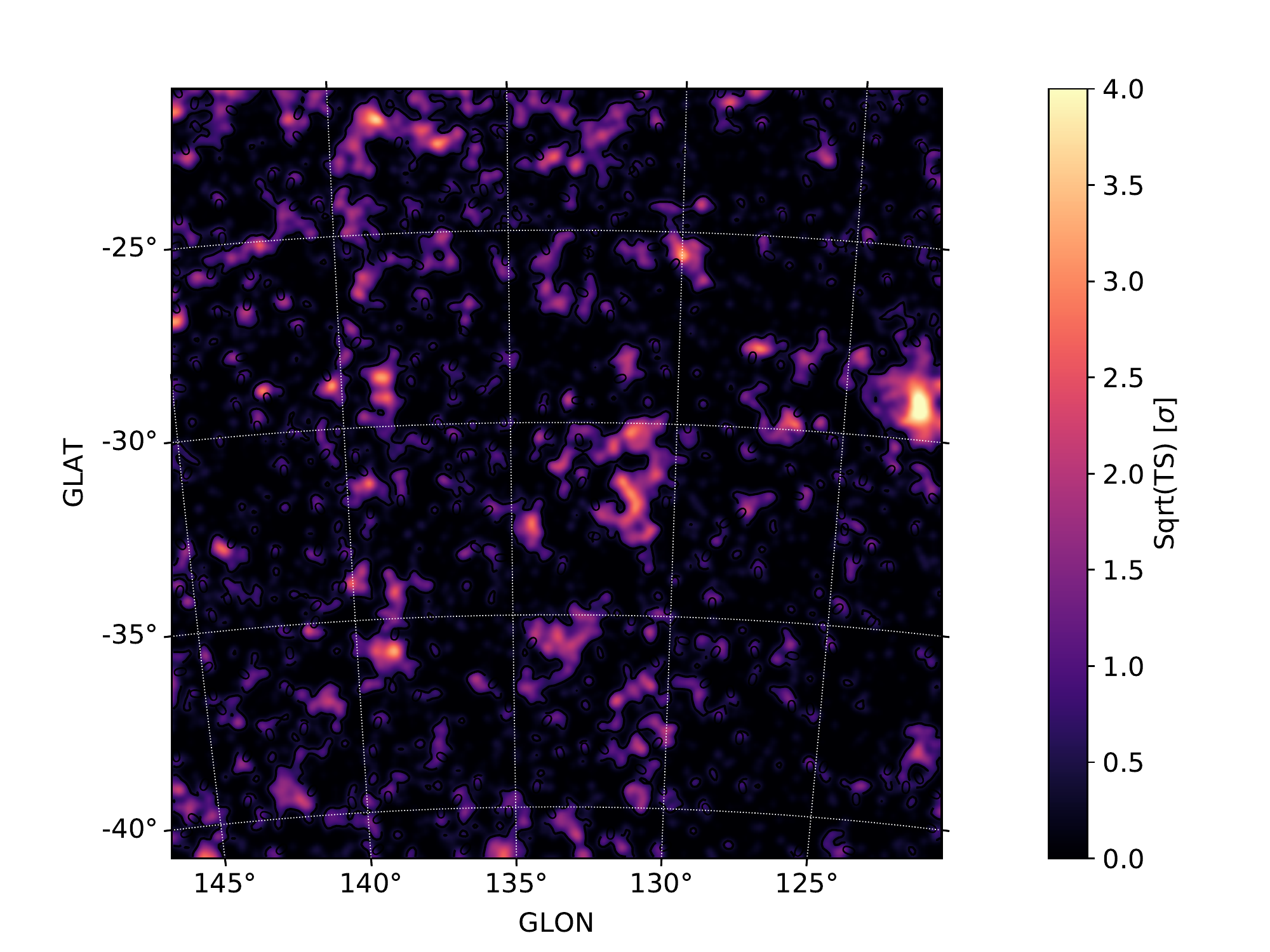} 
\caption{Left panels: count spectrum of all the component in the model together with the fractional residuals between the data and the best-fit model. Top left panel is for M31 while bottom left is for M33. Right panels: $TS$ maps for the fit to M31 (top panels) and M33 (bottom panels) ROIs. In these fits the SED parameters of point sources, the normalization of the isotropic template and the normalization and slope of the IEM are free to vary. We use here the Off IEM model and correspondent isotropic template.}
\label{fig:annuliallfree}
\end{figure}

We show in Fig.~\ref{fig:annuliallfree} the count spectrum with the residuals and the $TS$ map for this fit to M31 and M33 ROIs using the Off IEM and isotropic templates.
The change in normalization of the isotropic template has a best fit value of 1.00 (0.914) while the change of normalization and slope of the IEM are 1.010 (1.048) and $-0.03$ (-0.02) for M31 (M33). Therefore, for both M31 and M33 the deviation of the isotropic and IEM SED parameters from their input values are minimal.

Now we add to the model the three annuli and we re-do the fit. 
The results are reported in Table \ref{tab:annuli} where we can see that the $TS$ for all the three annuli is very small.
We run the same analysis for M31 also with the 8 Alt IEMs finding even tighter constraints.
The $TS$ is $\approx 0$ for all the three annuli with upper limits for the flux of $3-4 \cdot 10^{-10}$ ph/cm$^2$/s, $2.5-3.5 \cdot 10^{-10}$ ph/cm$^2$/s and $2.5-3.5 \cdot 10^{-9}$ ph/cm$^2$/s for the first, second and third annulus, respectively, and depending on the Alt IEM considered.

\begin{figure}[ht]
\includegraphics[scale=0.43]{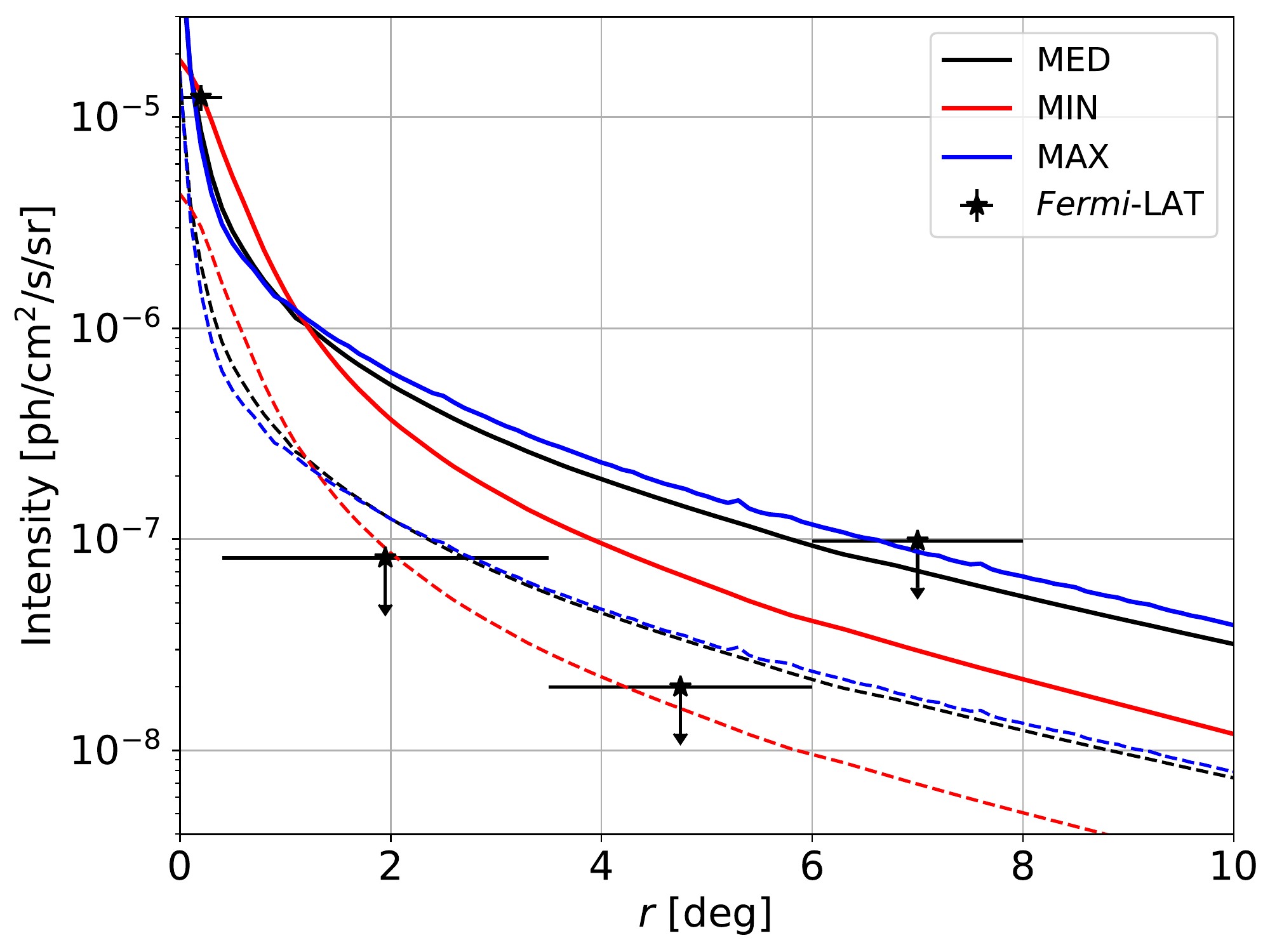}
\includegraphics[scale=0.43]{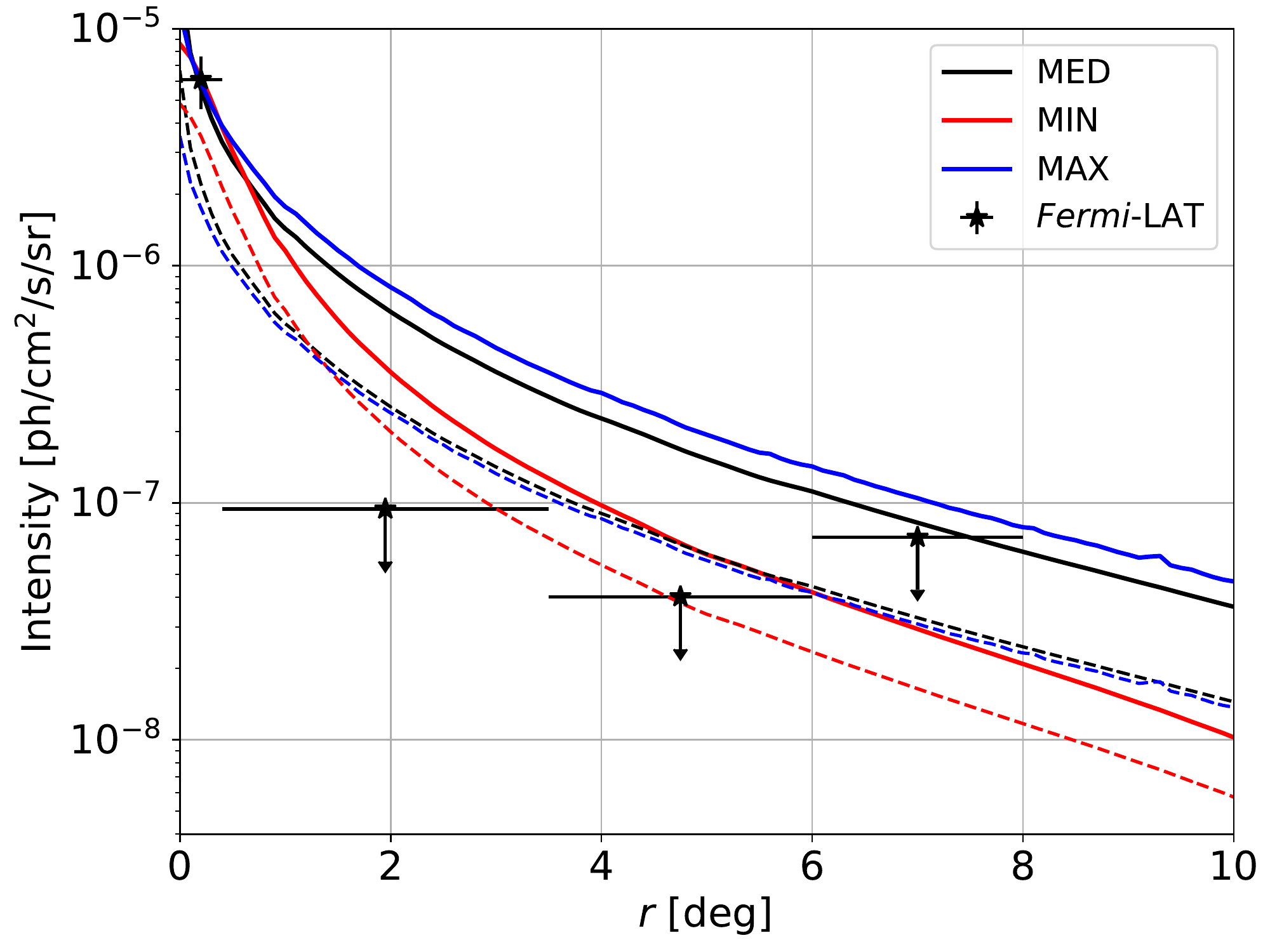} 
\caption{Intensity upper limits (black data) derived from our analysis in annuli of the M31 (left panel) and M33 (right panel) ROIs. The data point in the inner $0\fdg4$ represents the intensity for the astrophysical emission of the sources taken as a disk template for M31 and a point source for M33. The solid lines are the intensity for DM normalized to fit the astrophysical emission of M31 and M33. The dashed lines have been derived from the upper limits for the DM flux derived in Sec.~\ref{sec:dmplusastro}.}
\label{fig:JmapULs}
\end{figure}

In Fig.~\ref{fig:JmapULs} we show the intensity (the flux divided by the solid angle) upper limits found above together with the intensity of the DM contribution for {\tt MAX}, {\tt MED} and {\tt MIN}  DM models. We report the case where the DM flux fits the astrophysical contribution from M31 taken as a disk template and M33 as a point source. 
These DM intensity profiles are in tension with the upper limits for the annuli derived in this section. This analysis shows that if DM contributes entirely to the $\gamma$-ray emission from these two galaxies we should be able to detect a signal also in the outer region of M31 and M33, i.e. for $r\geq 2^{\circ}$, which would be absorbed by the annuli.
These results might change considering different assumptions for the DM distribution in M31 and M33.
We also display the DM intensity profiles for the upper limits of the flux reported in Sec.~\ref{sec:dmplusastro}. These contributions are compatible with the upper limits found in this section since they have to be considered as upper limits.

A possible weak point of this analysis is that we are leaving free to vary also the normalization of the isotropic template. 
If residuals are present in the ROI and their spatial distribution is mostly isotropic they can be absorbed by the isotropic template. Therefore, we decide to change the analysis made before by fixing the isotropic normalization in the fit to the value found in a control region. We choose the center of the control region to be at the same longitude of the M31 and M33 ROIs but at a latitude $20^{\circ}$ below. In this way the contribution of the isotropic is higher and we are able to constrain effectively its normalization.

\begin{table}[ht]
\begin{tabular}{c|ccc|ccc}
\hline \hline
  ISO free     		& annulus 1 		& annulus 1 		& annulus 3 		& annulus 1 		& annulus 1 		& annulus 3	\\ \hline
 TS      			& 0				&    0				&   	8			& 0				&    0				&	0		 \\ \hline
 Flux [ph/cm$^2$/s]  & $9.45 \cdot 10^{-10}$   &   $4.53 \cdot 10^{-10}$  &   $4.62 \cdot 10^{-9}$    & $1.09 \cdot 10^{-9}$   &   $9.12 \cdot 10^{-10}$  &   $1.91 \cdot 10^{-9}$	\\
\hline \hline
  ISO fixed     		& annulus 1 		& annulus 1 		& annulus 3 		& annulus 1 		& annulus 1 		& annulus 3	\\ \hline
 TS      			& 5				&    0				&   	20			& 0				&    0				&	0		 \\ \hline
 Flux [ph/cm$^2$/s]  & $1.82 \cdot 10^{-9}$   &   $8.05 \cdot 10^{-10}$  &   $6.56 \cdot 10^{-9}$    & $1.09 \cdot 10^{-9}$   &   $1.05 \cdot 10^{-9}$  &   $2.41 \cdot 10^{-9}$	\\
\end{tabular}
\caption{Summary table for the $TS$ of detection ($TS$) and flux in our analysis of {\it Fermi}-LAT data in the M31 and M33 ROI (in the left and right side) with 3 uniform annuli as explained in the text. The top (bottom) part of the table is with the isotropic template free to vary (fixed to the control region).}
\label{tab:annuli}
\end{table}

We find a best fit value for the isotropic normalization of 0.866 (0.889) for M31 (M33) in the control regions and we apply these values to the M31 and M33 ROIs fixing this parameter in the fit.
Even if we fix the isotropic in the fit, the residuals in the count map are still basically compatible to 0 in the entire energy range, similar to what is shown in Fig.~\ref{fig:annuliallfree}. 
Also the ROI $TS$ maps remain almost unchanged with no significant larger-scale residuals.

The results with the three uniform annuli are reported in Table \ref{tab:annuli}.
The $TS$ for the annuli is 5 (0) for the inner, 0 (0) for the second and 20 (0) for the outermost. 
The upper limits for the flux are slightly larger than in the previous case where we leave the isotropic template free to vary.
We run the same analysis for M31 also with the 8 Alt IEMs finding even tighter constraints.
The $TS$ is $\approx 0$ for the inner two annuli and between $5-10$ for the outermost with upper limits for the flux of $5-10 \cdot 10^{-10}$ ph/cm$^2$/s, $4-7 \cdot 10^{-10}$ ph/cm$^2$/s and $3-5 \cdot 10^{-9}$ ph/cm$^2$/s for the first, second and third annulus, respectively.

\subsection{Systematics in the DM results}
\label{sec:sys}

\begin{figure}[ht]
\includegraphics[scale=0.42]{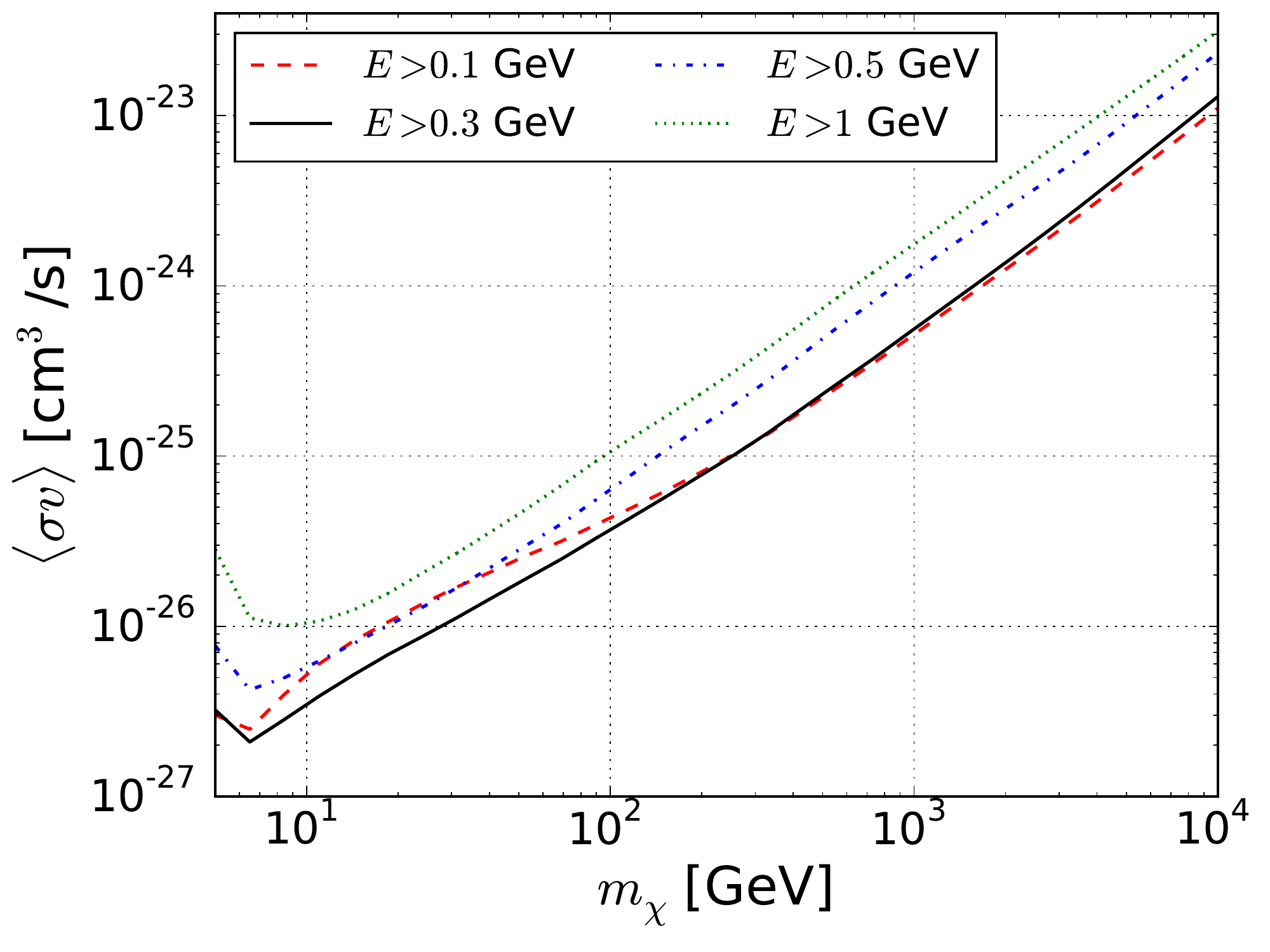} 
\includegraphics[scale=0.42]{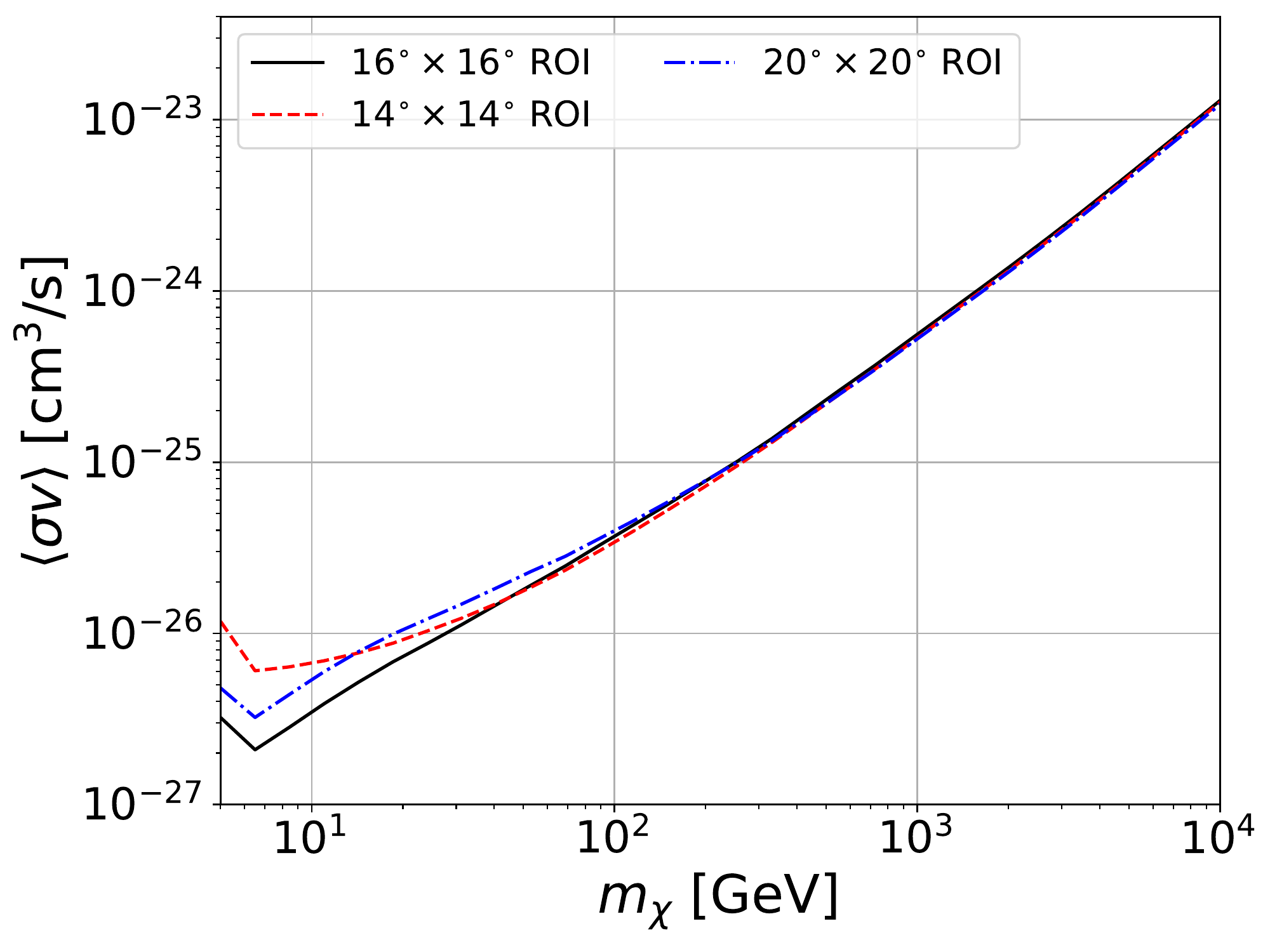}
\includegraphics[scale=0.42]{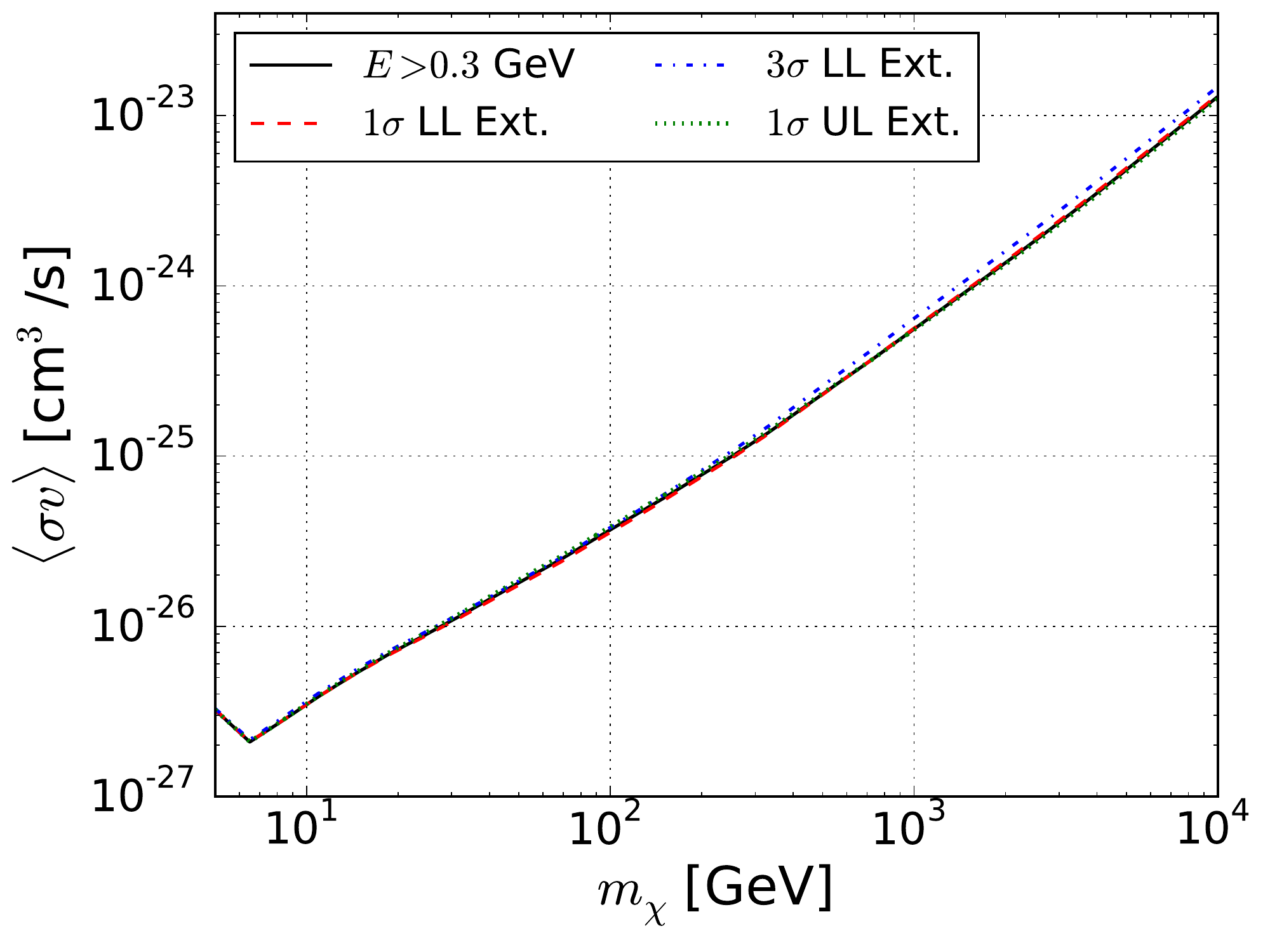}
\includegraphics[scale=0.42]{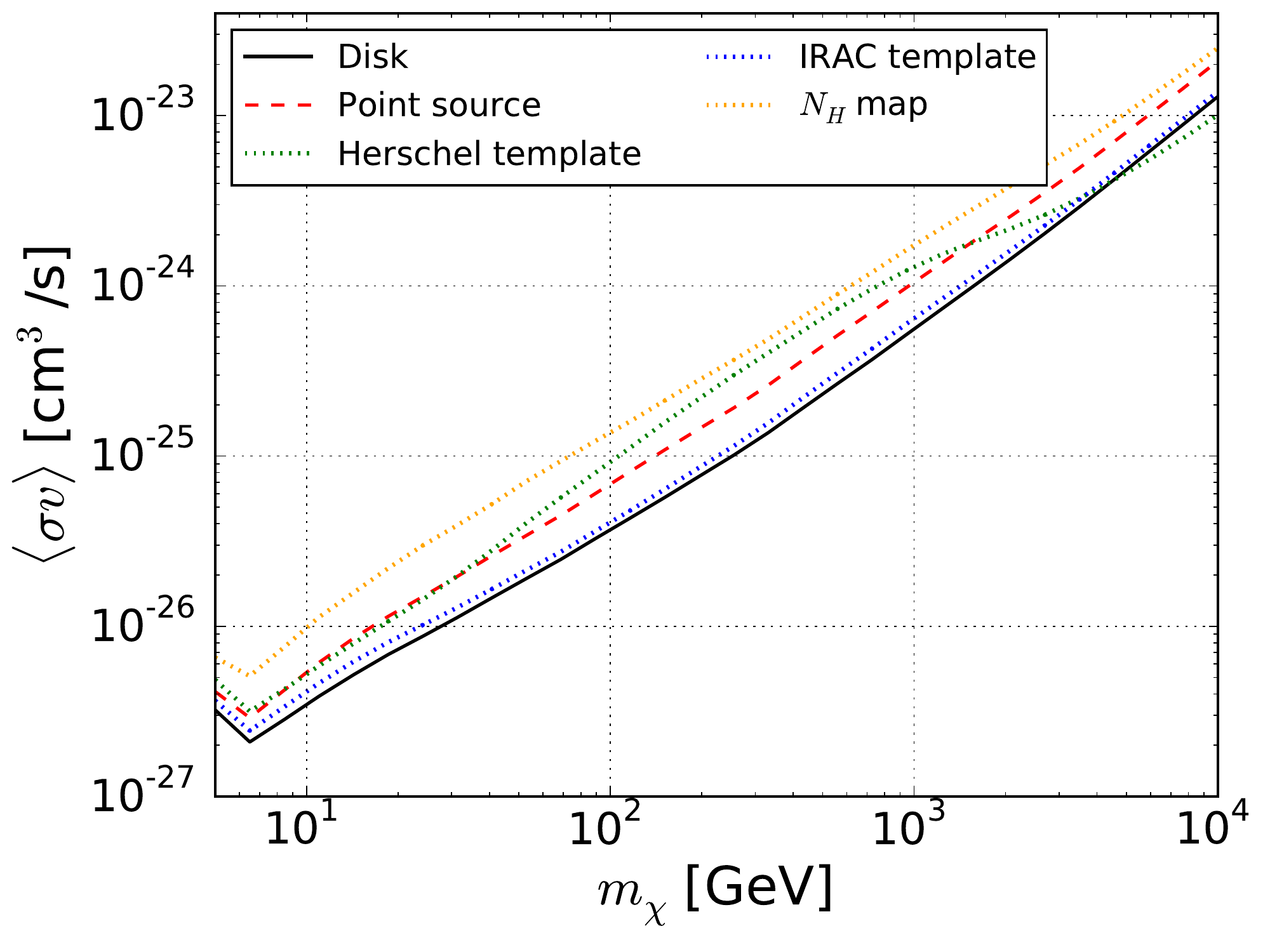}
\includegraphics[scale=0.42]{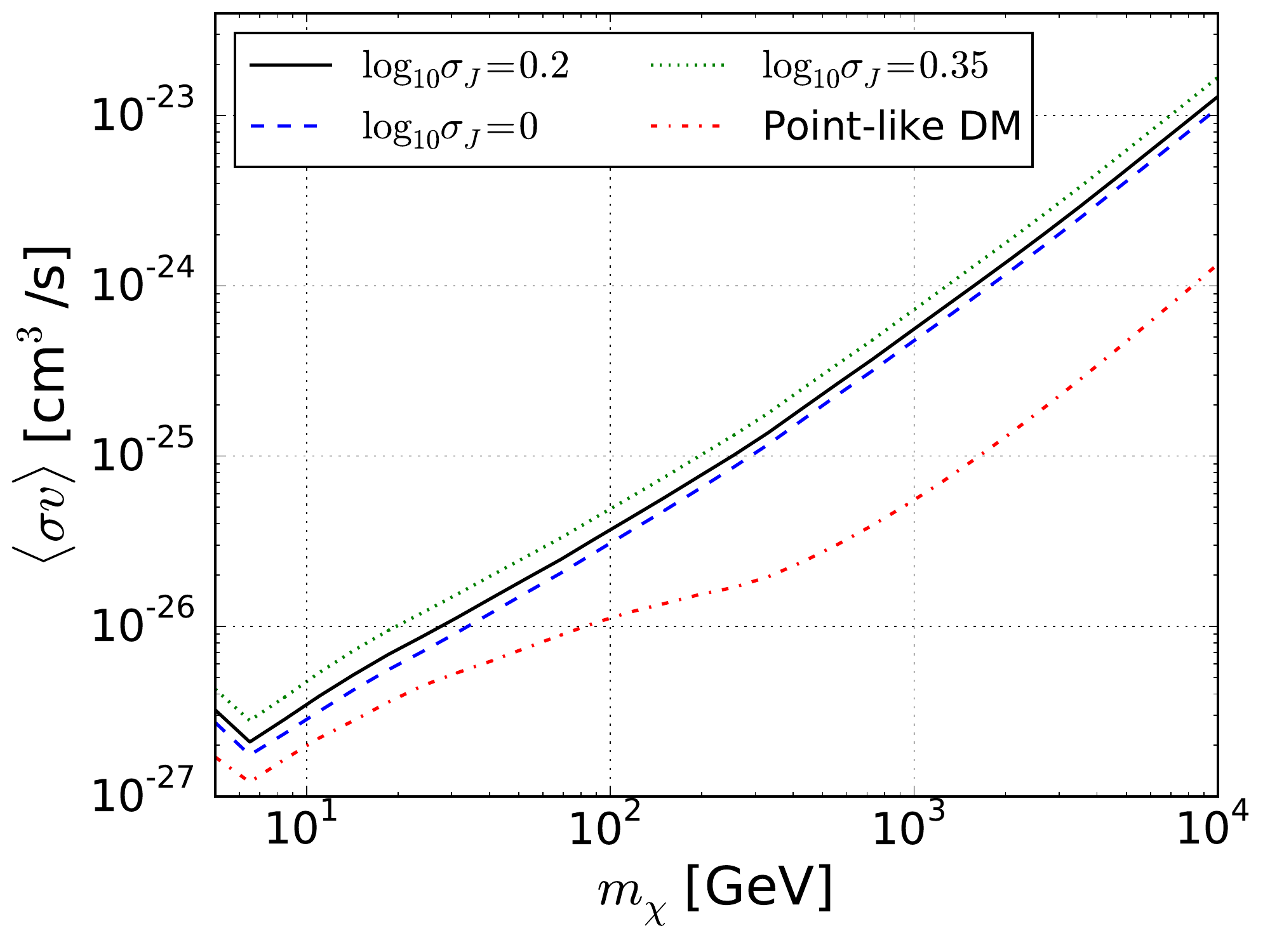} 
\includegraphics[scale=0.42]{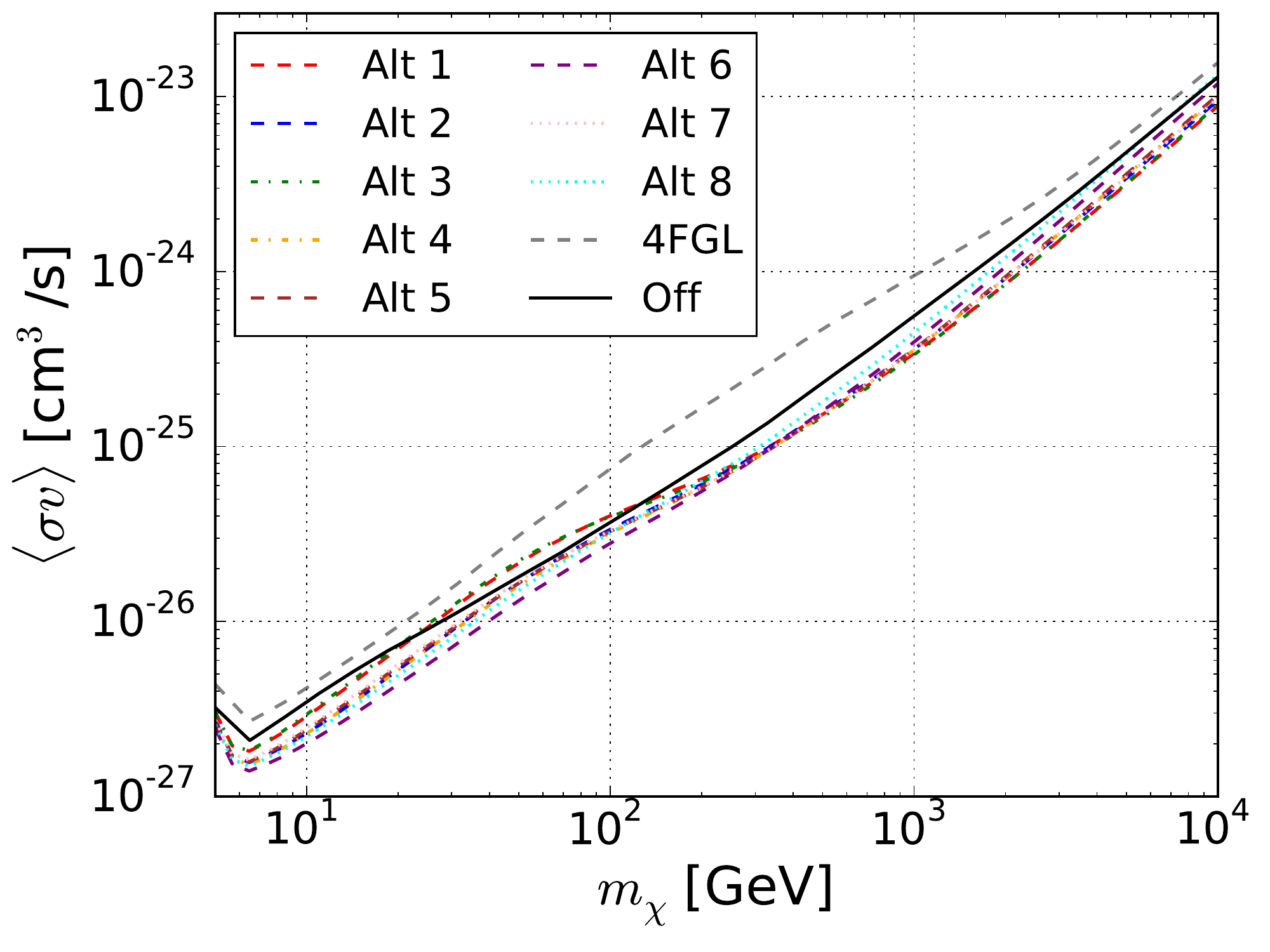} 
\caption{$95\%$ C.L. upper limits for $\langle \sigma v \rangle$ as a function of $m_{\chi}$ for the following cases. The benchmark case reported with a solid black line in these plots is for analysis of data at $E>0.3$ GeV, with the Off IEM, an ROI width $16^{\circ}\times16^{\circ}$, with a disk template for the astrophysical emission and with an uncertainty of the $J$ factor of $\log_{10}\sigma_{J}=0.2$. 
Top left: limits derived for different choices of the lower bound of the analysis: $E>0.3$ GeV (black solid line), $E>0.1$ GeV (red dashed line), $E>0.5$ GeV (blue dot-dashed line) and $E>1$ GeV (green dotten line). 
Top right: limits derived for an ROI width $16^{\circ}\times16^{\circ}$ (black line), $14^{\circ}\times14^{\circ}$ (red dashed line) and $20^{\circ}\times20^{\circ}$ (blue dot-dashed line).
Center left: limits derived for different sizes of the disk template: best-fit value (black solid line), $1\sigma_{\rm{EXT}}$ lower (upper bound) with red dashed (green dotted) line and $1\sigma_{\rm{EXT}}$ lower limit (with blue dot-dashed line). 
Center right: limits derived for different choices of the astrophysical emission: disk template (black solid), point source (red dashed), {\it Herschel}/PACS map (green dotted), {\it Spitzer}/IRAC map (blue dotted) and the atomic gas column density $N_{H}$ map (orange dotted).
Bottom left: limits derived for spatially extended DM model with the {\tt MED} model with $\log_{10}\sigma_{J}=0.2$ (black solid line), $\log_{10}\sigma_{J}=0$ (blue dashed line), $\log_{10}\sigma_{J}=0.35$ (green dotted line) and with a point-like DM spatial distribution (red dot-dashed line).
Bottom right: limits derived with the Alt IEMs used in \cite{Acero:2015prw}, the newest 4FGL catalog and IEM and isotropic templates \cite{Fermi-LAT:2019yla} and the Off IEM.}
\label{fig:sysM31}
\end{figure}

Different choices in the details of the data analysis can modify at different extent the results for our DM search.
In this section we report the study of the change in the results for the upper limits on $\langle \sigma v \rangle$ with respect to the lower bound of the energy range and ROI width considered in our analysis, size and model of the astrophysical emission of M31 and M33, prior on $J$ and spatial distribution for the DM component. Other assumptions might change the results by negligible factors.
The results are collected in Fig.~\ref{fig:sysM31} for M31 (for M33 we find very similar results).
The differences are calculated with respect to the baseline setup which is given by the following choices: energy range $E\in [0.3,1000]$ GeV, ROI width $16^{\circ}\times16^{\circ}$, disk template with size $0\fdg33$, uncertainty for the $J$ factor $\log_{10}\sigma_{J}=0.2$ and Off IEM.

The choice of the lower bound of the energy range considered in the analysis affects the limits for $m_{\chi}>100$ GeV as an overall normalization. For these masses, the results for $E>0.1$ and $>0.3$ GeV differ only by a factor of about 15\% with the limits for the former which are better.
For $m_{\chi}<100$ GeV the limits found with $E>0.1$ GeV become larger than the one for $E>0.3$ GeV because very small residuals at low energies are absorbed by the DM template.
On the other hand the limits for $\langle \sigma v \rangle$ derived assuming $E>0.5$ and $>1$ GeV are worse by about a factor 2 and 3, respectively.
Since the limits found with $E>0.3$ GeV are the tightest and the PSF and acceptance are much better at 0.3 GeV than at 0.1 GeV we decide to use this energy range in the rest of the analysis. This choice is also motivated by the fact that the significance for the detection of M31 and M33 and the $TS_{\rm{EXT}}$ is about the same order for $E>0.1$ and $>0.3$ GeV (see Table \ref{tab:ext}).

We found in Sec.~\ref{sec:fitting} that in the baseline model M31 is extended with a size for a uniform disk template of $0\fdg33 \pm 0\fdg04$.
During the search of a DM contribution in our pipeline we do not vary the size of the disk component.
We run our search for DM with a size of the disk template modified by $\pm 1\sigma_{\rm{EXT}}$ or decreasing it by $3\sigma_{\rm{EXT}}$ from the best-fit and assuming that $\sigma_{\rm{EXT}}$ is the $1\sigma$ error for the extension size.
These changes in the size of the astrophysical emission of M31 affects by a negligible contribution the upper limits for $\langle \sigma v \rangle$.

The disk template is a phenomenological model created to fit the $\gamma$-ray emission from M31 and is tuned directly on {\it Fermi}-LAT data. Employing this geometrical template could hide part of the DM emission. We perform the analysis by substituting the disk template with the following templates motivated by observations of M31 in other wavelengths: the {\it Herschel}/PACS map at 160 $\mu$m, {\it Spitzer}/IRAC map at 3.6$\mu$m, and the atomic gas column density $N_{H}$ map from \cite{2009ApJ...695..937B}. 
The results for the DM search is that we find no evidence for a DM contribution in any annihilation or decay channel and the upper limits on $\langle \sigma v \rangle$ are larger for the case with the $N_{H}$ ({\it Herschel}/PACS) map by a factor of about $2-3$ ($1.5-2$) for $m_{\chi} \in [10,1000]$ GeV. This is due to the fact that by using the $N_{H}$ or the {\it Herschel}/PACS templates the $TS$ of the astrophysical component decreases and more residuals remain in the M31 region that are partially absorbed by the DM component (see Fig.~\ref{fig:TSmapalt}).
On the other hand the results found with the {\it Spitzer}/IRAC template are very similar to the baseline model. Indeed, the $TS$ derived with this infrared map is close to the one obtained with the disk template (see Table \ref{tab:altmodel}).

Our benchmark DM spatial templates include a spatially extended map with an uncertainty for the $J$ factor that is 0.20 in $\log_{10}$ units.
We test how much the limits for DM change by assuming a point-like DM template and with $\log_{10}\sigma_{J}=0$ and $\log_{10}\sigma_{J}=0.35$.
These cases embed the uncertainty in the $J$ factor of the main halo of M31 and M33 found by assuming different functions for the DM distribution, which are $\log_{10}\sigma_{J}=0.35$ for M31 and $\log_{10}\sigma_{J}=0.25$ for M33 \cite{2012A&A...546A...4T,Fune:2016uvn}.
The upper limits decrease (increase) by a factor of about 15\% (25\%) using $\log_{10}\sigma_{J}=0$ ($\log_{10}\sigma_{J}=0.35$).
These differences are subdominant with respect to the ones reported above (e.g. the spatial template for M31).
Changing the DM template into a point-like morphology strengthens the limits by a maximum of about a factor of 8 for $m_{\chi}>300$ GeV and of about a factor of 4 for $m_{\chi} \sim 10$ GeV.
However, we know that a point-like DM template is not physically motivated for such close galaxies.
Choosing $\log_{10}\sigma_{J}=0$ decreases the limits by a normalization factor of the order of 15\%.

Moreover, changing the ROI width to $14^{\circ}\times14^{\circ}$ or $20^{\circ}\times20^{\circ}$ changes by a negligible amount the limits for $m_{\chi}>100$ GeV while at smaller masses the choice of a smaller ROI width can give significantly larger limits because of residuals generated by edging effects that are absorbed as DM signal at low energy.

Finally, we run the analysis with the Off, the Alt IEMs and using the newest 4FGL catalog and IEM and isotropic templates \cite{Fermi-LAT:2019yla}. The results change by at most $30\%$ between the Off and Alt IEMs while the 4FGL gives a difference of at most 50\% at $m_{\chi}=200$ GeV.
Even if the limits with the 4FGL catalog are higher than the other cases the significance for the presence of DM is still negligible.

Similar conclusions are also valid for the $\tau^+\tau^-$ annihilation channel and for the decay case.













\subsection{Null and injected signal simulations}
\label{sec:sim}

The pipeline that we employ in this paper can also be used to perform simulations.
In particular simulations are generated by {\tt Fermipy} that take the source model and randomize it with Poisson statistics.
This method is much faster than the tool {\tt gtobssim} which is included in the {\tt fermitools}\footnote{https://fermi.gsfc.nasa.gov/ssc/data/analysis/scitools/help/gtobssim.txt} and it is usually used for the same scope.
We consider here two types of simulations to validate our analysis pipeline: null and injected signal simulations.

\begin{figure}[ht]
\includegraphics[scale=0.43]{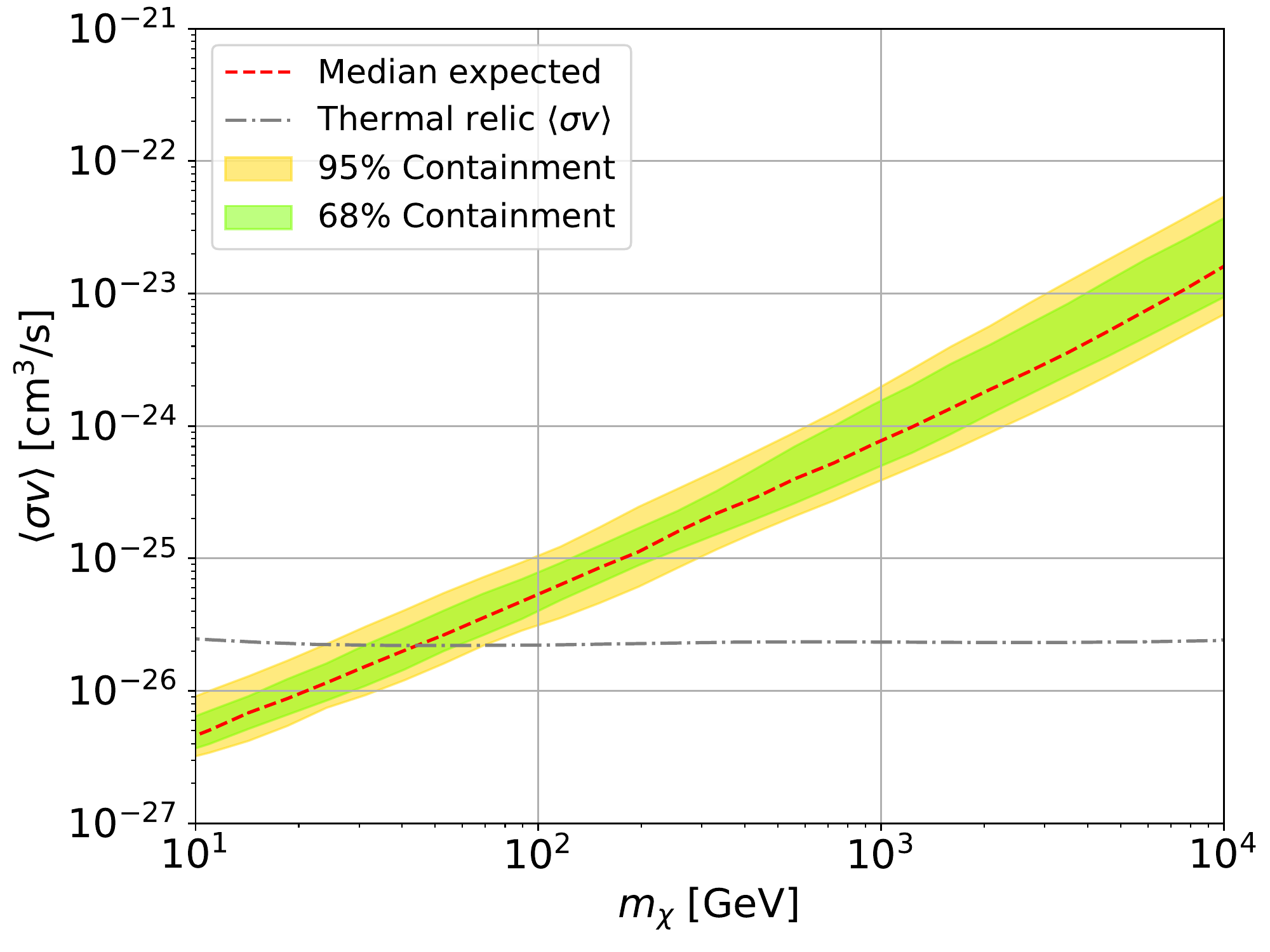}
\includegraphics[scale=0.43]{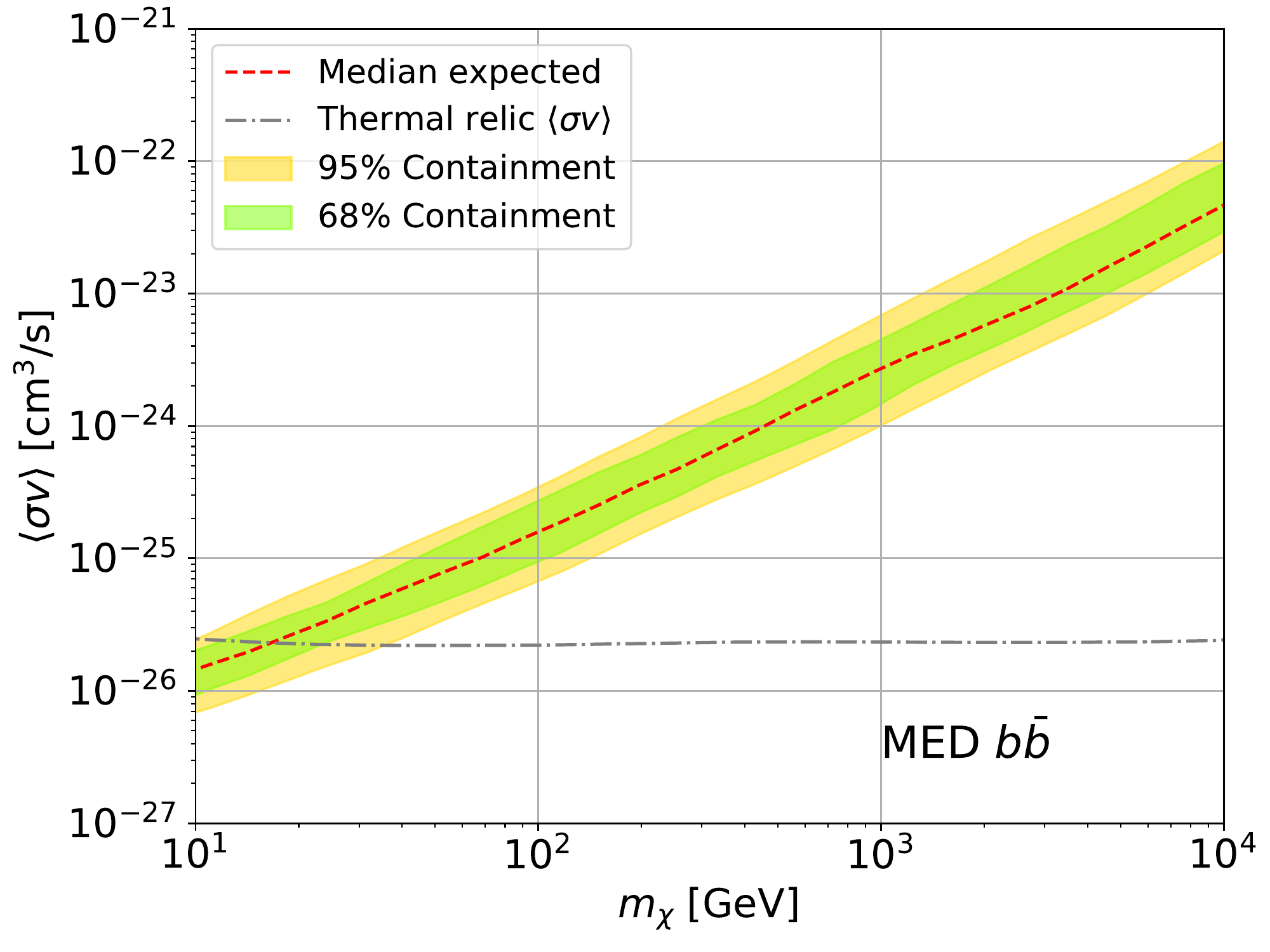} 
\caption{Results for 100 simulations of the null signal (see the text for further details). Upper limits for $\langle \sigma v \rangle$ for M31 (left panel) and M33 (right panel) for the $b\bar{b}$ annihilation channel and for the {\tt MED} DM model. The median (red dashed line) and the $95\%$ (yellow band) and $68\%$ (green band) containment bands over the 100 simulations are shown. The canonical thermal relic cross section is also reported \cite{Steigman:2012nb} (grey dot-dashed).}
\label{fig:siminjected}
\end{figure}

Null simulations are made by taking the model from the baseline fit, i.e.~without the DM contribution, and simulating the ROI.
Then we run the search for DM on the simulated data. 
The goal of these simulations is to calculate the expected limits in the absence of any DM signal.
We run the null signal simulations on both M31 and M33, for the $b\bar{b}$ and $\tau^+\tau^-$ channels and for {\tt MIN}, {\tt MED} and {\tt MAX} DM models.
We show in Fig.~\ref{fig:siminjected} the upper limits for $\langle \sigma v \rangle$ that we derive for 100 simulations of the null signal for M31 and M33. This is done for the $b\bar{b}$ annihilation channel and for the {\tt MED} model. 
The plots show the median upper limits and the $95\%$ and $68\%$ containment bands over the 100 simulations. As expected, the search for DM with these simulations gives a $TS\sim 0$. 
Therefore, we show the results in form of upper limits for $\langle \sigma v \rangle$.
The median is well contained in the $68\%$ containment bands and the limits rule out the thermal cross section for $m_{\chi}< 50$ GeV for M31 and $m_{\chi}< 20$ GeV for M33.

\begin{figure}[ht]
\includegraphics[scale=0.42]{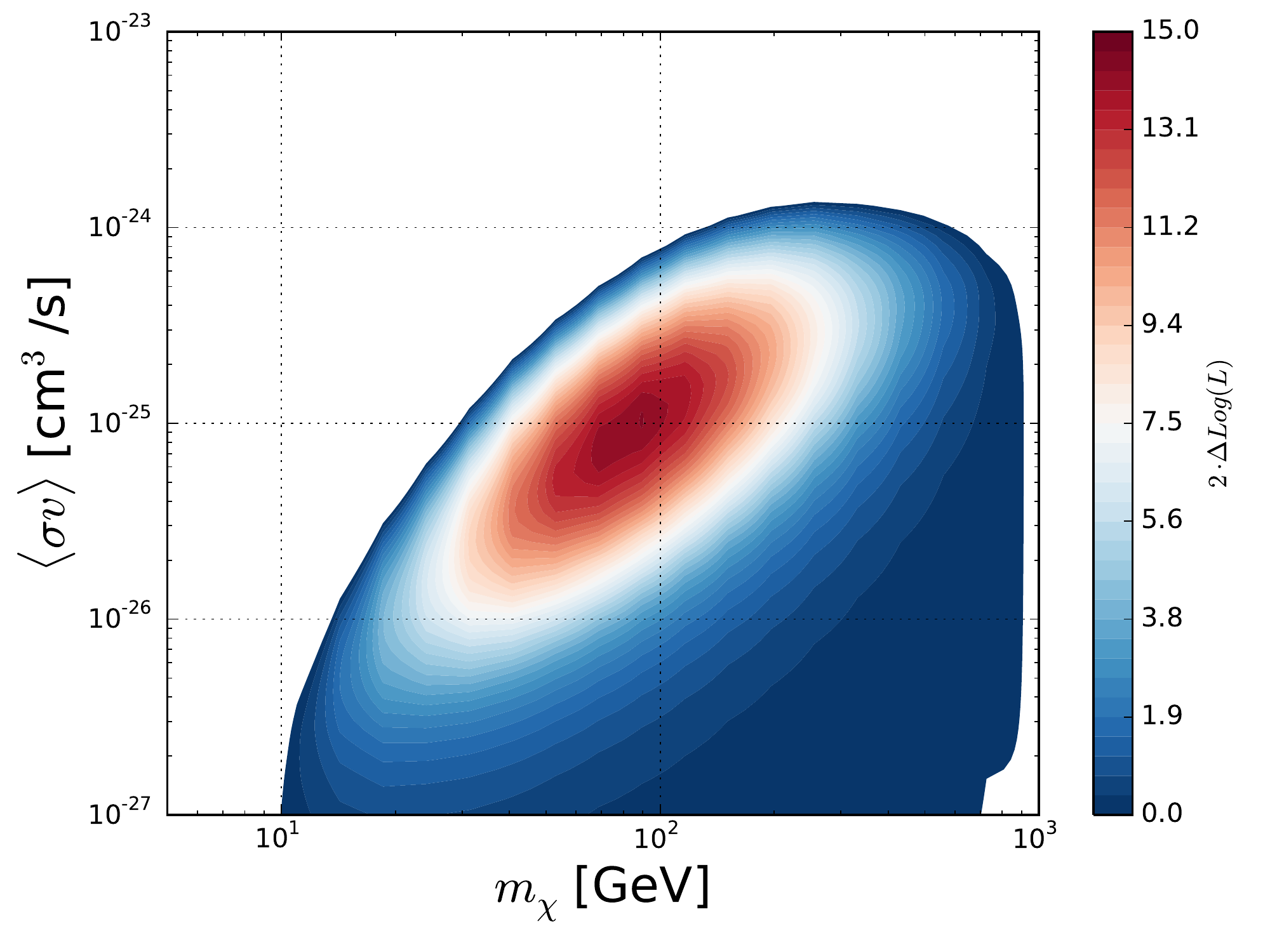}
\includegraphics[scale=0.42]{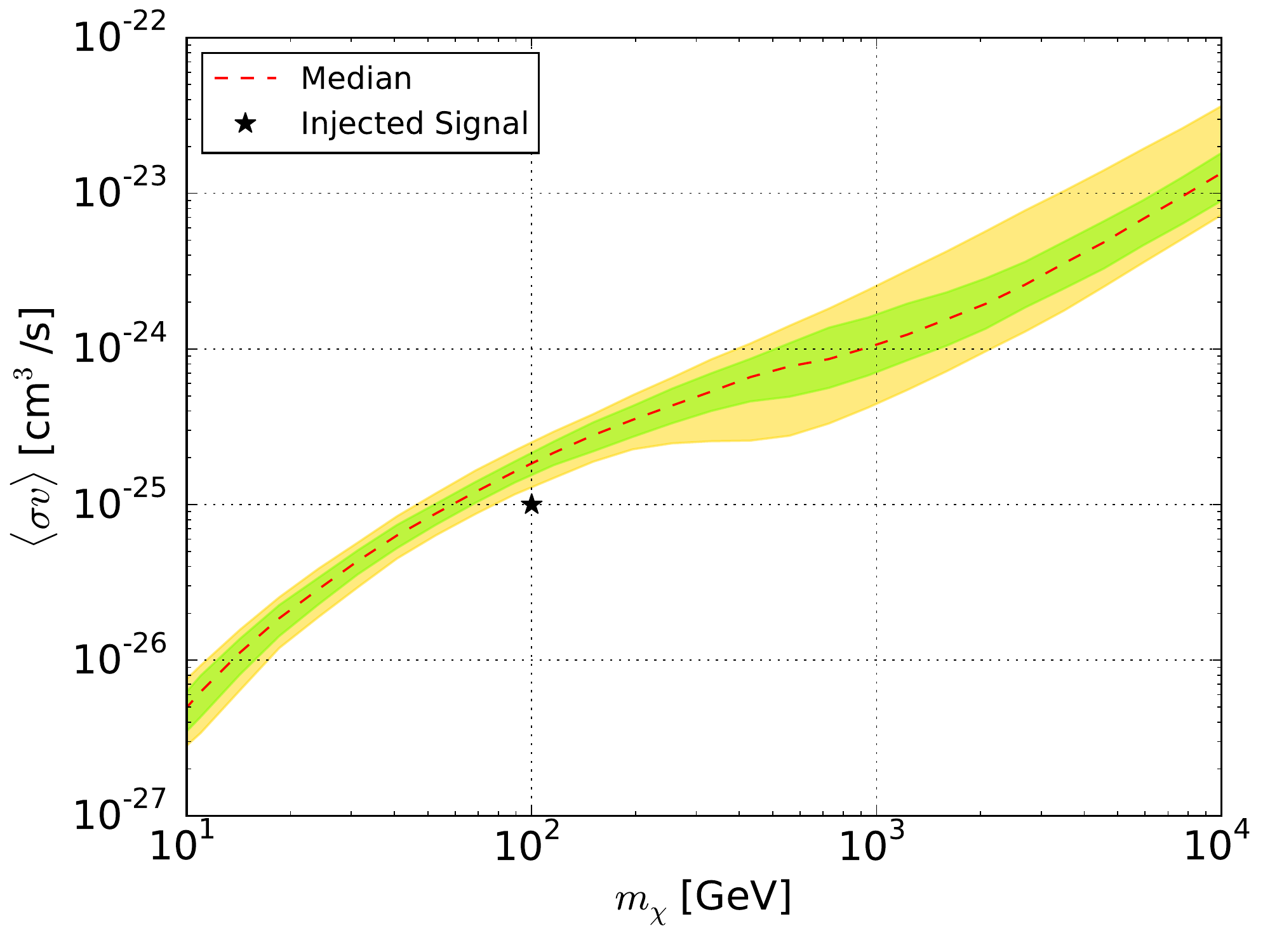} 
\caption{Results of the analysis of 100 simulations with an injected signal in the M31 ROI with $m_{\chi}=100$ GeV, $\langle \sigma v \rangle = 10^{-25}$ cm$^3$/s, for an annihilation channel $b\bar{b}$ and for the {\tt MED} DM model. Left panel: contour plot for $m_{\chi}$ and $\langle \sigma v \rangle$. The color bar represents the $TS$ for the signal. Right panel: upper limits for $\langle \sigma v \rangle$ as a function of the DM mass. The median (red dashed line) and the $95\%$ (yellow band) and $68\%$ (green band) containment bands over the 100 simulations are shown.}
\label{fig:siminjected2}
\end{figure}

On the other hand, for the injected signal simulations a DM signal for a specific annihilation channel, cross section and mass is added to the model. Then, {\tt Fermipy} generates the simulated data, which is analyzed in the same way as the actual data.
We perform these simulations to verify that our pipeline is able to recover an injected signal.
We choose to inject in the M31 ROI a DM signal with $m_{\chi}=100$ GeV, $\langle \sigma v \rangle = 10^{-25}$ cm$^2$/s and for an annihilation channel $b\bar{b}$.
We run 100 simulations and we use the {\tt MED} DM distribution.
This signal can be detected at most with a $TS$ of 15. The contour plot for the cross section and DM mass is reported for this simulation in Fig.~\ref{fig:siminjected}. The best fit cross section and DM mass is perfectly compatible with the characteristics of the injected signal.
Since the $TS$ for detection is below 25 we decide to calculate upper limits for each simulation. In Fig.~\ref{fig:siminjected2} we show the median and the $95\%$ and $68\%$ containment bands over the 100 simulations.
These limits are consistent with the cross section of the injected signal demonstrating once again that our pipeline is able to recover an injected DM signal.

\subsection{DM-only interpretation}
\label{sec:dminterp}

In this section we make the assumption that the $\gamma$-ray emission in the direction of M31 and M33 is entirely given by DM particle interactions.
In order to do so, we remove from the source model the astrophysical model for the $\gamma$-ray emission from M31 and M33, we include the DM template and we run the pipeline described in Sec.~\ref{sec:analysis}.

Table \ref{tab:DMsignM31} and \ref{tab:DMsignM33} contain the best fit and $1\sigma$ errors for the DM mass and cross section for $b\bar{b}$ annihilation channel and for M31 and M33 respectively.
We report these results using only the main DM halo (SH) or including also DM substructures (SHS) and for each of these cases we try the {\tt MIN}, {\tt MED} and {\tt MAX} models for the DM distribution.

We first focus on the results found for M31.
The $TS$ for the presence of DM for the $b\bar{b}$ annihilation channel, is 27 for the {\tt MED} DM model in the SHS case and 55 for the Einasto profile in case of SH.
This DM candidate has a mass of 30 (20) GeV and a cross section of $3.2\cdot 10^{-26}$ cm$^2$/s ($5.2\cdot 10^{-26}$ cm$^2$/s) for SHS (SH) case.
The {\tt MIN} and {\tt MAX} DM distributions for SHS and the Burkert and Adiabatic for the SH case provide similar significances and DM masses but larger and lower values for the best-fit cross sections.
On the other hand, for the $\tau^+\tau^-$ annihilation channel the best fit mass is at the lower limit of the DM mass considered in our analysis, i.e.~5 GeV, and a cross section of about $1.5 \cdot 10^{-26}$ cm$^3$/s for the {\tt MED} model.
The decay scenario provides a very low significance for the Burkert and Einasto DM profiles and both $b\bar{b}$ and $\tau^+\tau^-$ channels with $TS$ values of the order of $5-8$.
The only case that provides a large significance is with the Adiabatic DM profile for which in the $b\bar{b}$ channel the best fit is $m_{\chi}=30$ GeV and $\tau=6.5\cdot 10^{24}$ s for a $TS=21$ while for the $\tau^+\tau^-$ channel $m_{\chi}\leq 5$ GeV and $\tau \leq 9.9\cdot 10^{24}$ s for a $TS=20$.
Only the Adiabatic DM profile provides a high significance because its spatial profile is similar to the $\gamma$-ray signal. On the other hand the Burkert and Einasto models have a spatial distribution that is much broader (see Fig.~\ref{fig:d-and-j-factors}).

\begin{table}[ht]
\begin{tabular}{c|ccc||c|ccc}
\hline \hline
    \multicolumn{4}{c||}{SHS}  & \multicolumn{4}{c}{SH}   \\ 
\hline
DM model & $m_{\chi}$ & $\langle \sigma v \rangle$ & $TS$  & DM model &  $m_{\chi}$ & $\langle \sigma v \rangle$ & $TS$  \\ \hline
{\tt MIN} & $25^{+40}_{-15}$ & $1.3^{+2.2}_{-0.6} \cdot 10^{-25}$ &  27  & Burkert &  $20^{+10}_{-5}$ & $1.4^{+1.0}_{-0.3} \cdot 10^{-25}$ &  43 \\
{\tt MED} & $30^{+35}_{-15}$ & $3.2^{+4.1}_{-1.7} \cdot 10^{-26}$ &  27  & Einasto &  $20^{+10}_{-5}$ & $5.2^{+2.9}_{-1.1} \cdot 10^{-26}$ &  55 \\
{\tt MAX} & $40^{+60}_{-20}$ & $1.1^{+2.4}_{-0.6} \cdot 10^{-26}$ &  20  & Adiabatic &  $20^{+10}_{-5}$ & $2.0^{+1.2}_{-0.5} \cdot 10^{-26}$  & 56 \\
\end{tabular}
\caption{Summary table for the $TS$, $m_{\chi}$ and $\langle \sigma v \rangle$ of DM in M31 ROI considering SHS (left side) or SH case (right side). We assume here a $b\bar{b}$ annihilation channel.}
\label{tab:DMsignM31}
\end{table}

The results for M33 for the DM annihilating into $b\bar{b}$ for the {\tt MIN}, {\tt MED} and {\tt MAX} models in case of SHS and for Einasto and Burkert DM profiles for the SH scenario are reported in Table \ref{tab:DMsignM33}.
The significance for the presence of DM for the $b\bar{b}$ annihilation channel, considering the {\tt MED} DM model, is 18 for the SHS case and 31 for the Einasto profile in the SH scenario.
This DM candidate has a mass of 30 (20) GeV and a cross section of $9\cdot 10^{-25}$ cm$^2$/s ($2.3\cdot 10^{-24}$ cm$^2$/s) for SHS (SH) case.
The {\tt MIN} and {\tt MAX} DM distributions provide similar significances and DM masses but larger and lower values for the best-fit cross sections.
The decay scenario with $b\bar{b}$ channel provides $TS$ values of the order of 12 for Einasto and 16 for the Burkert profile with $m_{\chi}=15$ GeV and $\tau=1.3\cdot 10^{25}$.
On the other hand the $\tau^+\tau^-$ channel gives $TS$ values of the order of 11 for Einasto and 14 for the Burkert profile with $m_{\chi}\leq 5$ GeV and $\tau\leq 2.1\cdot 10^{25}$.

\begin{table}[ht]
\begin{tabular}{c|ccc||c|ccc}
\hline \hline
    \multicolumn{4}{c||}{SHS}  & \multicolumn{4}{c}{SH}   \\ 
\hline
DM model & $m_{\chi}$ & $\langle \sigma v \rangle$ & $TS$  & DM model &  $m_{\chi}$ & $\langle \sigma v \rangle$ & $TS$  \\ \hline
MIN & $20^{+50}_{-10}$ & $5^{+8}_{-3} \cdot 10^{-24}$ &  23  & Burkert &  $50^{+60}_{-30}$ & $2.6^{+3.3}_{-1.8} \cdot 10^{-24}$ &  25 \\
MED & $30^{+60}_{-20}$ & $9^{+16}_{-7} \cdot 10^{-25}$ &  18  & Einasto & $90^{+110}_{-60}$ & $2.3^{+2.8}_{-1.6} \cdot 10^{-24}$ &  31 \\
{\tt MAX} & $25^{+45}_{-20}$ & $7^{+20}_{-5} \cdot 10^{-27}$ &  13  & &   &   &  \\
\end{tabular}
\caption{Same as Table \ref{tab:DMsignM31} but for M33. }
\label{tab:DMsignM33}
\end{table}

\begin{figure}[ht]
\includegraphics[scale=0.42]{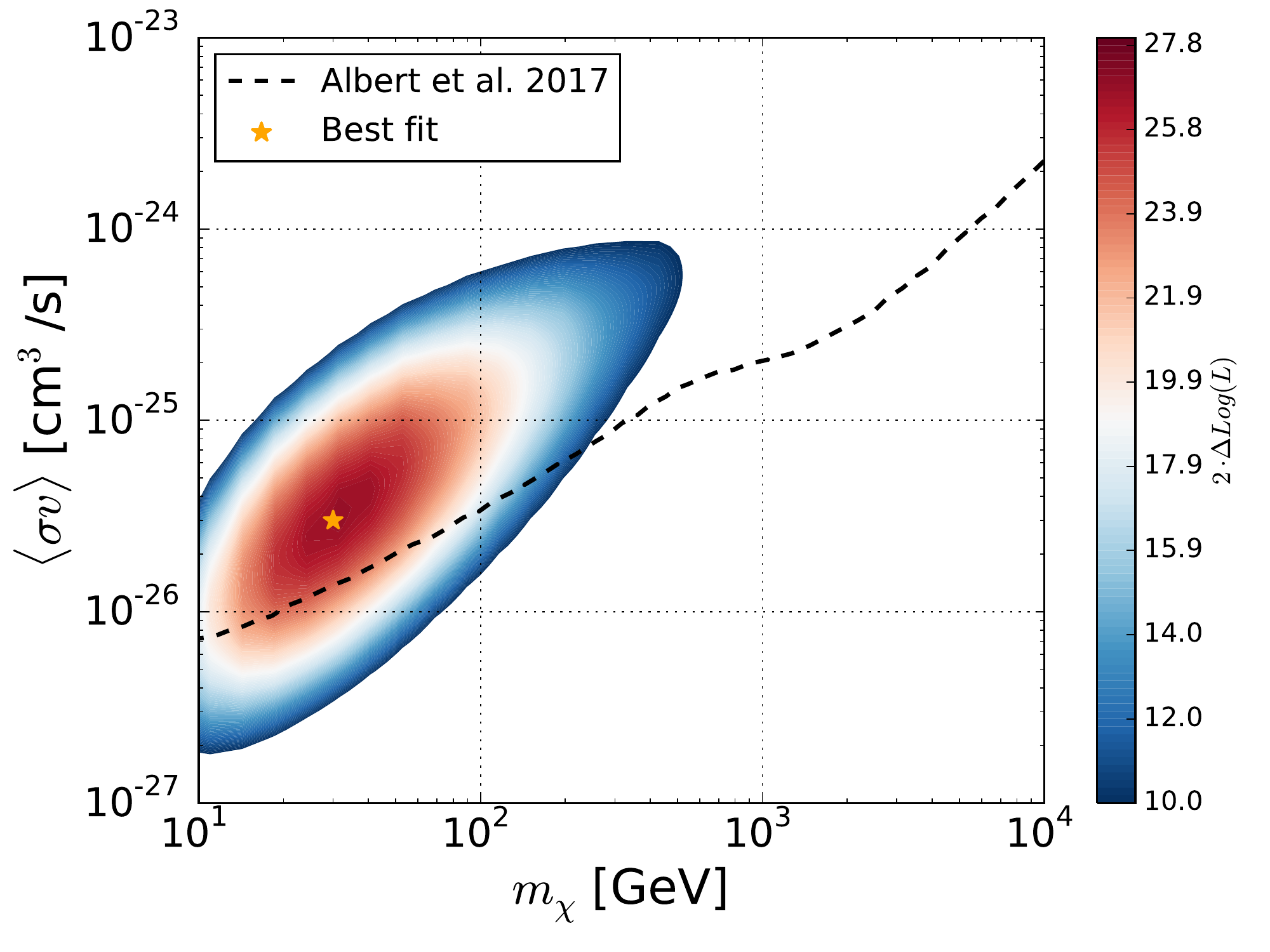}
\includegraphics[scale=0.42]{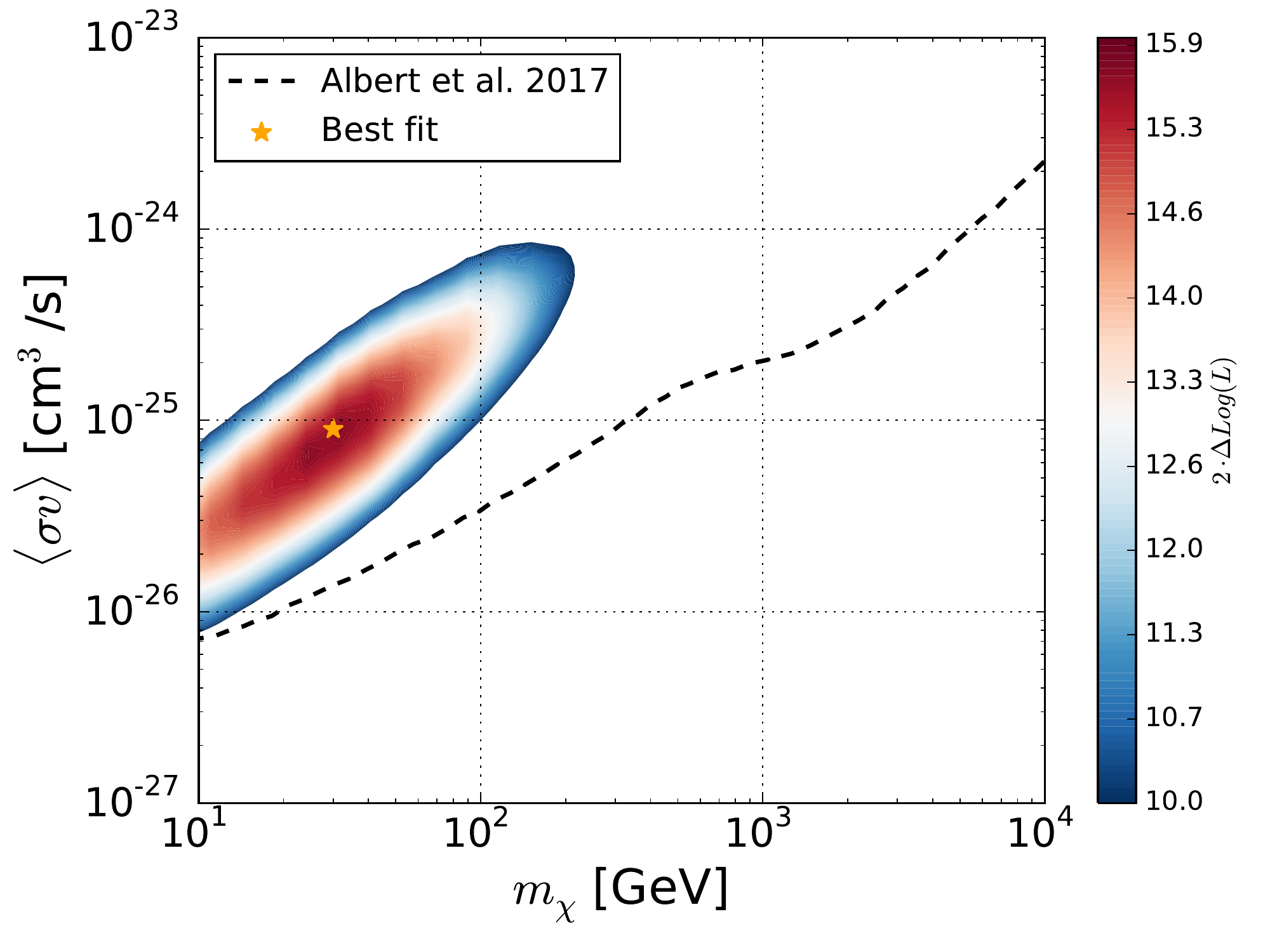}
\caption{$TS$ for the presence of DM ($2\Delta \log{L}$) as a function of DM mass and cross section for $b\bar{b}$ annihilation channel and {\tt MED} DM model for M31 (left panel) and M33 (right panel). We also display the upper limits for $\langle \sigma v \rangle$ derived for dwarf spheroidal galaxies in \cite{Fermi-LAT:2016uux}.}
\label{fig:bestDMonly}
\end{figure}

The DM candidates for either $b\bar{b}$ and $\tau^+\tau^-$ annihilation channels are in strong tension with limits found from dwarf spheroidal galaxies \cite{Fermi-LAT:2016uux} (see Fig.~\ref{fig:bestDMonly}). This tension is present for the {\tt MIN} and {\tt MED} DM models while it is alleviated for the {\tt MAX} DM distribution.

\subsection{DM plus astrophysical emission interpretation}
\label{sec:dmplusastro}

In this section we make the more realistic assumption that in addition to a putative DM signal there is also an astrophysical contribution from the galaxy itself.
We use the astrophysical emission from M31 and M33 as given in the model found with the baseline fit, thus using an extended source with a disk template with size $0\fdg33$ for M31 and a point source for M33.
Moreover, we include in the analysis the correlations between the SED parameters of DM with the ones of M31 and M33 and the other background sources (see, e.g., Sec.~\ref{sec:analysis}). 

We do not find any significant emission when we include the astrophysical contribution of M31 and M33. Indeed, the $TS$ for the presence of DM is very close to 0 for both annihilation and decay and $b\bar{b}$ and $\tau^+\tau^-$ channels.

The disk template for M31 is a phenomenological model tuned directly on $\gamma$-ray data and can hide a possible DM contribution. Therefore, we calculate the DM $TS$ also with the following templates: the {\it Herschel}/PACS map at 160 $\mu$m, the {\it Spitzer}/IRAC map at 3.6$\mu$m, and the atomic gas column density $N_{H}$ map from \cite{2009ApJ...695..937B}.

We find no evidence for a DM contribution since the $TS$ is at most of a few considering all the annihilation or decay channels.

We therefore set upper limits on the annihilation cross section $\langle \sigma v \rangle$ or lower limits for the decay time $\tau$.

\begin{figure}[ht]
\includegraphics[scale=0.42]{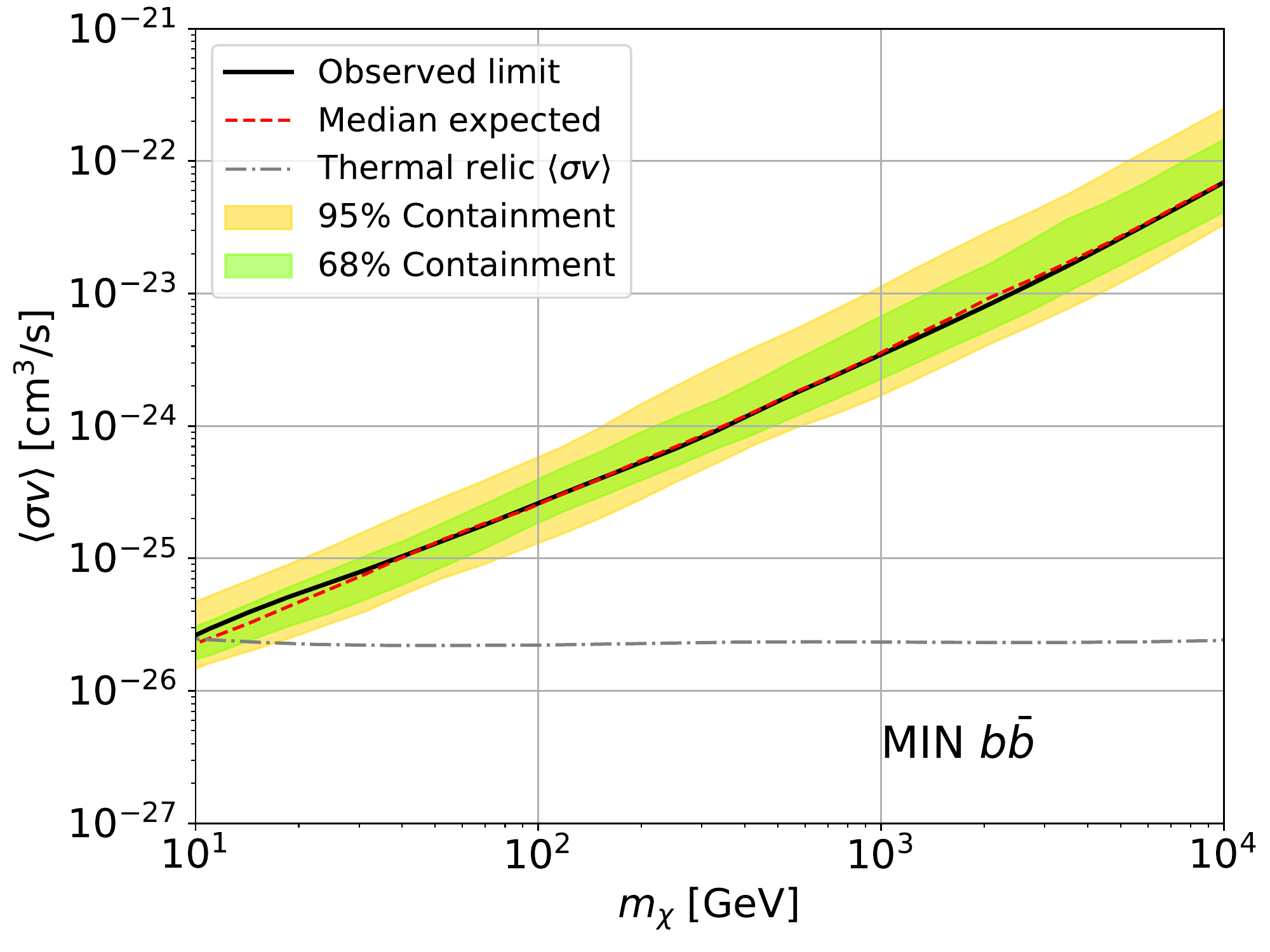}
\includegraphics[scale=0.42]{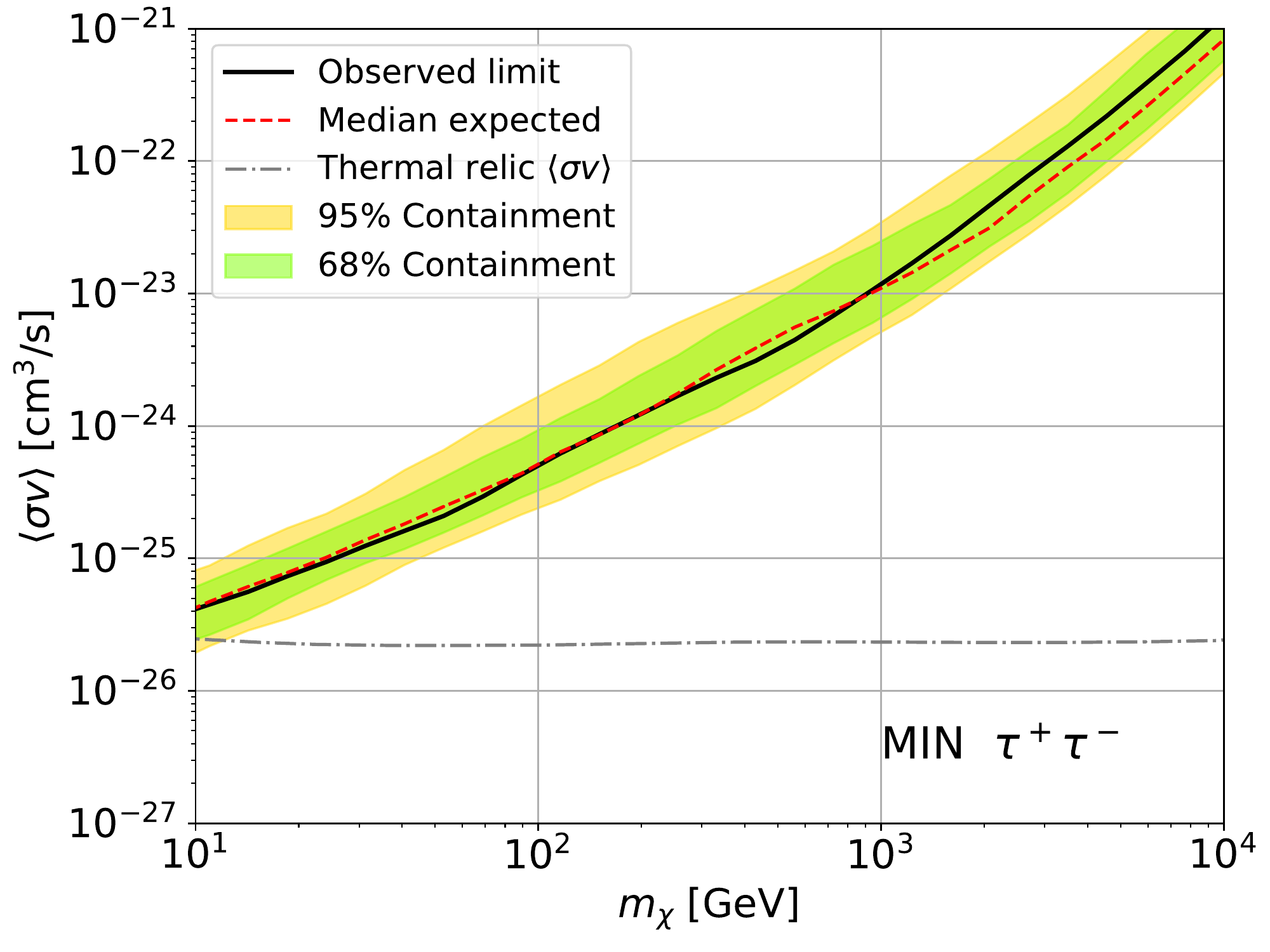}
\includegraphics[scale=0.42]{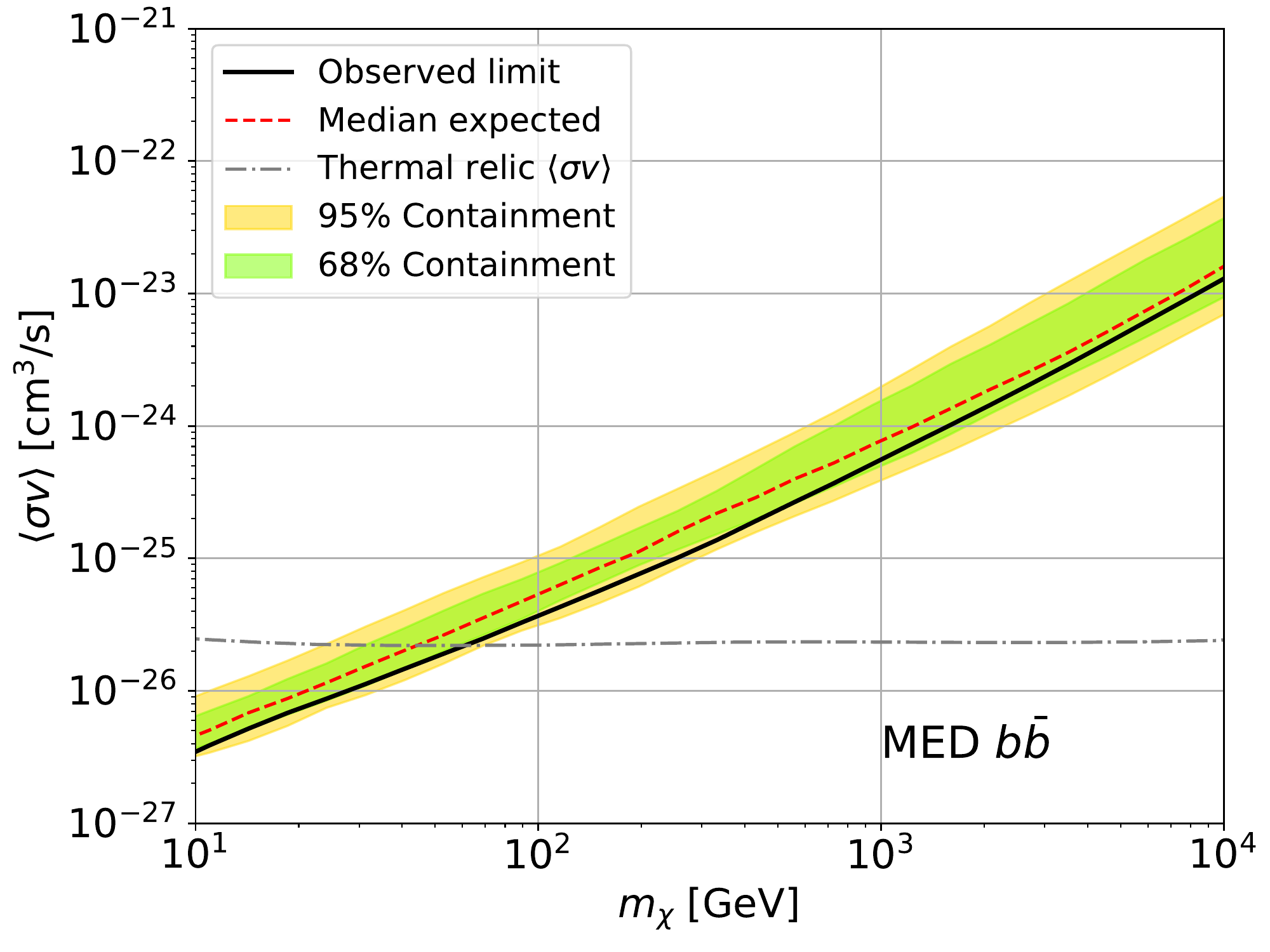}
\includegraphics[scale=0.42]{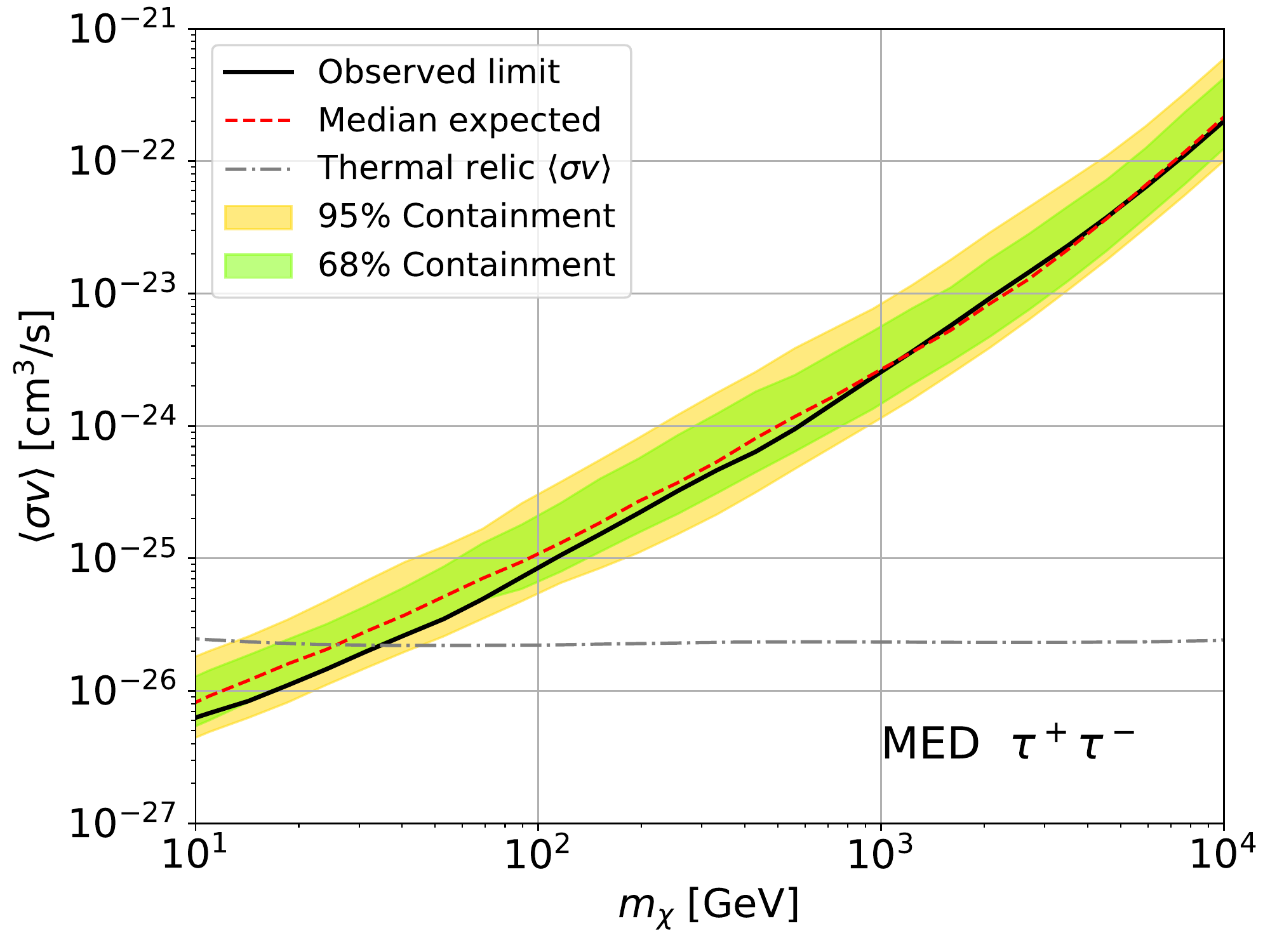}
\includegraphics[scale=0.42]{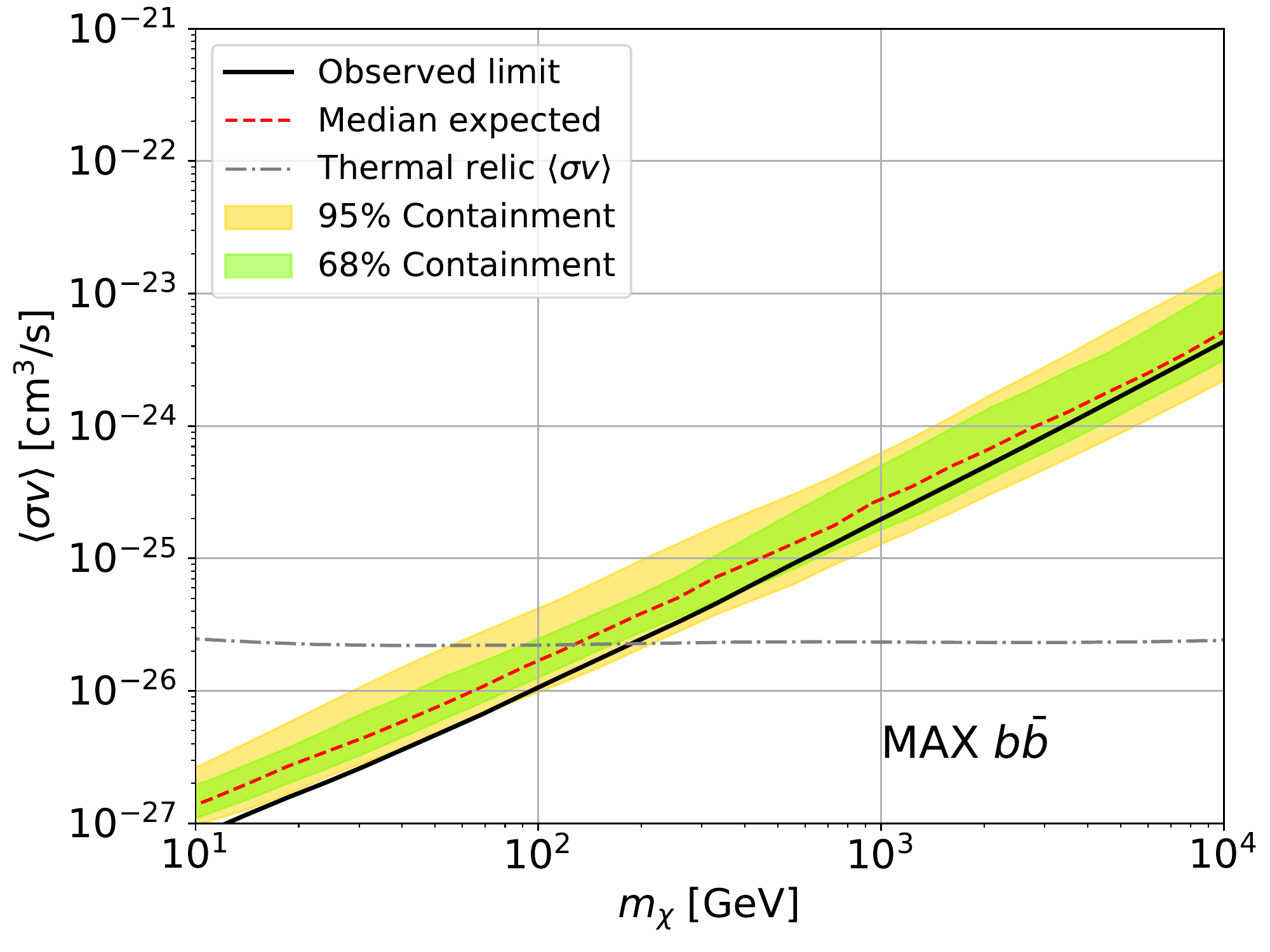}
\includegraphics[scale=0.42]{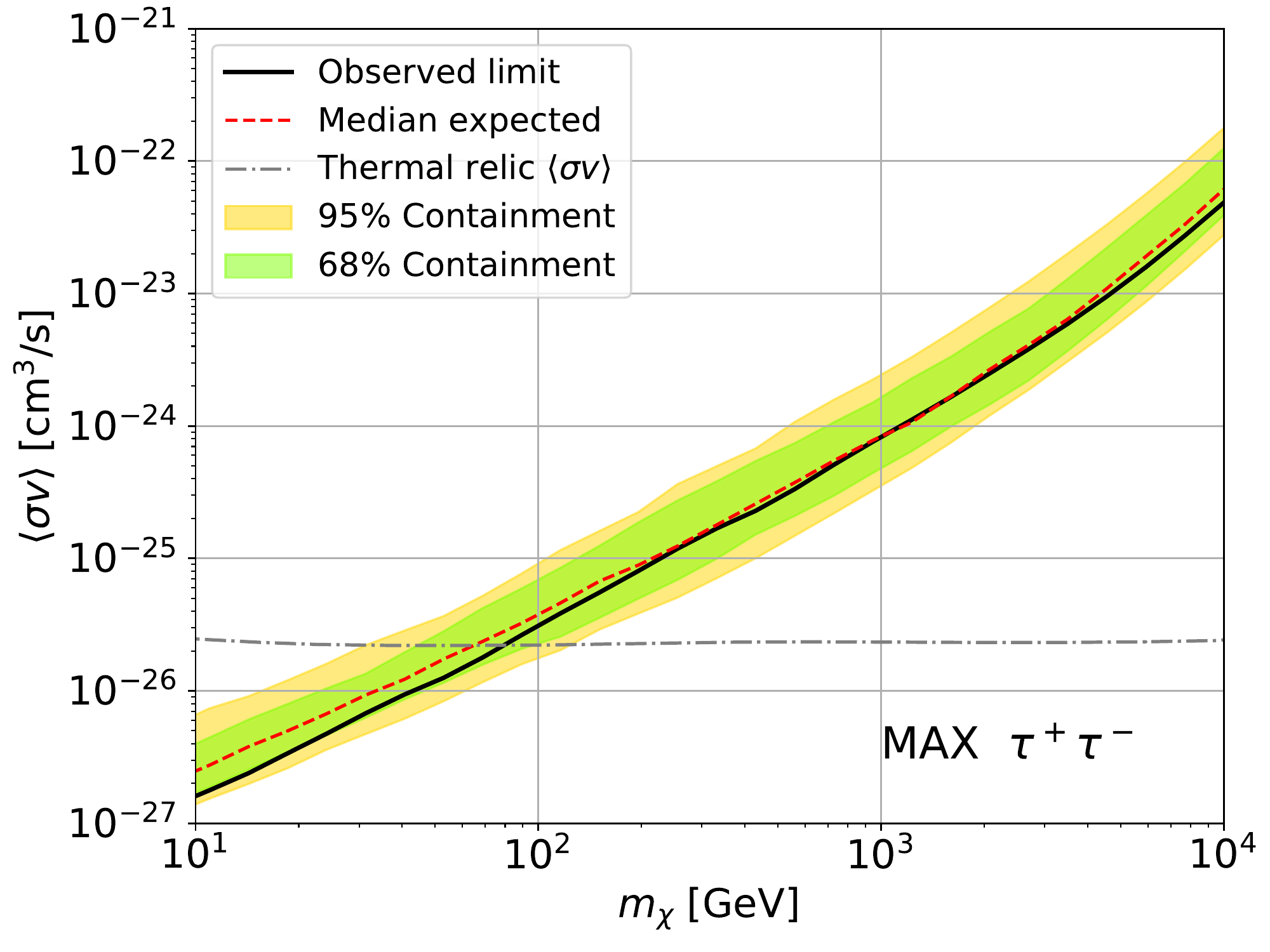}
\caption{Upper limits for the annihilation cross section of DM for M31. On the left (right) side we show the limits for the $b\bar{b}$ ($\tau^+\tau^-$) annihilation channel. The first/second/third row is for the {\tt MIN}/{\tt MED}/{\tt MAX} DM distribution model. The horizontal dashed line shows the canonical thermal relic cross section~\cite{Steigman:2012nb}}
\label{fig:limitsM31ann}
\end{figure}

\begin{figure}[ht]
\includegraphics[scale=0.42]{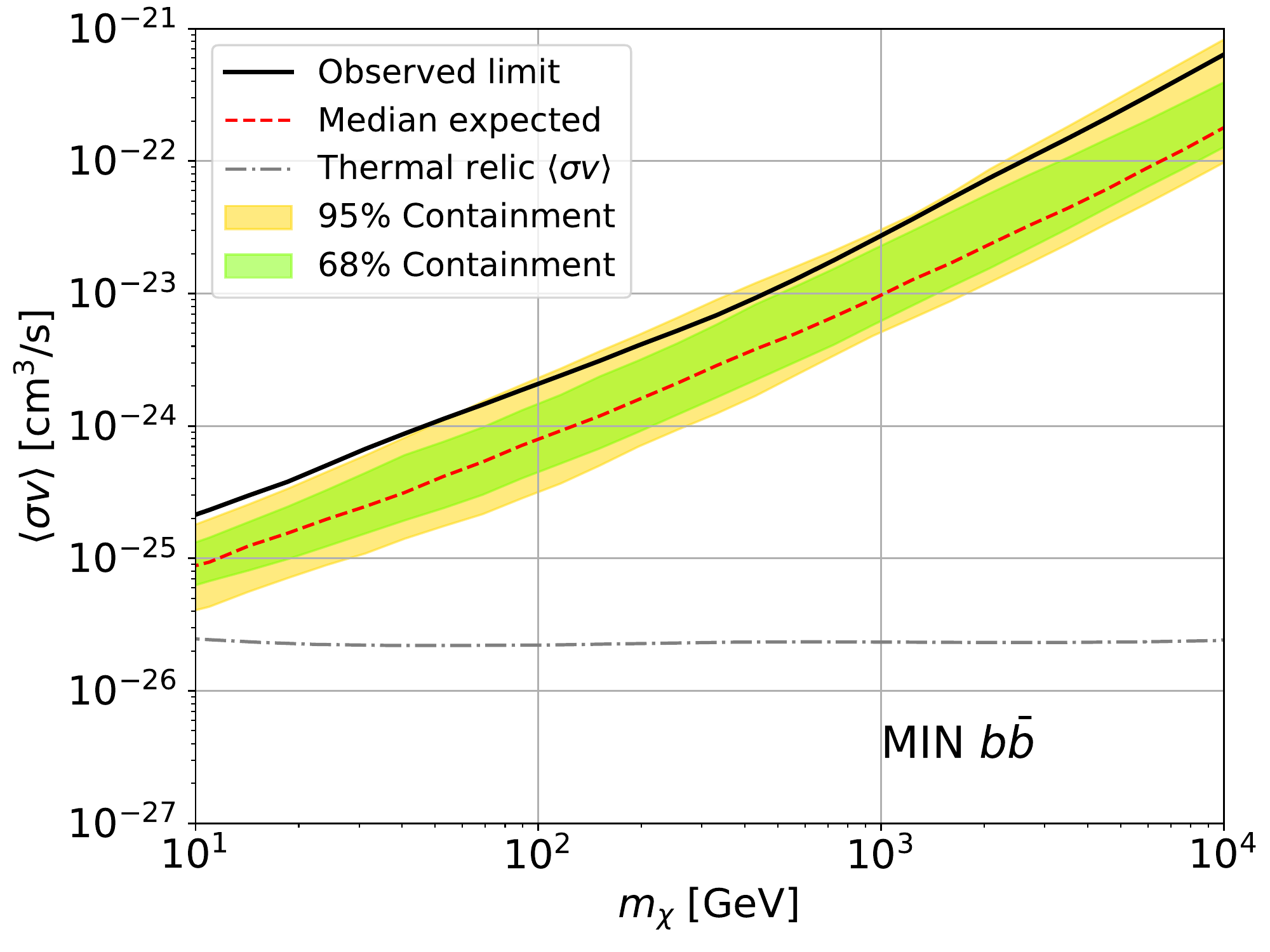}
\includegraphics[scale=0.42]{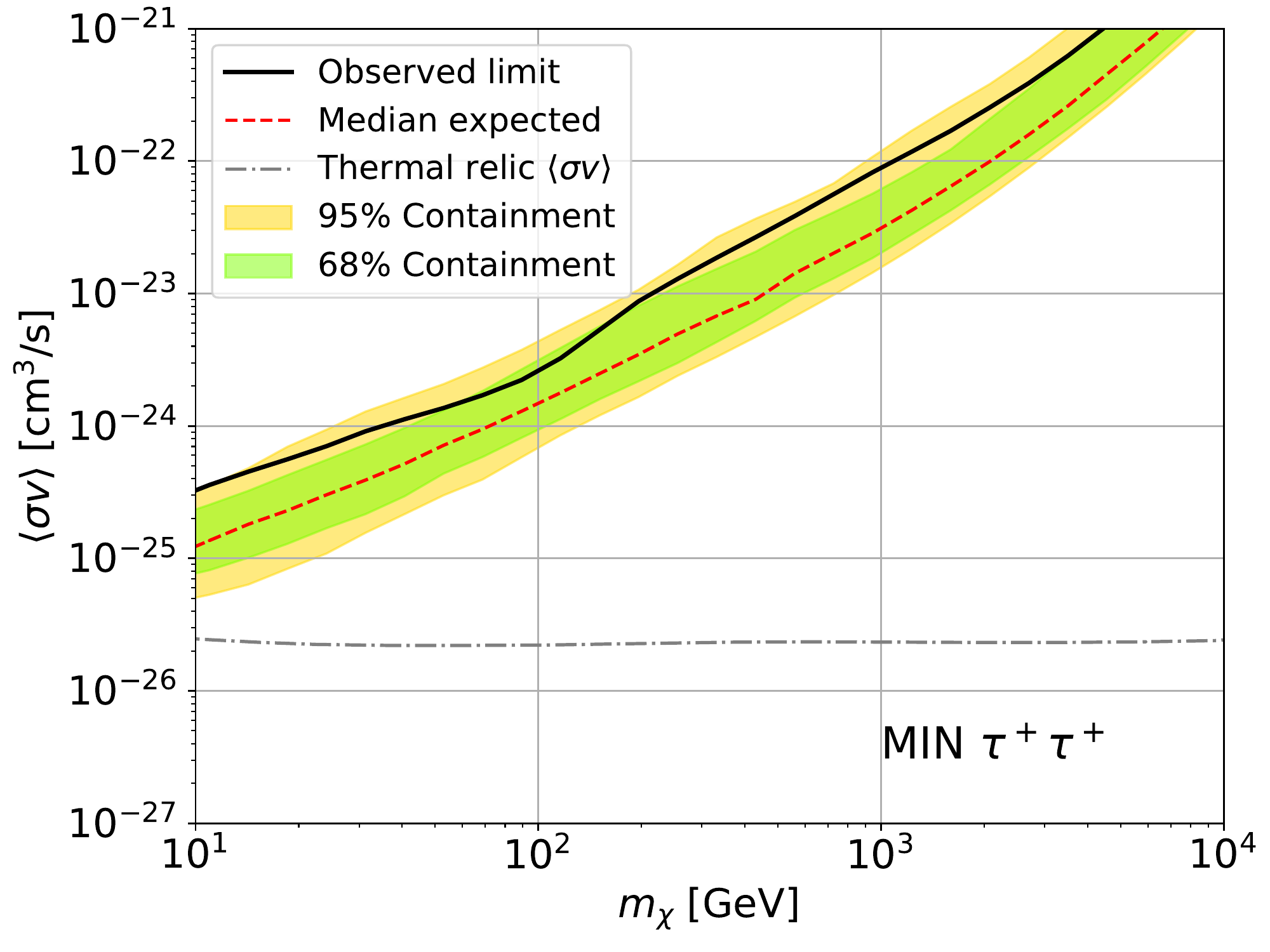}
\includegraphics[scale=0.42]{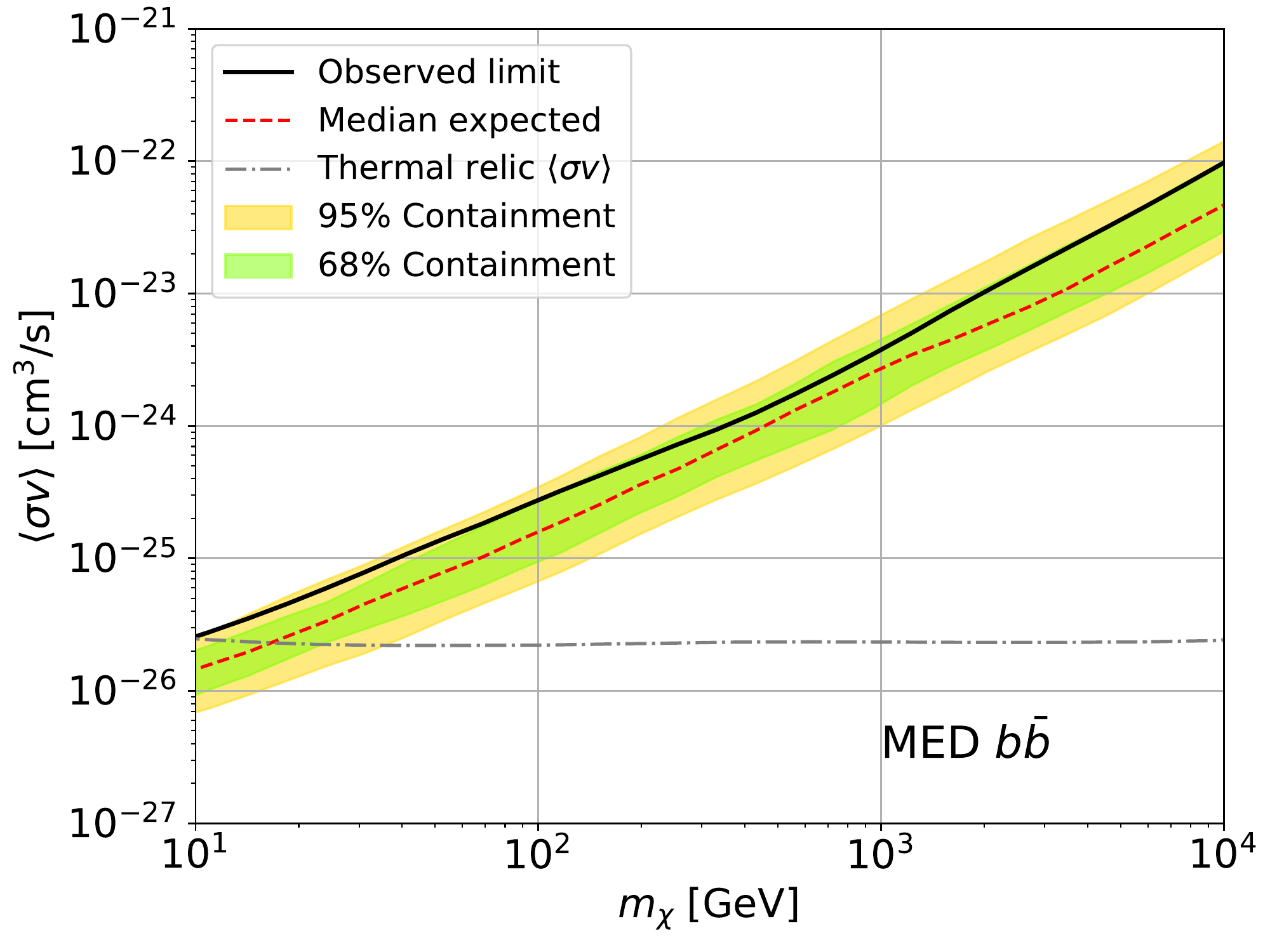}
\includegraphics[scale=0.42]{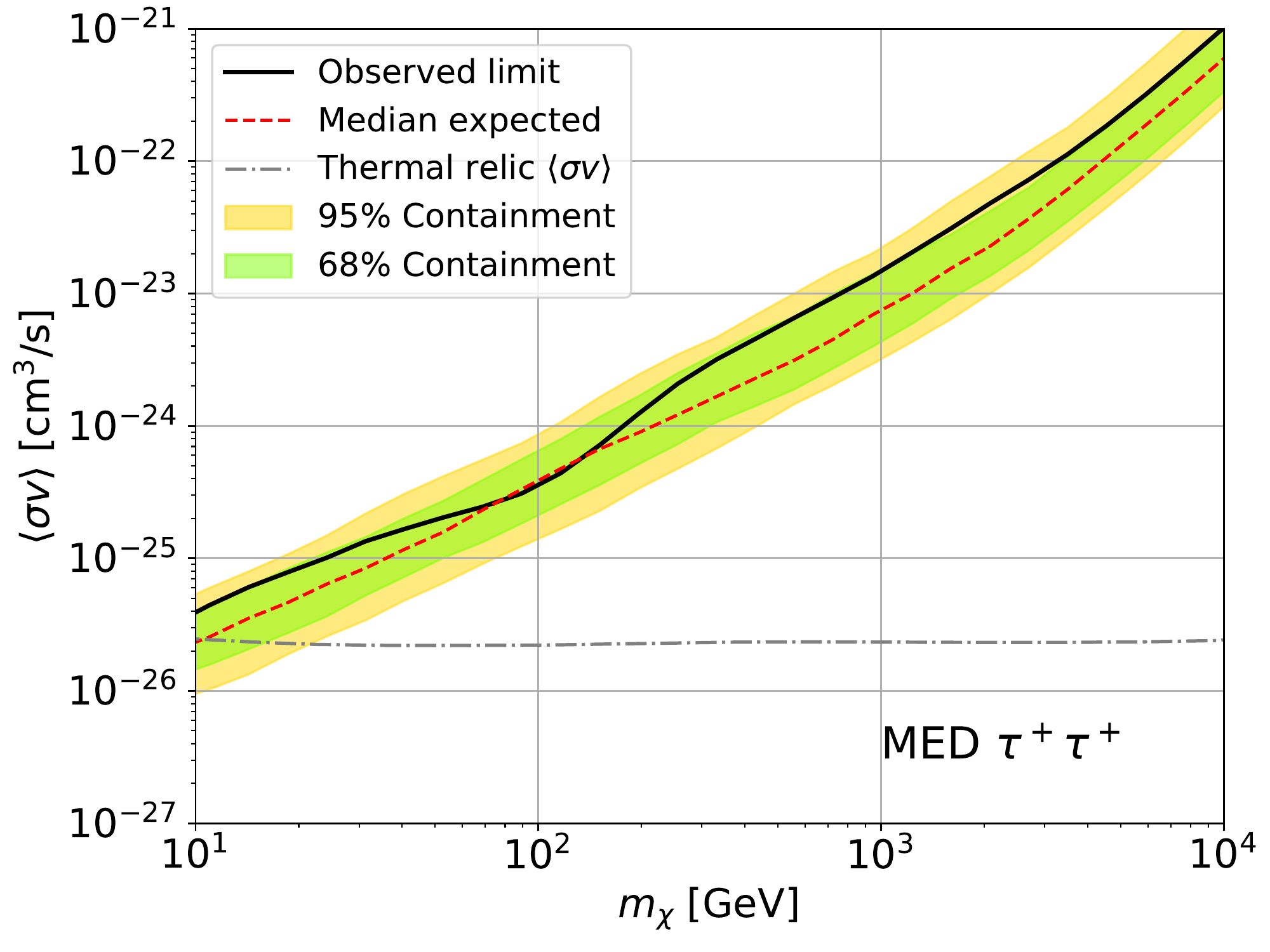}
\includegraphics[scale=0.42]{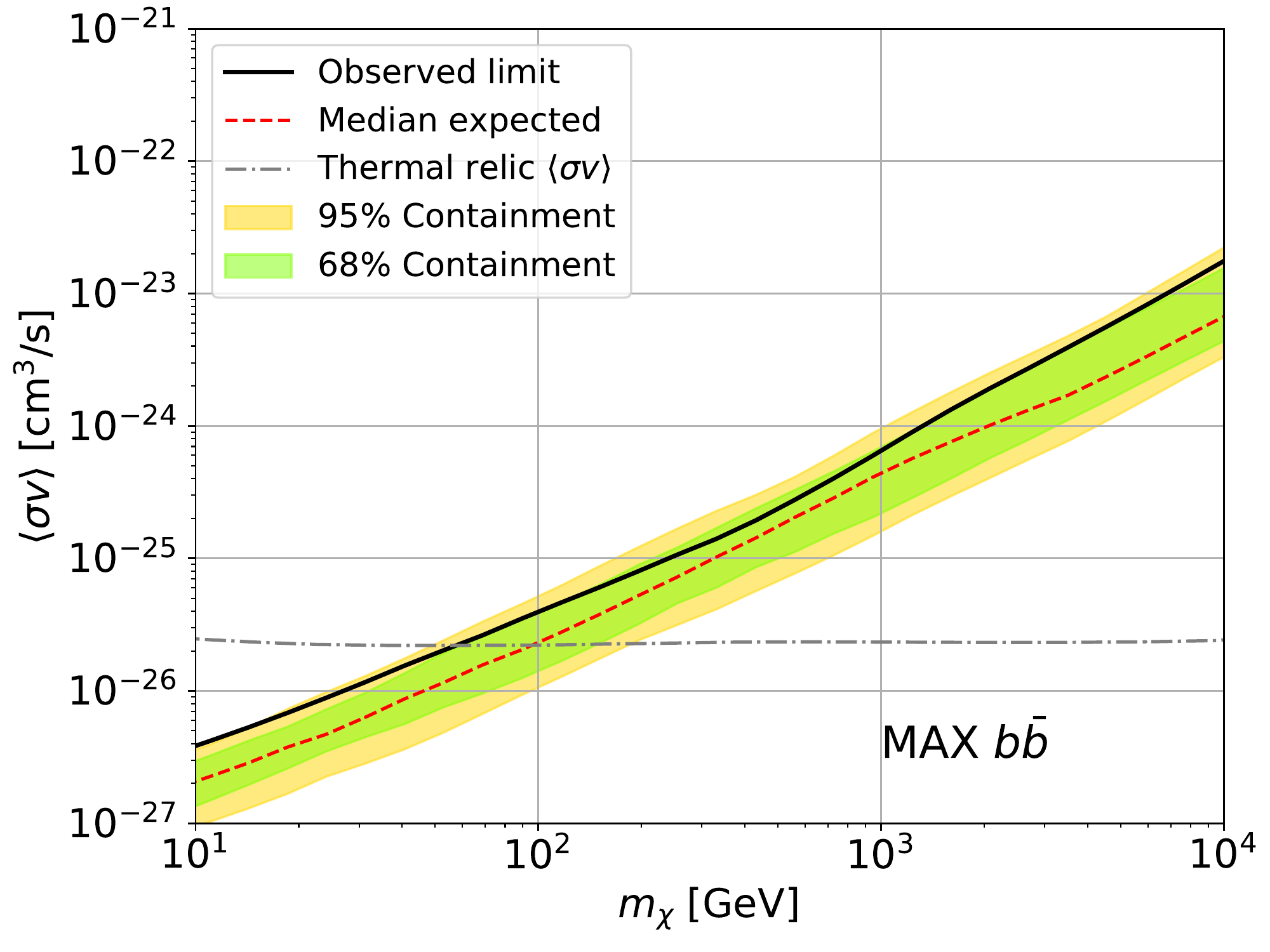}
\includegraphics[scale=0.42]{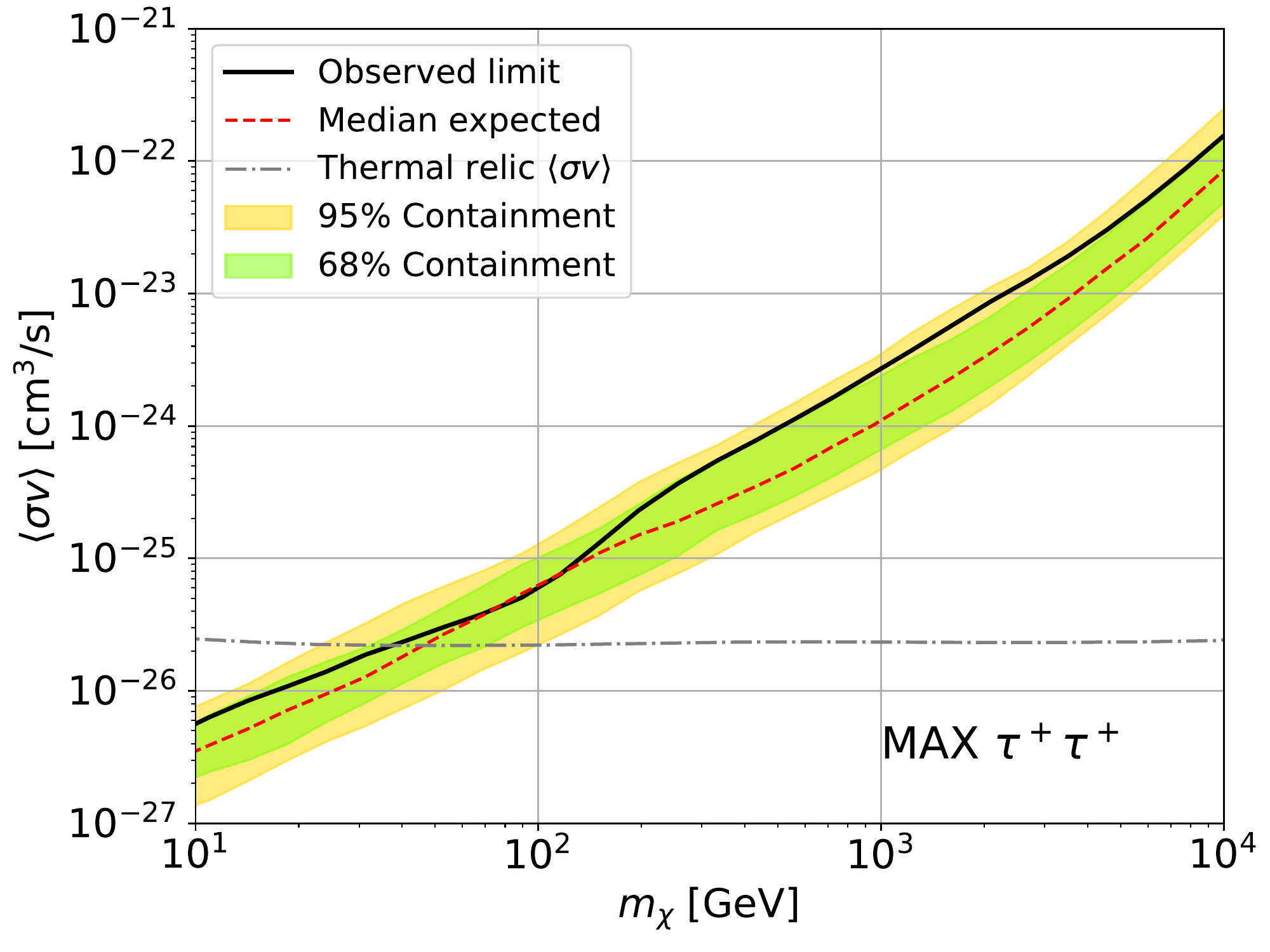}
\caption{Same as Fig.~\ref{fig:limitsM31ann} but for M33.}
\label{fig:limitsM33ann}
\end{figure}

\begin{figure}[ht]
\includegraphics[scale=0.42]{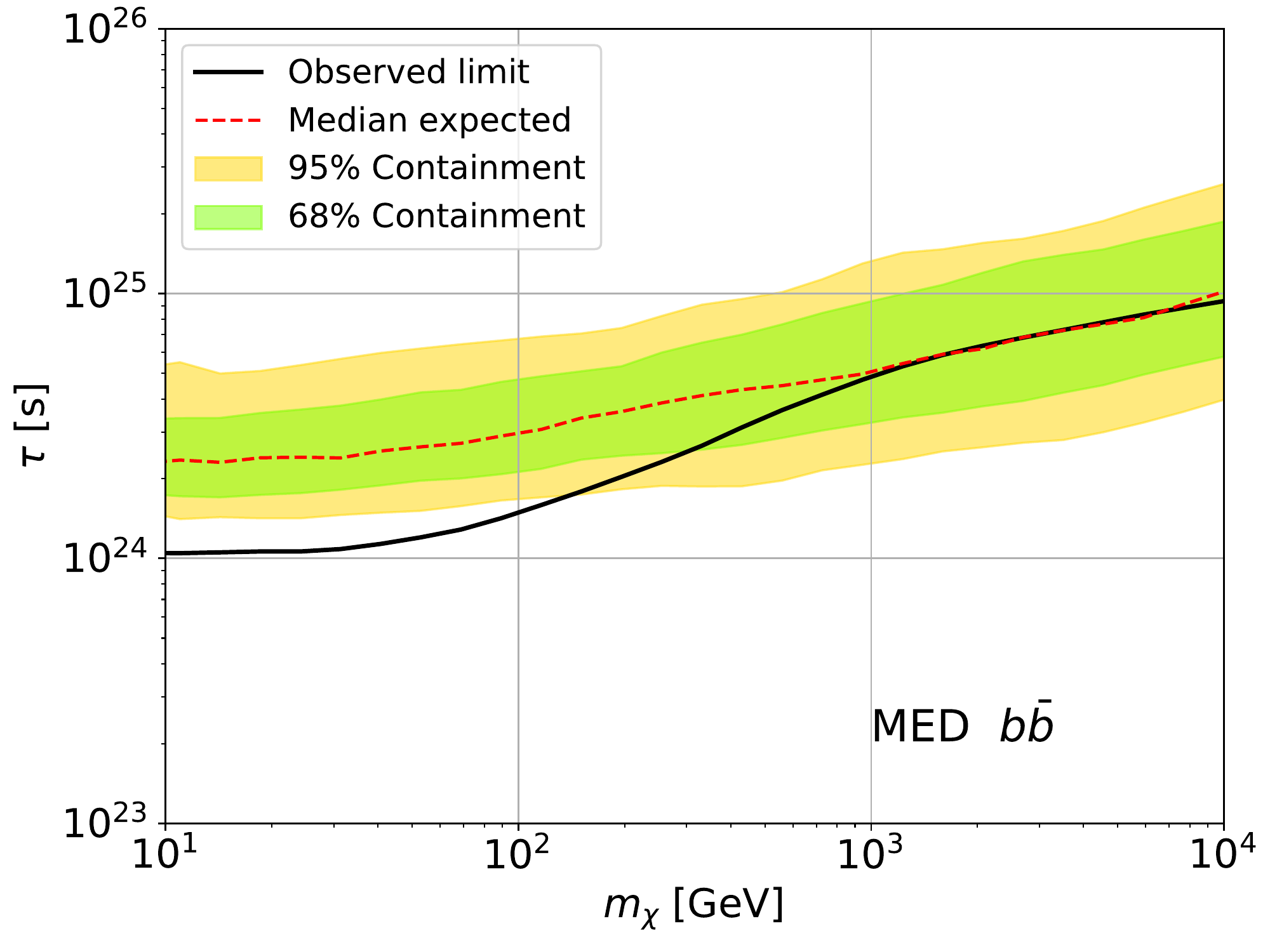}
\includegraphics[scale=0.42]{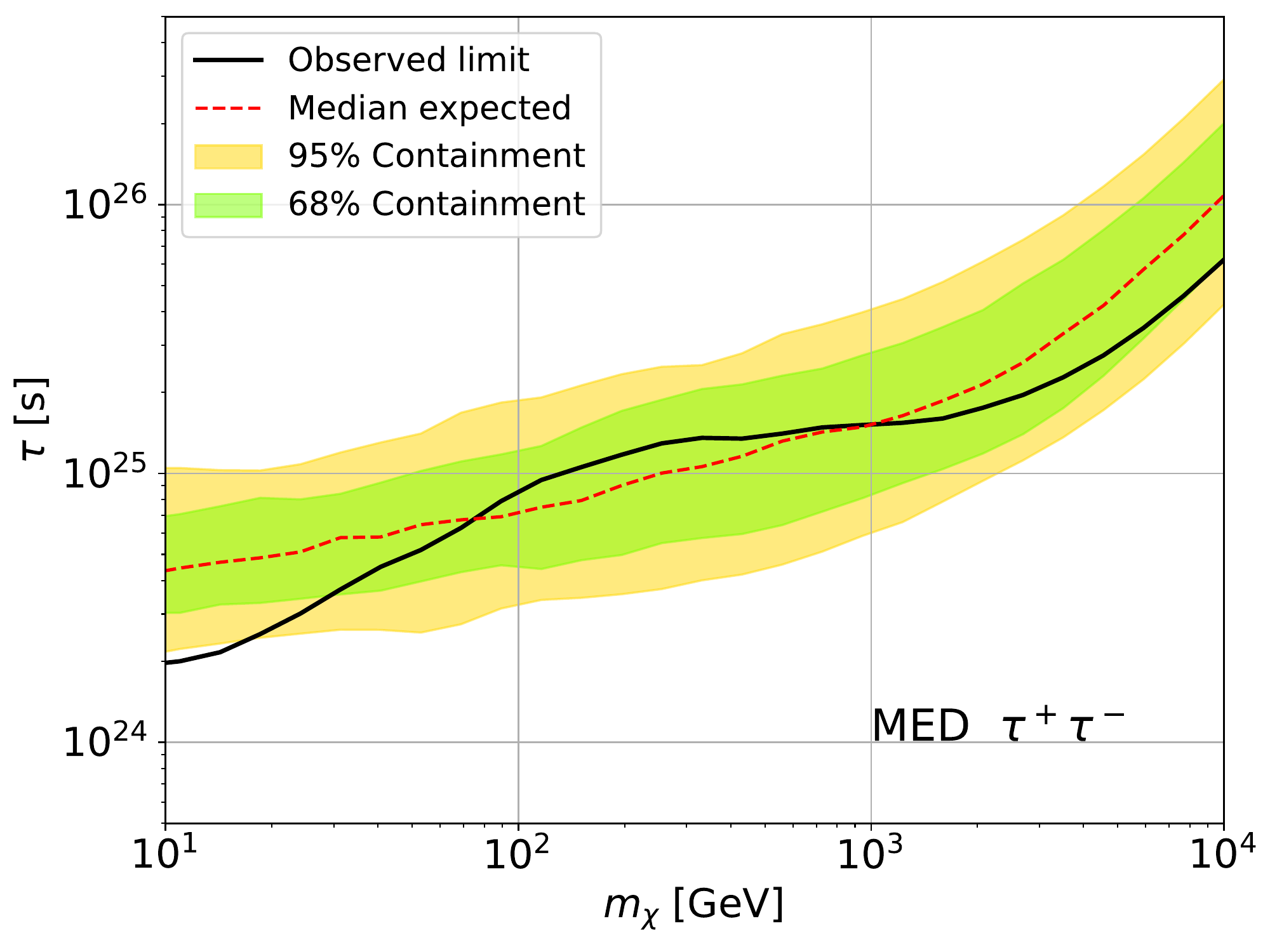}
\includegraphics[scale=0.42]{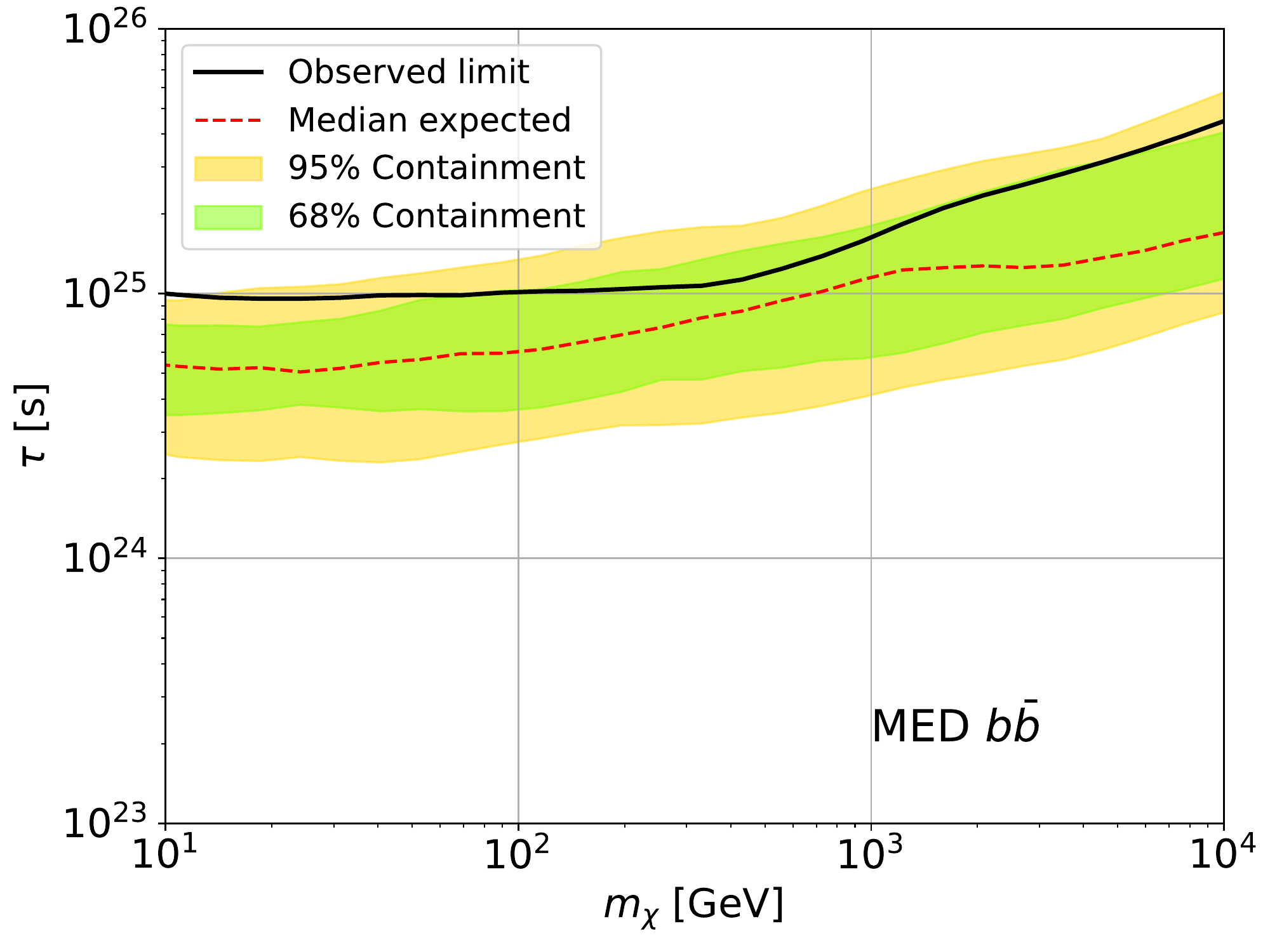}
\includegraphics[scale=0.42]{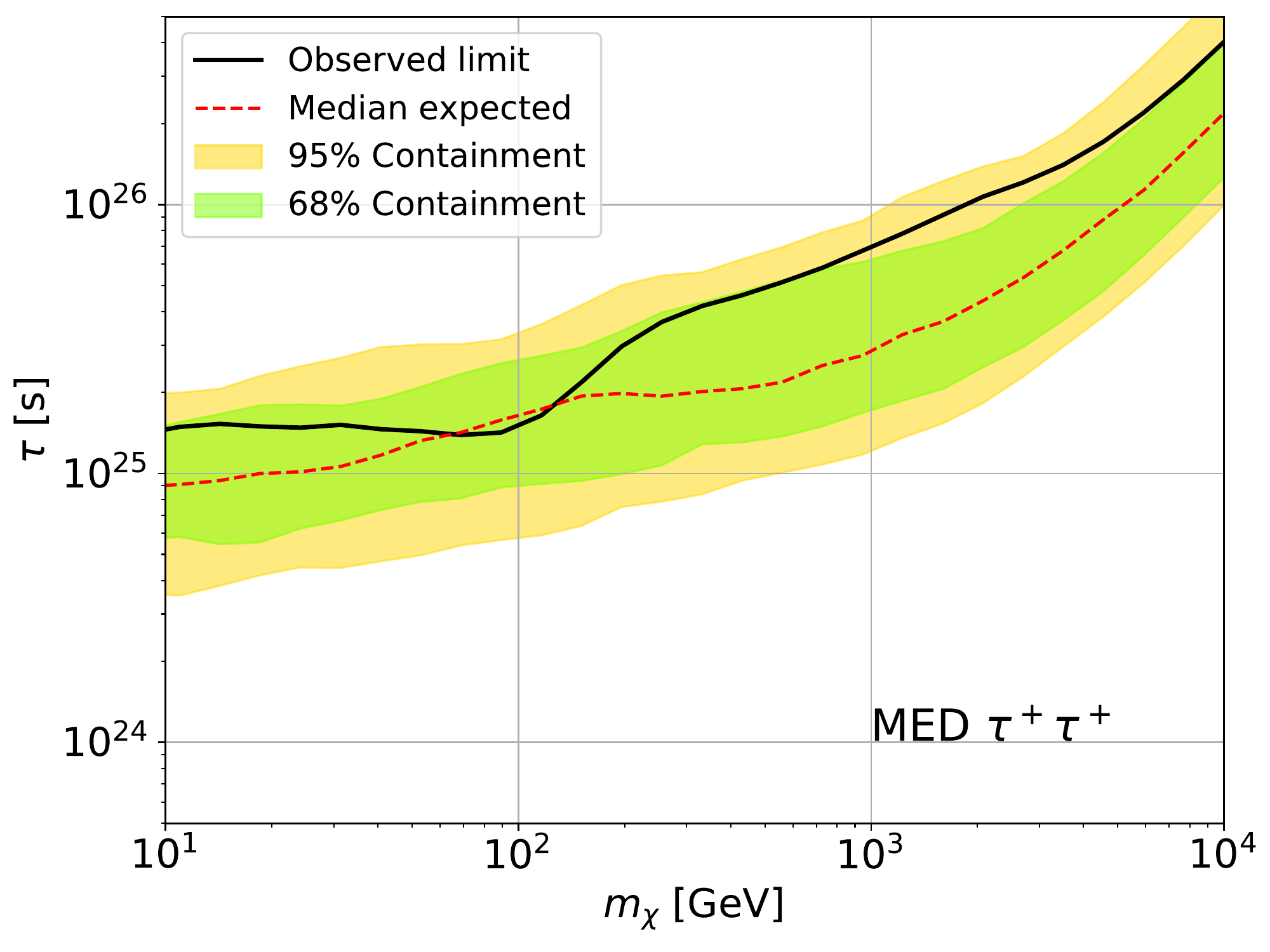}
\caption{Lower limits for the decay time of DM for M31 (top panel) and M33 (bottom panel). On the left (right) side we show the limits for the $b\bar{b}$ ($\tau^+\tau^-$) decay channel. These results have been derived with the {\tt MED} DM distribution model.}
\label{fig:limitsM31M33decay}
\end{figure}

In Fig.~\ref{fig:limitsM31ann} and \ref{fig:limitsM33ann} we show the upper limits for the annihilation cross section $\langle \sigma v \rangle$ and in Fig.~\ref{fig:limitsM31M33decay} we show the lower limit for the decay time $\tau$.
Together with the observed limits we also report the expected limits in case there is no signal (see Sec~\ref{sec:sim}).
First of all we note that the observed limits for the DM annihilation scenarios are in all cases included in the $95\%$ containment bands. 
For M33 the observed limits are systematically larger than the median expected limits from the null simulations but since for all DM scenarios they are included in the $95\%$ containment band this difference is not significant.
Second, the limits derived with the {\tt MAX} DM distribution model are the strongest for both M31 and M33. 
This is expected because the $J$ factor for this model is higher than the {\tt MED} and {\tt MIN} models.
In case of the {\tt MED} DM model the limits found for M31 constrain the thermal cross section up to about 50 GeV while in case of M33 only the {\tt MAX} DM model is able to reach the thermal cross section.
In the case of decay of DM particles all the limits are well included in the $95\%$ containment bands except for M31 and the case with $b\bar{b}$ channel. This is probably due to a local fluctuation present in the observed limits that is not seen in the simulations but the discrepancy is not significant.

These are the limits found using our benchmark case for the data analysis (energy range and ROI width), astrophysical template for M31 and M33 (disk template for M31 and point source for M33), IEM and isotropic templates.
The limits slightly change assuming a different choice for the above cited parameters and models and we reported in Sec.~\ref{sec:sys} the magnitude of these changes.

\section{Discussion and Conclusions}\label{sec:conclusion}

We performed a systematic study of the $\gamma$-ray emission from M31 and M33 galaxies with a particular focus on a possible DM contribution.
We first used our analysis to find the best geometrical model that explains the $\gamma$-ray flux from these sources.
For M31 the best model is a uniform disk with a size $0\fdg33$ and $TS_{\rm{EXT}}=13$ while for M33 is a point-like source.

We also fit $\gamma$-ray emission with templates derived from other wavelengths:  (far-)infrared and gas column density maps. These templates provide worse fits for M31 and M33, with respect to data-driven templates, leaving more residuals in the model.
The templates that best explain the data are infrared maps that trace the emission from the stellar bulge, meaning that most of the $\gamma$-ray flux that we observe comes probably from this component.

If we interpret the flux from M31 and M33 using only DM we have DM candidates with a mass around 20-50 GeV and a cross section that is close to the thermal one for M31 and is around $10^{-25}$ cm$^2$/s for M33 and the {\tt MED} DM model. All the DM candidates found with the {\tt MED} and {\tt MIN} DM models and for $b\bar{b}$ and $\tau^+\tau^-$ annihilation and decay channels are ruled out by the current limits found from the MW dwarf spheroidal galaxies \cite{Fermi-LAT:2016uux}.

Finally we made the more realistic assumption that the flux from M31 and M33 comes at least partially from the galaxy. We use in this case the disk template for M31 and the point source morphology for M33 or the templates from infrared or Hydrogen gas column densities for M31.
We do not find any excess for the presence of DM for all these cases so we put limits on the annihilation cross section or the decay time that for the {\tt MAX} and {\tt MED} DM models constrain the thermal cross section up to 200 GeV and 70 GeV (50 GeV and 10 GeV) for M31 (M33), respectively.

\begin{figure}[ht]
\includegraphics[scale=0.6]{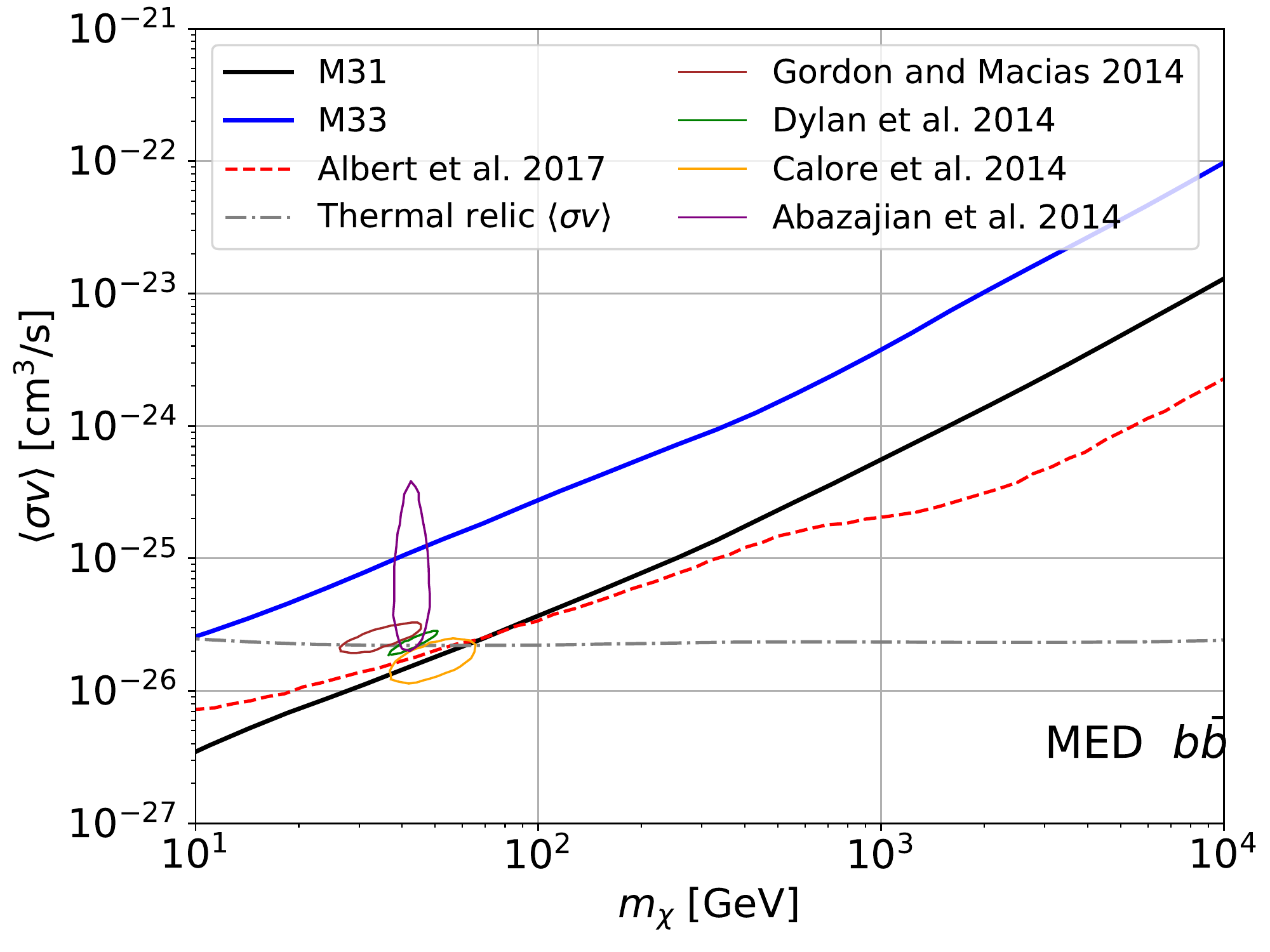}
\caption{Comparison between the 95\% CL upper limits from M31 (black solid line) and M33 (blue solid line) presented in this paper with the limits found from dwarf spheroidal galaxies~\cite{Ackermann:2015zua} (red dashed line). We also show the confidence regions for cross section and mass determined by analyses of the Galactic center excess \cite{Gordon:2013vta,Daylan:2014rsa,Calore:2015,Abazajian:2014fta}. The M31 and M33 upper limits are based on the {\tt MED} DM model (see Sec.~\ref{sec:dm}). The horizontal dashed line shows the thermal relic cross section~\cite{Steigman:2012nb}.
\label{dmCompare}}
\end{figure}

In Figure \ref{dmCompare}, we compare our results with different limits set by other studies for the $b\bar{b}$ channel.
We see that the {\tt MED} DM model is able to constrain the DM interpretation of the GC excess and that our limits for M31 are similar to the ones derived with Dwarf Spheroidal Galaxies up to about 1 TeV.

\begin{acknowledgments}

The \textit{Fermi} LAT Collaboration acknowledges generous ongoing support
from a number of agencies and institutes that have supported both the
development and the operation of the LAT as well as scientific data analysis.
These include the National Aeronautics and Space Administration and the
Department of Energy in the United States, the Commissariat \`a l'Energie Atomique
and the Centre National de la Recherche Scientifique / Institut National de Physique
Nucl\'eaire et de Physique des Particules in France, the Agenzia Spaziale Italiana
and the Istituto Nazionale di Fisica Nucleare in Italy, the Ministry of Education,
Culture, Sports, Science and Technology (MEXT), High Energy Accelerator Research
Organization (KEK) and Japan Aerospace Exploration Agency (JAXA) in Japan, and
the K.~A.~Wallenberg Foundation, the Swedish Research Council and the
Swedish National Space Board in Sweden.
 
Additional support for science analysis during the operations phase is gratefully
acknowledged from the Istituto Nazionale di Astrofisica in Italy and the Centre
National d'\'Etudes Spatiales in France. This work performed in part under DOE
Contract DE-AC02-76SF00515.

X.H. acknowledges supports by the National Natural Science Foundation of China (NSFC-11503078, NSFC-11661161010 and NSFC-11673060) and the “Light of West China” Program of the Chinese Academy of Science.

Resources supporting this work were provided by the NASA High-End Computing (HEC) Program through the NASA Advanced Supercomputing (NAS) Division at Ames Research Center.

\end{acknowledgments}


\bibliography{M31_paper}

\end{document}

%% file: fermiStyle.tex
\newcommand{\irf}[1]{\texttt{#1}}

\newcommand{\Fermi}{{\textit{Fermi}}}
\newcommand{\fermi}{\Fermi}

\input{thisPaperStyle.tex}

%% file: thisPaperStyle.tex
\usepackage{color}

\ifdefined\bwfigures
\else
\fi

\newcommand\fdg{\mbox{$~.\!\!^\circ$}} 
\newcommand\dg{\mbox{$~\!\!^\circ$}}